\documentclass[twocolumn]{aastex63}

\submitjournal{ApJ}
\shorttitle{Sonora Cholla}
\shortauthors{Karalidi et al.}
\usepackage{lineno}
\begin{document}

\title{The Sonora Substellar Atmosphere Models. II. \\
Cholla: A Grid of Cloud-free, Solar  Metallicity Models in Chemical Disequilibrium for the \emph{JWST} Era}

\author{Theodora Karalidi}
\affil{Department of Physics, University of Central Florida, 4111 Libra Dr, Orlando, FL 32816, USA \textit{tkaralidi@ucf.edu}}

\author{Mark Marley}
 \affil{Lunar \& Planetary Laboratory, University of Arizona, Tucson, AZ, 85721, USA}
\author{Jonathan J. Fortney}
\affil{Department of Astronomy, University of California Santa Cruz, 1156 High Street, Santa Cruz, CA 95064, USA}
\affil{Other Worlds Laboratory, University of California, Santa Cruz, CA 95064, USA}

\author{Caroline Morley}
 \affil{Department of Astronomy, University of Texas at Austin, Austin, TX 78712, USA}
 \author{Didier Saumon}  
 \affil{Los Alamos National Laboratory, PO Box 1663, Los Alamos, NM 87545, USA} 
\author{Roxana Lupu}
 \affil{BAER Institute, NASA Ames Research Center, Moffett Field, CA 94035, USA}
\author{Channon Visscher}
  \affil{Dordt University, Sioux Center IA; Space Science Institute, Boulder, CO}

\author{Richard Freedman} 
  \affil{SETI Institute, NASA Ames Research Center, Moffett Field, CA 94035, USA}

\begin{abstract}
Exoplanet and brown dwarf atmospheres commonly show signs of disequilibrium chemistry. In the \emph{James Webb Space Telescope} era high resolution spectra of directly  imaged exoplanets will allow the characterization of their atmospheres in more detail, and allow systematic tests for the  presence of chemical species that deviate from thermochemical equilibrium in these atmospheres. Constraining  the presence of disequilibrium chemistry in these atmospheres as a function of parameters  such as their effective temperature and surface gravity will allow us to  place better constrains in the physics governing these atmospheres.  This paper is part of a series of works presenting the Sonora grid of  atmosphere models \citep[][]{marley19, morley19}. In this paper we present a grid of cloud-free, solar metallicity  atmospheres for brown dwarfs and wide separation giant planets with key molecular species such as CH$_4$, H$_2$O, CO and NH$_3$  in disequilibrium. Our grid covers atmospheres with $T_\mathrm{eff}  \in$[500 K,1300 K], $\log g$ $\in [3.0,5.5]$ (cgs) and an eddy diffusion parameter  of $\log K_{zz}=2$, 4 and 7 (cgs). We study the effect of different parameters within the grid  on the temperature and composition profiles of our atmospheres. We discuss their effect  on the near-infrared colors of our model atmospheres and the detectability of CH$_4$, H$_2$O, CO and NH$_3$ using the \emph{JWST}.  We compare our models against existing MKO and Spitzer observations of brown dwarfs and verify the importance of disequilibrium chemistry for T dwarf atmospheres.  Finally, we discuss how our models can help constrain the vertical structure and chemical composition of these atmospheres.

\end{abstract}


\section{Introduction}

With thousands of exoplanets and brown dwarfs detected to date, ``substellar science'' has turned its focus from the mere detection to the characterization 
of these objects. The characterization of the atmospheres of exoplanets and 
brown dwarfs is of prime interest as it holds 
key information on the composition and evolution of the atmosphere and,  
indirectly, of the protoplanetary disk from which it formed. 
The characterization of exoplanet and brown dwarf atmospheres is done
either by detailed comparison of observations to grids of self-consistent 
forward models \citep[e.g.][]{marley12, allard12,allard14}, or by 
MCMC-driven retrievals on these spectra \citep[e.g.][]{line15, burningham17, kitzmann19}.  

Models of atmospheres with equilibrium chemistry 
can provide a good fit to a large number of atmosphere spectra
\citep[e.g.,][]{buenzli15,yang16,kreidberg14}. However, 
a number of exoplanet and brown dwarf spectra suggest that their atmospheres 
are in chemical disequilibrium, with species such as CH$_4$, CO and NH$_3$
being enhanced (CO) or subdued (CH$_4$ and NH$_3$) in comparison to their equilibrium abundance 
profiles \citep[e.g.,][]{saumon06,moses11,barman15}. 
Hotter atmospheres tend to be governed by chemical equilibrium, 
while cooler atmospheres can be in chemical equilibrium in deeper, hotter 
layers and in disequilibrium on the higher, cooler, visible layers due to quenching 
\citep[see review by][]{madhu16}.

Chemical disequilibrium due to quenching was first suggested in the atmosphere 
of Jupiter \citep[][]{prinn77} and later brown dwarfs \citep[][]{fegley96}. Chemical disequilibrium 
has since been observed in the atmospheres of a number of exoplanets and is shown to be ubiquitous for cooler brown dwarfs 
\citep[e.g.,][]{noll97,geballe2001,burgasser06,saumon06,leggett2007b,moses11, miles20}. 
Quenching happens in an atmosphere when transport processes, like convection 
or eddy diffusion, transport molecules higher up in cooler atmospheric layers 
where the chemical reaction times are slower. Due to the rate of transport of the 
molecules being faster than the chemical reaction rates, the bulk composition 
of the atmosphere will still be representative of the deeper, hotter layers, such 
that the abundances of higher, cooler layers will be out equilibrium given local 
pressure and temperature conditions.

There are two particular atmospheric pairs for quenched species that involve 
important molecules in exoplanet and brown dwarf atmospheres: CO-CH$_4$ 
and N$_2$-NH$_3$. An overview of the chemical reactions and intricacies of 
each cycle can be found, for example, in \citet[][]{visscher11,zahnle14} and 
\citet[][and references therein]{madhu16}. Previous studies have 
showed the importance of quenching and the resulting chemical disequilibrium
for a number of observations. 
A previous work that was similar in scope to our own is that of 
\citet[][]{hubeny07}. They studied the effect of disequilibrium chemistry on the 
spectra of L and T dwarfs as a function of gravity, eddy diffusion coefficient, 
and the speed of the CO/CH$_4$ reaction and found that all these parameters 
influence the magnitude of the departure from chemical equilibrium in their 
model atmospheres. Following up on those studies
\citet[][]{zahnle14} performed a theoretical study of CH$_4$/CO and 
NH$_3$/N$_2$ chemistry and highlighted the importance that surface gravity should 
play in the quenching of exoplanet and brown dwarf atmospheres. \citet[][]{zahnle14} also visited the 
influence of the atmospheric scale height and atmospheric structure on quenching. 

A number of papers have modeled quenched atmospheres to study 
the effect of quenching on atmosphere spectra and/or fit observations of 
brown dwarf or imaged exoplanet spectra. Most of these however, calculate the 
spectra of exoplanet atmospheres using atmospheric temperature--pressure 
and composition profiles that are calculated independently of each other (i.e., the 
change of the composition profile does not inform and, potentially, alter the 
temperature--pressure profile of the atmosphere, which in turn may affect the 
composition profile). Models that use a self-consistent scheme to study the effect of 
quenching on atmospheres are still rare. 
\citet[][]{hubeny07} showed that not calculating the atmospheric profiles in
a self--consistent way could result in errors of up to $\sim$100K for a given pressure. Recently, \citet{phillips2020} presented an update of the 1D radiative-convective model 
ATMO that calculates the 
atmospheric profiles of an atmosphere in a self-consistent way.
\citet{phillips2020} applied their updated code to cool T and Y dwarfs and 
showed that quenching affects the atmospheric spectra and proposed that the 
3.5-5.5$\mu$m window can be used to constrain the eddy diffusion of these cool
atmospheres.  Even in the case of hot Jupiters, \citet{drummond2016} showed that for strong chemical disequilibrium cases, not using a self--consistent scheme can also lead to errors of up to $\sim$100K for a given pressure.

\emph{HST} and ground-based observations of exoplanet and brown dwarf atmospheres have allowed us to 
get a first glimpse of their atmospheric composition. \citet{miles20}, e.g., 
presented low resolution ground-based observations of seven late T to Y 
brown dwarfs and showed how  
their observations can constrain the $\log K_{zz}$ of these 
atmospheres when compared with theoretical models. In the \emph{JWST}  era 
the number of exoplanets with high quality spectra will 
increase significantly. \emph{JWST}  observations, in addition to observations with
forthcoming Extremely Large Telescopes (ELTs) on the ground will allow 
the community to study in more detail atmospheric compositions and 
to test for the existence of disequilibrium chemistry in 
atmospheres as a function of atmospheric properties, such as effective 
temperature, surface gravity, metallicity, insolation etc, and allow us 
to place significantly better constrains in exoplanet and brown dwarf 
atmospheric physics. 
The long-wavelength coverage of \emph{JWST} observations will allow 
for the first time the \emph{simultaneous} characterization of multiple 
pressure layers in these atmospheres and enable us to constrain 
changes in atmospheric chemistry and cloud composition with pressure. 
However, the accuracy of our retrievals will depend on the accuracy of our 
models. 

To prepare for this era a grid of model spectra and composition 
profiles is needed that can be used for the characterization of 
exoplanet and brown dwarf atmospheres. Here, we expand on the work of 
\citet[][]{zahnle14} to study the effect of quenching on model 
atmospheres, via the calculation of self-consistent temperature--pressure and 
composition profiles. This paper is part of a series of studies that present
the Sonora grid of atmosphere models. \citet[][]{marley19} presented the
first part of Sonora: a grid of cloud-free, set of
metallicities ([M/H]= -0.5 to + 0.5) and C/O ratios 
(from 0.5 to 1.5 times that of solar) atmosphere models and spectra named Sonora Bobcat. 
\citet[][]{morley19} will present the extension of the Sonora grid to cloudy 
atmospheres. 
In this paper we present the extension of the Sonora grid to cloud-free, solar 
metallicity atmospheres in chemical disequilibrium.
We have adapted our well-tested radiative-convective equilibrium atmospheric structure code that is previously described in, e.g., 
\citet[][]{marley02,fortney05,fortney08,marley12,morley12}, to model atmospheres in
chemical disequilibrium due to quenching. As our follow on to our first generation grid , Sonora Bobcat (Marley et al., submitted), we name this grid Sonora Cholla.

In Cholla we focus on these important and interconnected atmospheric abundances: 
CO-CH$_4$-H$_2$O-CO$_2$ and NH$_3$-N$_2$-HCN to be in disequilibrium in 
our model atmospheres. To define the disequilibrium volume mixing ratios 
of our atmosphere we followed the treatment of \citet{zahnle14} 
(for more details see Sect.~\ref{sect:code}). 
We present models for atmospheres with 
effective temperatures ranging from 500~K to 1300~K, log $g$=[3.0,5.5] (cgs) and 
for various values for the eddy diffusion coefficient. The extension of the 
grid to cloudy atmospheres and different metallicities 
will be part of future work. Our grid models and spectra 
are given open-access to the community as an extension of the Sonora 
grid \citep[][]{marley18}.

This paper is organized as follows: In \S \ref{sect:code} we discuss the
updates to our code and its validation. In \S \ref{sect:thegrid}
we present results from our grid of quenched atmospheres. We then discuss the 
effect of quenching on the temperature--pressure profile
(\S \ref{sec:kzz_ts}), composition profiles (\S \ref{sect:kzz_cmps}), and near-infrared colors 
(\S \ref{sect:colors}) of our model atmospheres. We follow by discussing the 
effect of quenching on the detection of CH$_4$, CO, H$_2$O and NH$_3$ by 
\emph{JWST}  (\S \ref{sect:detect}) and the colors of our model atmospheres (\S \ref{sect:colors}). 
Finally, in \S \ref{sect:discussion} we discuss the importance of quenching in exoplanet and brown dwarf atmospheres and how JWST will improve our understanding 
of chemistry changes in atmospheres and present our conclusions.

\section{The code} \label{sect:code}

We used the atmospheric structure code of M.~S.~Marley and collaborators, as 
described in a variety of works  
\citep{marley96,marley02,marley12,fortney05,fortney08,morley12,marley19} which follows 
an iterative scheme to calculate self-consistently the temperature--pressure profile, 
composition profiles, and emission spectra of a model atmosphere. The code uses the 
correlated k-method to calculate the wavelength dependent gaseous absorption of the 
atmosphere \citep[e.g.,][]{goody89}. 
In its original version the k-absorption tables the code uses are `premixed',
i.e., the mixing ratio of different species is specified for a given
(pressure, temperature) point. Modeling an atmosphere in chemical 
disequilibrium with such a scheme requires the 
re-calculation of k-absorption tables for every quenched 'premixed' atmosphere.

\subsection{Adaptations of the code}

We adapted the radiative-transfer code to follow 
the random overlap method with resorting-rebinning of k-coefficients as 
described in \citet[][]{amundsen17}. The advantage of this method is that it 
allows the calculation of the k-absorption coefficients of a variable mix 
of species, with the abundance of every species (and thus its influence 
to the total absorption of a layer) being determined at run time. 
Changes in the temperature--pressure and composition profiles 
of an atmosphere directly inform changes in the absorption of an atmospheric 
layer, and allow the calculation of properties of quenched atmosphere in a 
self-consistent scheme. This in turn requires iteration for the atmosphere 
to converge.

We then allowed the following species to be in chemical 
disequilibrium due to quenching in our model atmospheres: CH$_4$, CO, NH$_3$, 
H$_2$O, CO$_2$, HCN and N$_2$. Quenching of these species was calculated following 
\citet[][]{zahnle14}. At any given point, i.e. for 
every temperature--pressure profile (hereafter TP profile),  
on the iterative scheme, the code calculates the quenching level for every  
species (CO, CH$_4$, H$_2$O, CO$_2$, NH$_3$, N$_2$, HCN)  following 
\citet[][]{zahnle14}, and quenches 
the composition profiles of each species accordingly.  
In particular, for every set of species we calculate for each layer of our 
model atmosphere the mixing timescale $t_\mathrm{mix}$ which depends on the eddy 
diffusion coefficient $K_{zz}$, and the chemical reaction rate of the 
species $t_\mathrm{species}$ which depends on the pressure and temperature 
of the layer (see \citet[][]{zahnle14}). The deeper layer for which $t_\mathrm{mix} 
< t_\mathrm{species}$ 
is set as the quenching level as above that the reaction rate is slower than 
convection and the latter takes over in the atmosphere. The composition of the 
species for each layer at pressures lower than the quenching level 
are kept constant at the value of the quench level. 
The quenched profiles are then used in the radiative transfer scheme for 
the next iteration. 
In the current version of the code we have used a simple profile for the 
eddy diffusion coefficient $K_{zz}$, keeping it constant at the noted 
model value for all levels. We have set the code up in a modular 
way such that in future iterations we will be able to use variable with 
pressure $K_{zz}$ profiles (see Sect.~\ref{sect:discussion}).

Finally, in its original version the code performs the 
radiative-transfer calculations using 196 wavelength bins, chosen 
appropriately to take into account the major 
spectral features in an atmosphere. As we discuss in 
Sect.~\ref{sec:res_influence}, we adapted the number of bins to 661 to 
increase the accuracy of our code whilst mixing molecules.

We note that members of our group have previously explored the effect of 
disequilibrium 
chemistry in the atmospheres of brown dwarfs and exoplanets \citep[e.g.,][]{saumon07, visscher11, burningham11, visscher11, moses2016}. 
For instance \citet{saumon07,burningham11} explored the effect of 
disequilibrium chemistry in the atmospheres of Gliese 570D, 2MASS J1217-0311,
2MASS J0415-0935 and Ross 458C. These papers used a simpler, but not fully self-consistent method described in 
\citet{saumon06} to calculate the quenched volume mixing ratios of 
an atmosphere.  Self-consistent radiative-convective atmosphere models were 
calculated, assuming equilibrium chemistry, and then non-equilibrium chemical 
abundances were explored later, varying the strength of vertical mixing.  
While this method proved extremely useful for exploring the phase of disequilibrium 
chemistry and often provided good fits to observations, the calculation of the 
quenched volume mixing ratios is not performed in a self--consistent way in 
the radiative transfer scheme. Meaning, the altered chemical abundances did 
not feed back into changes in the atmospheric radiative-convective 
equilibrium TP profile. Here, we updated our code to calculate the properties
of an atmosphere with disequilibrium chemistry in a fully self-consistent way.

Finally, we note that various teams have explored different ways to define 
the quenching levels in an atmosphere and their effect on atmospheric chemistry \citep[e.g.,][]{visscher11,visscher2012, drummond2016,tsai2018}. 
While using a full chemical kinetics network might be more accurate for 3D 
general circulation models of irradiated exoplanets \citep[e.g.,][]{tsai2018}, \citet[][]{zahnle14} 
showed that for the atmospheres of interest in this paper the approximation we use is faster than a chemical network and remains accurate. In particular, \citet[][]{zahnle14} compared the 
quenching levels retrieved from a full kinetics model, that uses the entire network 
of chemical reactions possible in these atmospheres, against the Arrhenius-like time scale
$t_\mathrm{species}$ we adopt in this paper. \citet[][]{zahnle14} showed that this  
approximation is valid for self-luminous atmospheres, but 
cautioned that it might not work for strongly irradiated exoplanets. In this paper 
we focus on self-luminous atmospheres and we adopted the \citet[][]{zahnle14} 
approximation. Our code is modular and allows for the future implementation of other quenching schemes to model highly irradiated atmospheres, for example.

%We have set the code up in a modular way such that in future iterations (e.g., if we are interested in modelling highly irradiated exoplanets) we can adopt different schemes to define the quenching levels in our model atmospheres.}

\subsection{Validating the code}\label{sect:validation}

\begin{figure}[]
\centering
\includegraphics[width=\linewidth]{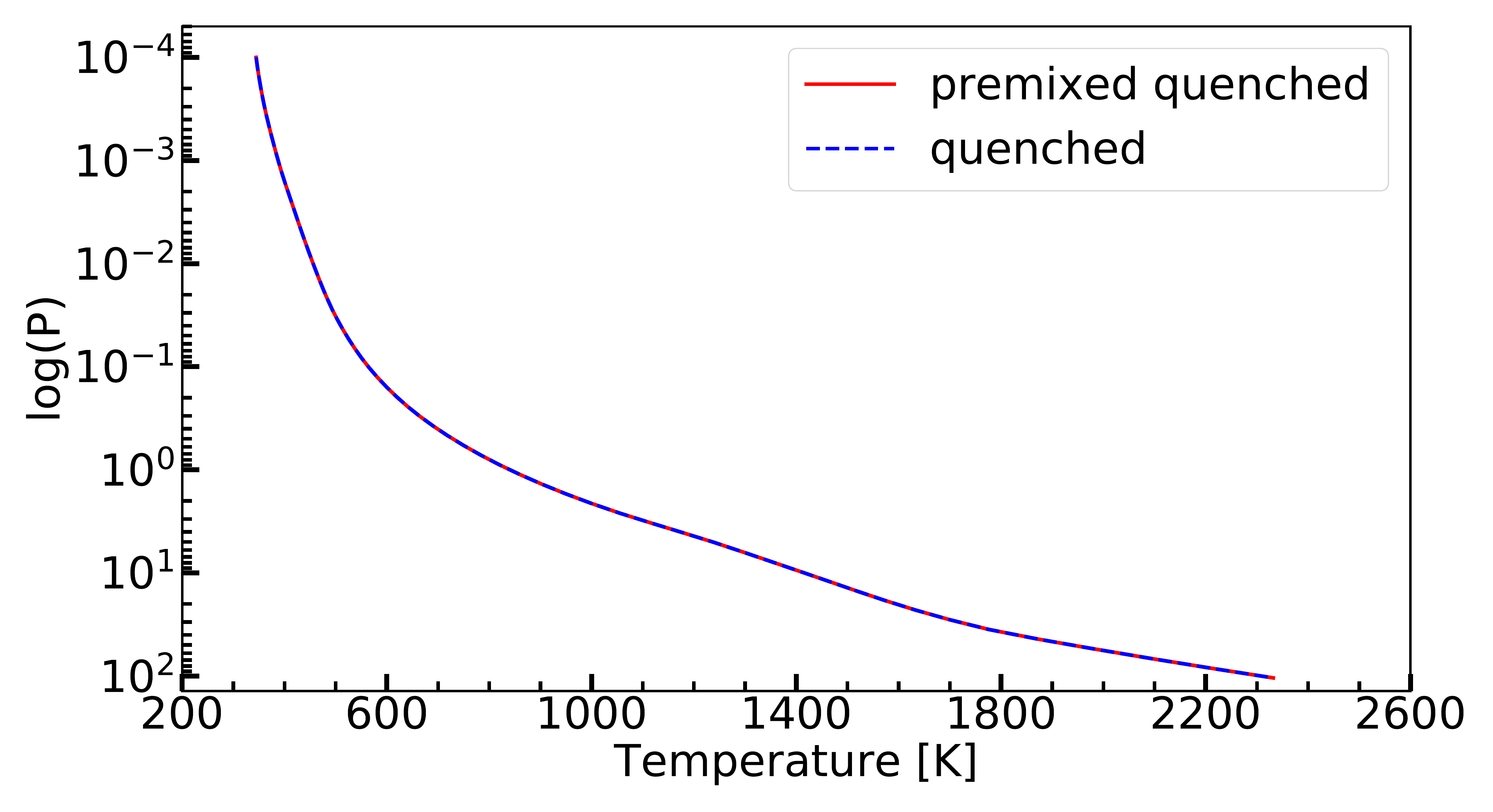}
\centering
\includegraphics[width=\linewidth]{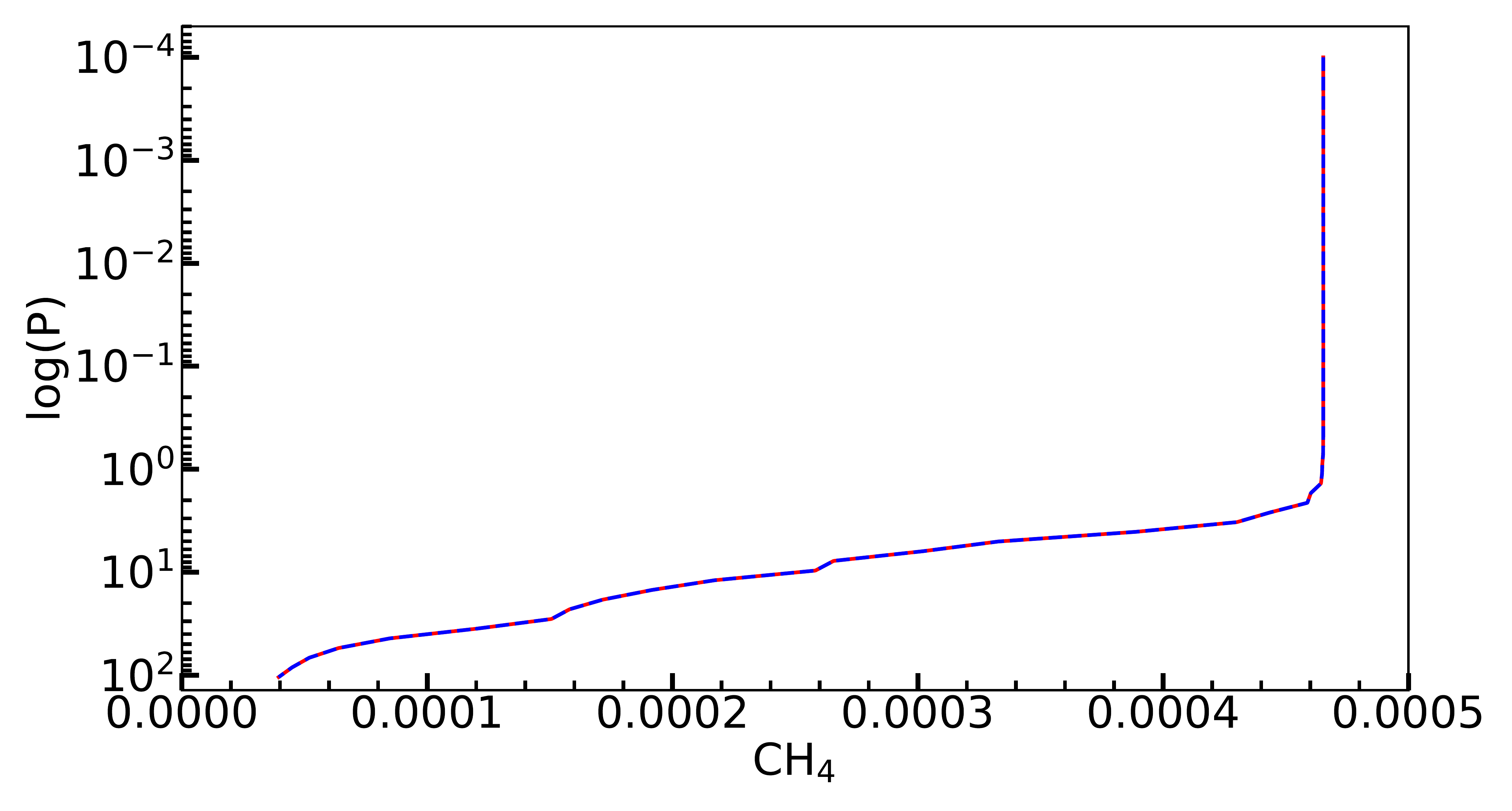}
\centering
\includegraphics[width=\linewidth]{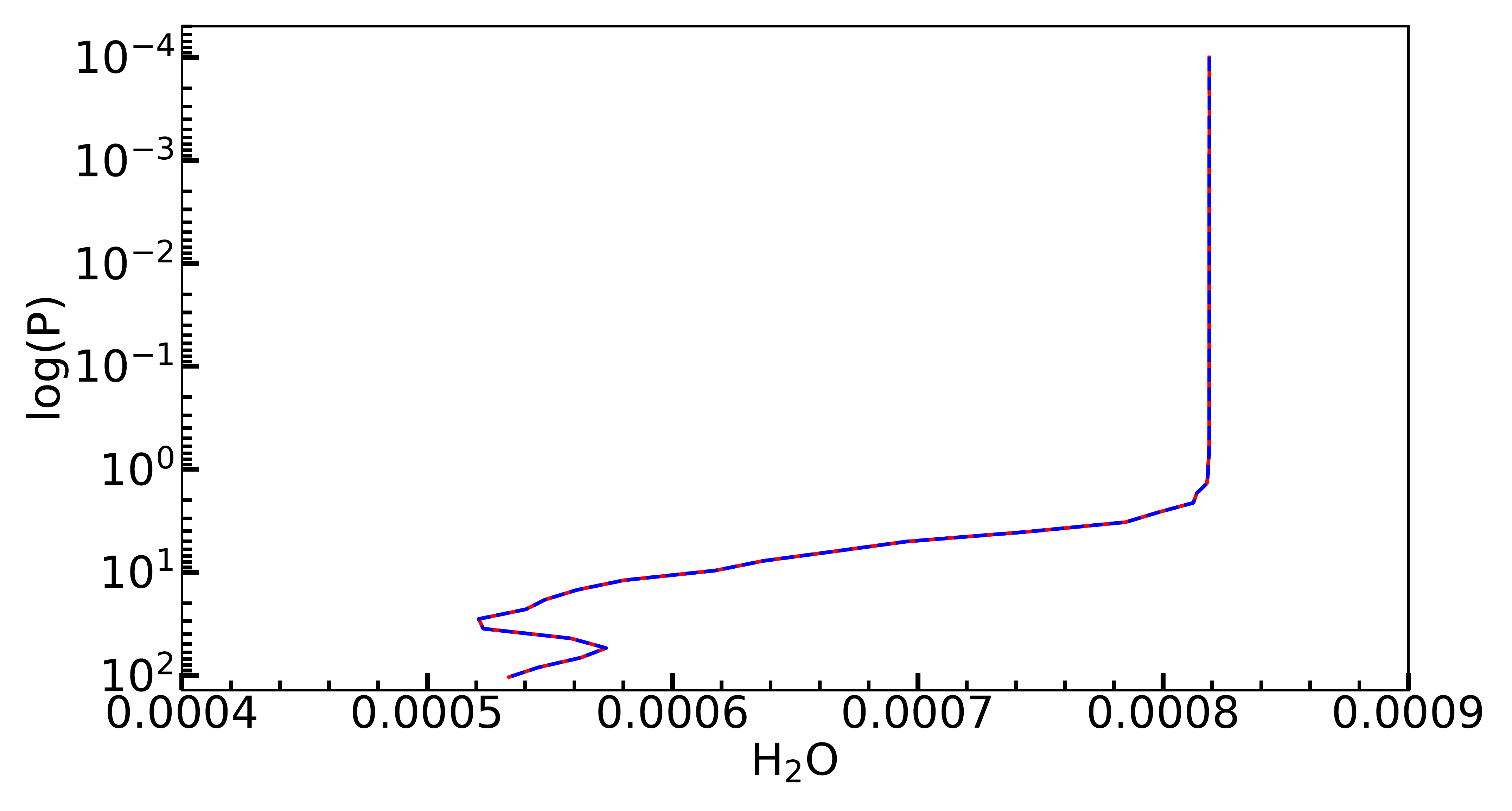}
\caption{Temperature-pressure (TP) (top panel), CH$_4$ mixing ratio (middle panel), and H$_2$O (bottom panel) mixing ratio profiles of a quenched model atmosphere with $T_\mathrm{eff}$=1000~K and 
$\log g$=5.0 calculated by our well-tested premixed code (red, dashed lines) and our new 
quench-enabled code (blue, solid lines). The relative errors between our premixed 
models and our new quenched models are $\lesssim10^{-4}$.}
\label{fig:tests_1}
\end{figure}

We validated the code against its well-tested premixed version for 
atmospheres with different compositions and at different effective
temperatures and gravities. 
We tested the effect of mixing the k-coefficients at run time on the TP profile versus the `premixed' k-coefficients. We calculated `premixed' k-coefficients where we kept the abundance of key species at a constant value above their quenching level. We then run the premixed version of the code with these k-coefficients and compared the resulting TP and composition profiles against those of the new code.
In Fig.~\ref{fig:tests_1} we show the TP profile and composition profiles 
for CH$_4$ and H$_2$O of a quenched atmosphere with $T_\mathrm{eff}$=1000~K and $g$=1000 m s$^{-2}$. 
The premixed quenched atmosphere profiles are plotted with the red, dashed lines and our new quenched atmosphere profiles run with the same chemistry with the blue, solid lines.  
The profiles match, with relative errors being $\lesssim10^{-4}$. Our 
quenched models were found to fit 
the well-tested premixed models, with the colder atmospheres having a relative 
larger error 
($\lesssim10^{-2}$ at 400~K) than the hotter atmospheres ($\lesssim10^{-4}$ at 
1000~K). The corresponding errors 
in the produced spectra were also dependent on the temperature of the 
model with the 
hotter quenched atmosphere models showing a better match to the premixed models 
(relative error $\lesssim$ 5\% at 1600~K) than the colder quenched models 
(relative error $\lesssim$10\% at 400~K; see also Fig.~\ref{fig:tests_2}).

Finally, we compared the $T_\mathrm{eff}$ 
of our mixed k-coefficient model atmospheres with the $T_\mathrm{eff}$ of the 
corresponding pre-mixed model atmosphere in the full parameter space 
covered by our models. The relative error in the $T_\mathrm{eff}$ on which our 
models converged was $\lesssim 0.001$\% across the parameter space. 
For example, the absolute error of $T_\mathrm{eff}$ for our 
$T_\mathrm{eff}$ = 650~K models was $\lesssim$0.2~K for all 
$\log g$ and for the $T_\mathrm{eff}$ = 1250~K models it was $\lesssim$0.1~K. 
Thus, the introduction of disequilibrium chemistry in a self-consistent scheme  
does not affect the effective temperature of our model atmosphere. %}

\subsubsection{Model spectral resolution and accuracy of the code}\label{sec:res_influence}

\begin{figure}[]
\centering
\includegraphics[width=\linewidth]{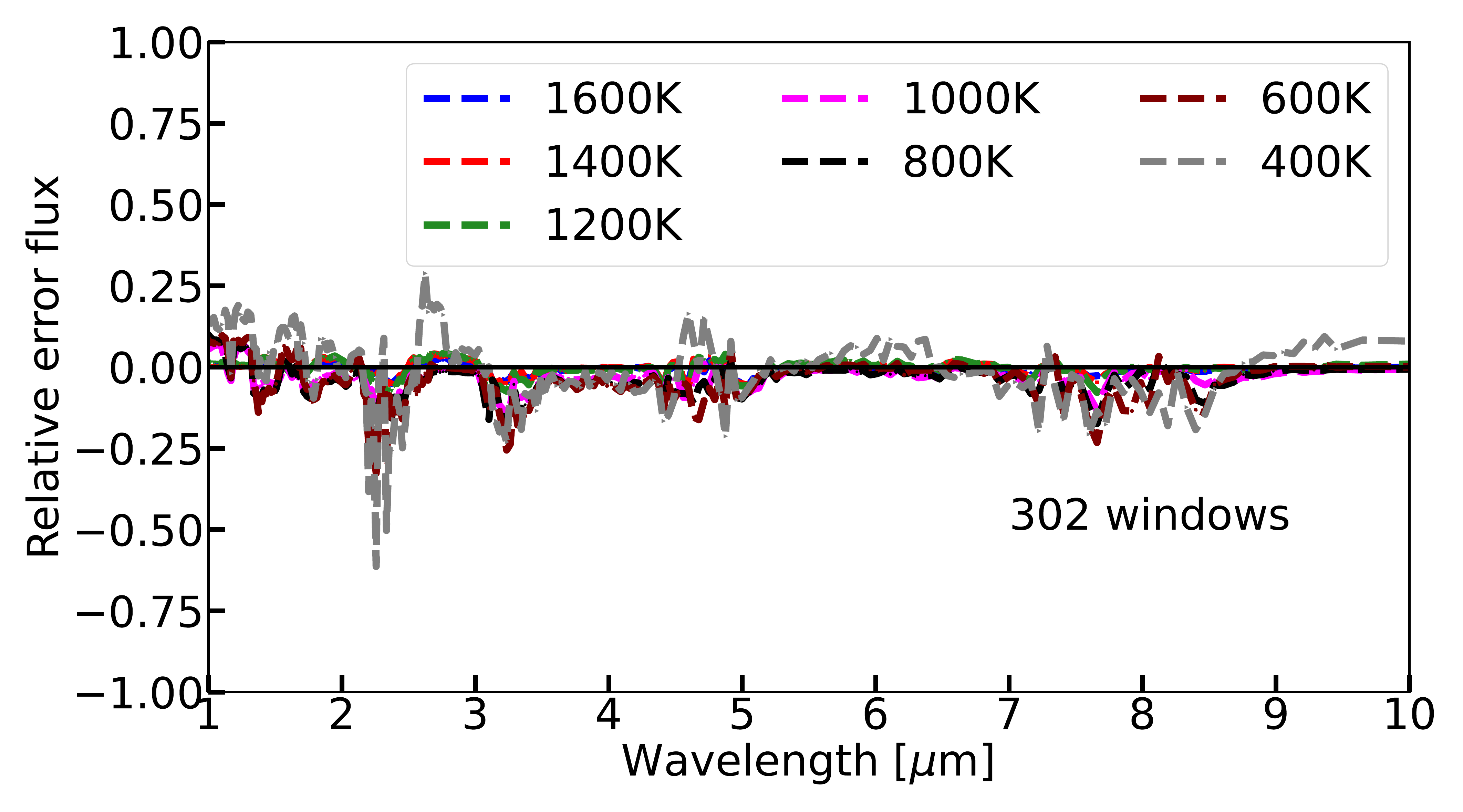}
\includegraphics[width=\linewidth]{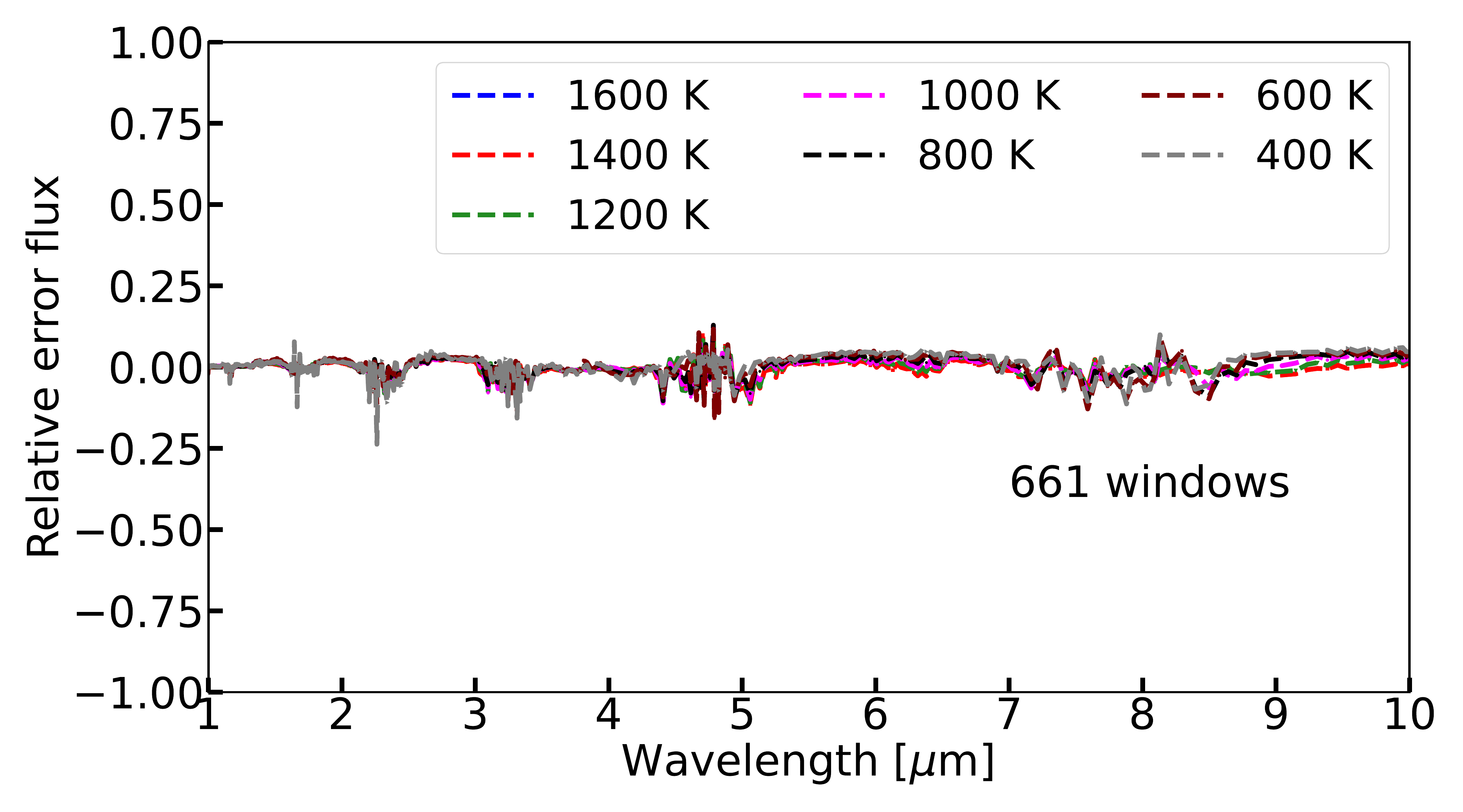}
\caption{Relative flux error between our well-tested premixed model and our 
new quenched model for atmospheres.  The top panel shows, at a fixed gravity (log $g$=5) 
$T_\mathrm{eff}$ values ranging from 400~K to 1600~K for 302 windows.  The bottom panel 
shows the same as the top panel but for 661 windows.
%a 1000 K model stepped through surface gravities from log $g$=4.00 to 5.5, with  661 windows . 
Increasing the number of windows from 302 to 661 windows decreased the average error in 
our model flux (by 0.001 at 1600~K and a factor of 2 at 400~K). 
} 
\label{fig:tests_2}
\end{figure}

\citet[][]{amundsen17} reported that the spectral resolution of the models 
could have an effect on 
the accuracy of the resorting-rebinning of k-coefficients method. Indeed, we found
that using our code's ``native'' resolution of 196 bins resulted in large 
discrepancies in the converged TP and composition profiles and spectra of 
our new code versus those of the well-tested premixed code. Increasing 
the number of bins progressively improved the accuracy (see 
Fig.~\ref{fig:tests_2}). 
Increasing the number of wavelength bins though, increases the time the model 
atmosphere code needs to converge to a solution. We ran a number of test cases at 
different resolutions ranging from 196 to 890 windows and compared the increase 
in accuracy of the converged TP and 
composition profiles and spectra 
with the increase in code running time. Following this procedure we chose to run our grid in a resolution of 661 windows, with the higher density of points typically needed where H$_2$O and CH$_4$ opacities were the largest. 
As can be seen in Fig.~\ref{fig:tests_2} the accuracy of comparison to the
premixed version depends on the temperature of the model atmosphere, 
with the colder quenched atmospheres ($\lesssim$ 600~K) showing a larger 
deviation from the premixed quenched atmospheres (top and middle panels) for the same chemistry.

\subsubsection{Validating our code against previously published results}

\begin{figure*}[]
\centering
\includegraphics[width=.65\linewidth]{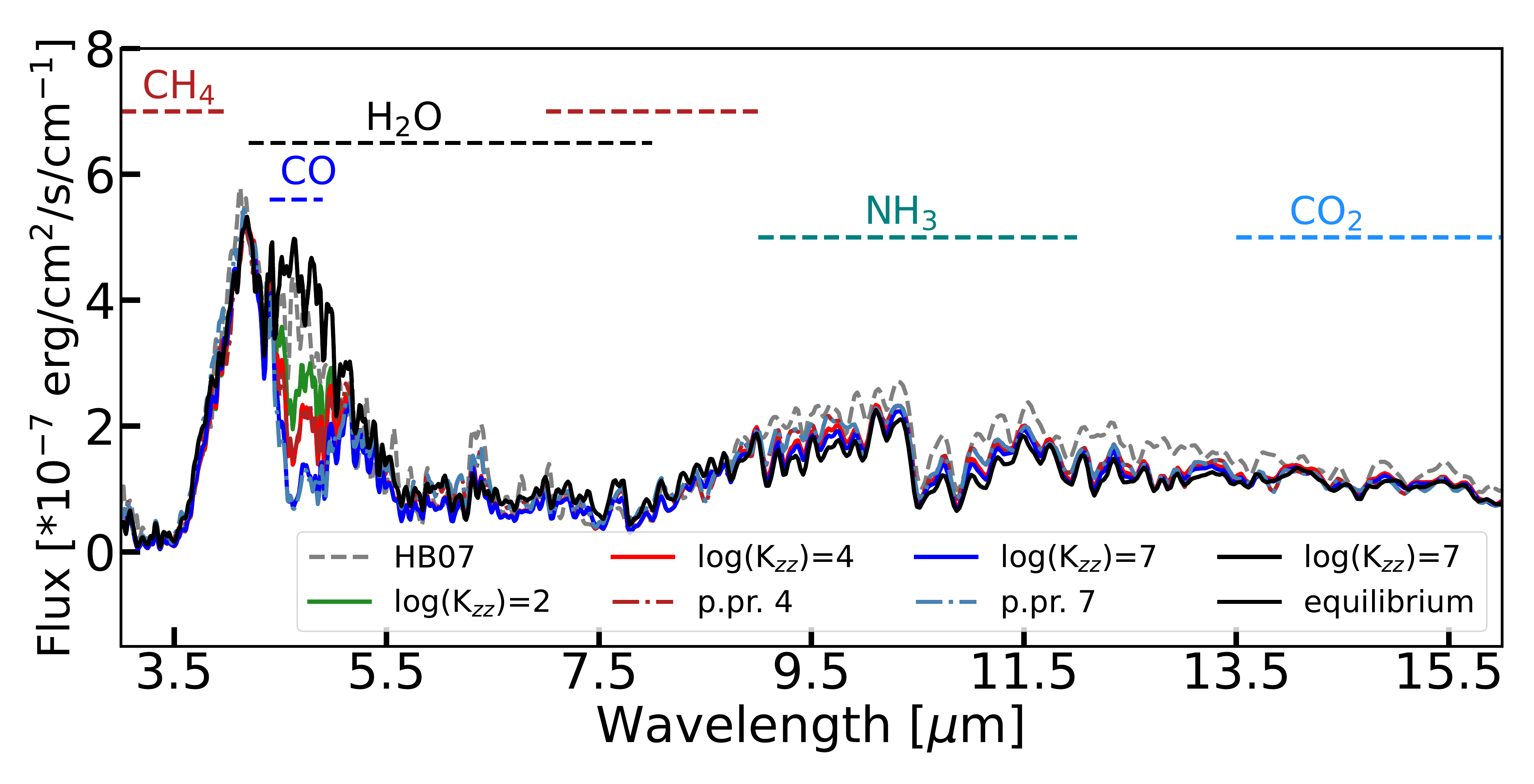}
\centering
\includegraphics[width=.65\linewidth]{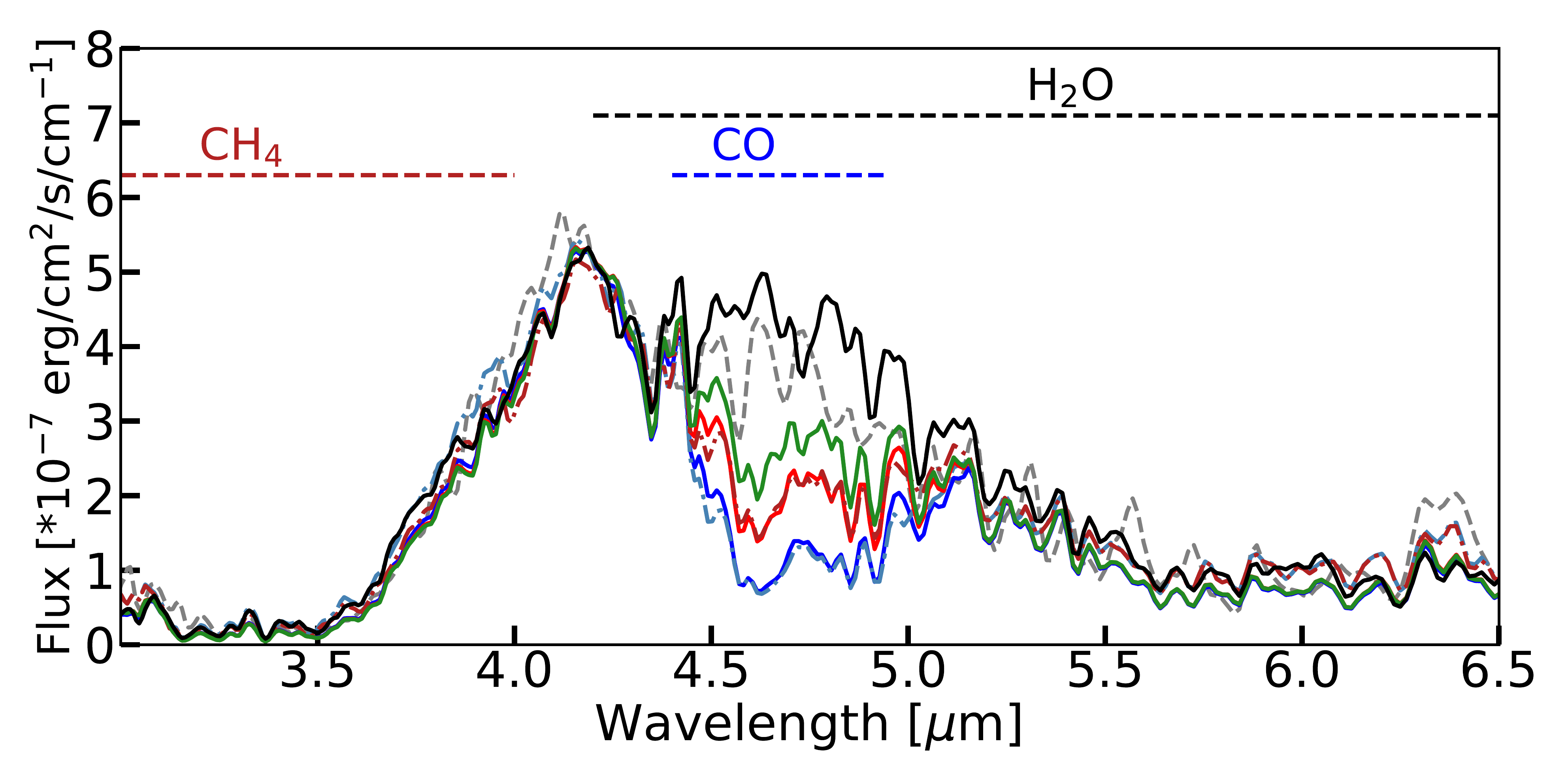}
\centering
\includegraphics[width=.5\linewidth]{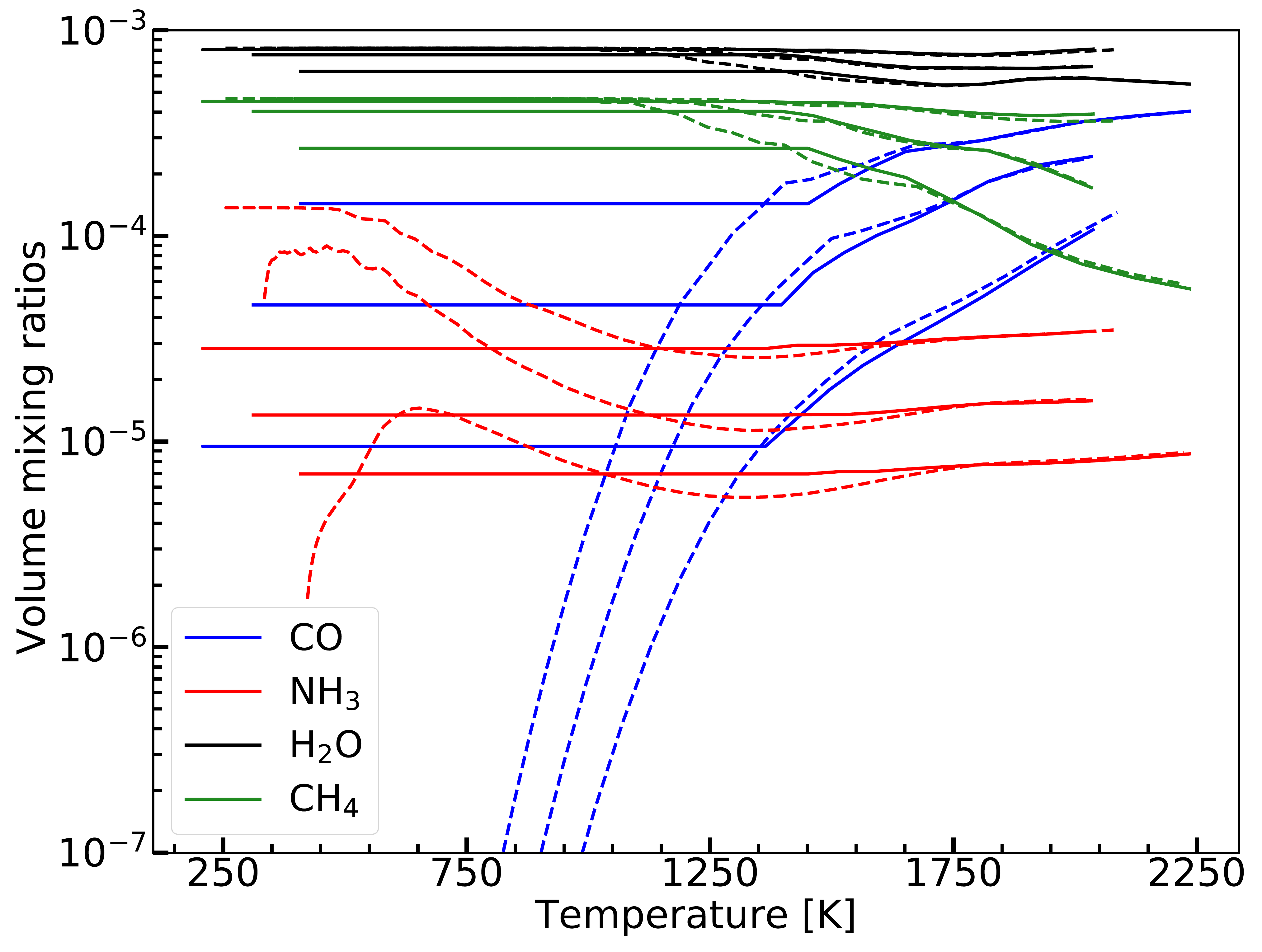}
\caption{Top panel: 3$\mu$m - 16$\mu$m spectra of our 900~K, log $g$=5.5 model 
atmospheres with equilibrium chemistry (black line) and with disequilibrium 
chemistry with $\log K_{zz}$=2 (green line), 4 (red line) and 7 (blue line). We also show the corresponding post-processed disequilibrium spectra for 
$\log K_{zz}$=4 (dark red dashed-dotted line) and 7 (light blue dashed-dotted line) for reference. The spectra were binned down to a resolution of 200 for plot clarity. 
Middle panel: Same as middle panel, but zoomed in 
to wavelengths between 3 and 6$\mu$m. Here, we overplotted the corresponding 
\citet{hubeny07} fast model from: 
\url{www.astro.princeton.edu/$\sim$burrows/non/non.html} (gray dashed line). 
Bottom panel:
Volume mixing ratios for CH$_4$(green lines), H$_2$O (black lines), 
NH$_3$ (red lines) and CO (blue lines) for our disequilibrium 
chemistry models with $\log K_{zz}$=4 (solid lines) 
and equilibrium chemistry models (dashed lines) for atmospheres with $\log g$=5.5 
and $T_\mathrm{eff}$=800~K, 1000~K and 1200~K. Note that the colder a model is,
the lower its minimum temperature is (so the further to the left the line 
extends).}
\label{fig:comp_hb}
\end{figure*}

We also compared our output against published results from other 
groups. In Fig.~\ref{fig:comp_hb} (top panel) we show spectra 
between 3~$\mu$m and 16~$\mu$m for our model atmospheres with
$T_\mathrm{eff}$=900~K, $\log g$=5.5 and for the equilibrium 
chemistry (black line) and disequilibrium chemistry with $\log K_{zz}$=2 
(green line), 4 (red line) and 7 (blue line) cases. For comparison, we also show post-processed disequilibrium spectra \citep[following the same method as in][]{geballe2009} for the $\log K_{zz}$= 4 (dark red dashed-dotted line) and 7 (light blue dashed-dotted line) cases. The post-processed spectra are obtained from dis-equilibrium chemistry computed with the TP structure of the atmosphere in equilibrium. 
The spectra were binned 
down to a resolution of 200 for plot clarity. The middle panel is a zoom-in 
in the 3~$\mu$m to 6.5~$\mu$m region, where we overplotted the  
900\,K, $\log g=5.5$ and $\log K_{zz}= 4$ `fast' model\footnote{\citet[][]{hubeny07} considered two different sets of rate constants for the CO-$\rm CH_4$ conversion; our reaction rates are most similar to their `fast' models.} of 
\citet[][]{hubeny07} (gray, dashed line) from: 
\url{www.astro.princeton.edu/~burrows/non/non.html}. 
These figures are comparable to Figs.12-13 of \citet[][]{hubeny07}. 
Finally, the bottom panel shows the volume mixing 
ratios of our model atmospheres with log $g$=5.5 and $T_\mathrm{eff}$=800~K, 
1000~K and 1200~K, and for the equilibrium chemistry models (dashed lines) 
and the disequilibrium chemistry models with $\log K_{zz}=4$ (in [$\mathrm{cm^2/s}$]; 
solid lines). This figure is comparable to Fig. 2 of \citet[][]{hubeny07}.

Our model atmospheres have comparable volume mixing ratio patterns to the 
model atmospheres of \citet[][]{hubeny07}. However, slight changes in the TP 
profiles during our iterative calculation of the TP and composition profiles 
resulted in slightly lower CO content than the equilibrium models even 
deeper than the CO quenching level, unlike the \citet[][]{hubeny07} models. 
The CO absorption in the 4.7~$\mu$m window (middle panel) increased with 
increasing $\log K_{zz}$ in a comparable way to \citet[][]{hubeny07}. 

Comparing the absolute values of our model spectra and composition profiles 
with the fast model of \citet[][]{hubeny07} (middle panel of our 
Fig.~\ref{fig:comp_hb} and comparison of bottom panel with their Fig. 2), 
we see that our models have more CO and less CH$_4$ than 
the \citet[][]{hubeny07}  fast model, as expected (see also Sect.~4.2 of \citet{zahnle14}). The different opacities used by \citet[][]{hubeny07} also cause differences in the spectra.
Finally, in agreement with  
these authors, and as first pointed out and explained in \citep[][end of section 5.3]{saumon06} for Gl 570D, we found that the NH$_3$ absorption at wavelengths 
$\gtrsim$10~$\mu$m is relatively insensitive to our $\log K_{zz}$, 
over the range of values investigated. 

Finally, comparing our model disequilibrium spectra with the post-processed disequilibrium spectra we note an overall agreement between 3~$\mu$m and 16~$\mu$m. The most prominent difference is in the $\sim$10.5~$\mu$m NH$_3$ feature where our absorption is 30\% deeper than for the post-processed spectra. In Fig.~\ref{fig:gl570d_a} (bottom panel) we show our disequilibrium model spectrum (blue lines) and the corresponding post-processed spectrum (cyan dashed line) for a $T_\mathrm{eff}$=800~K, $\log g$=5.0 model. In this case our $\sim$10.5~$\mu$m NH$_3$ feature is in agreement with the post-processed model. Both our models are cooler than the equilibrium models (and thus the post-processed models) by 30~K to 120~K throughout the atmosphere. At the pressure range probed by the NH$_3$ feature the change in temperature and NH$_3$ content of our atmosphere for the (900~K, 5.5) model is smaller than the change for the (800~K, 5.0) model. This results in a deeper NH$_3$ feature for our (900~K, 5.5) model, which is closer to the equilibrium feature than the post-processed model. Decoupling the temperature-pressure from the atmospheric chemistry calculation as in the post-processed spectra could thus lead to changes in the spectra for some $T_\mathrm{eff}$-$\log g$ combinations and impact our atmospheric characterization.

\subsection{Gliese 570D}\label{sect:gl570d}

Finally, we tested the code against observations of late T-dwarf Gliese 570D. 
The T7.5 dwarf Gliese 570D (hereafter Gl570D) was observed by
\citet{geballe01}, \citet{burgasser04}, \citet{cushing06}, and  \cite{geballe2009} and was 
shown to exhibit evidence of disequilibrium chemistry. Gl570D is now 
considered the archetypal cloud-free brown dwarf atmosphere in disequilibrium. 
\citet[][]{saumon06} fitted the observations of Gl570D and retrieved a 
temperature of 800-820~K and log $g$=5.09-5.23. Using models of atmospheres with 
disequilibrium chemistry \citet[][]{saumon06} obtained a good fit of 
the spectrum of Gl570D, including the 10.5$\mu$m NH$_3$ feature. 
\citet[][]{saumon06} tested different possible sources for the NH$_3$ 
depletion in the atmosphere of Gl570D and concluded that it must  
be due to disequilibrium chemistry. \citet[][]{hubeny07} also showed that 
a 800~K, $\log g$=5.0, $\log K_{zz}$=4 model gave a good fit to the 6~$\mu$m - 
12~$\mu$m window observations.
\citet[][]{line15} performed a retrieval on the Gl570D spectra and 
retrieved a $T_\mathrm{eff}$ of 714$^{+20}_{-23}$K and log $g$=4.76$^{+0.27}_{-0.28}$.

\begin{figure}[]
\centering
\includegraphics[width=\linewidth]{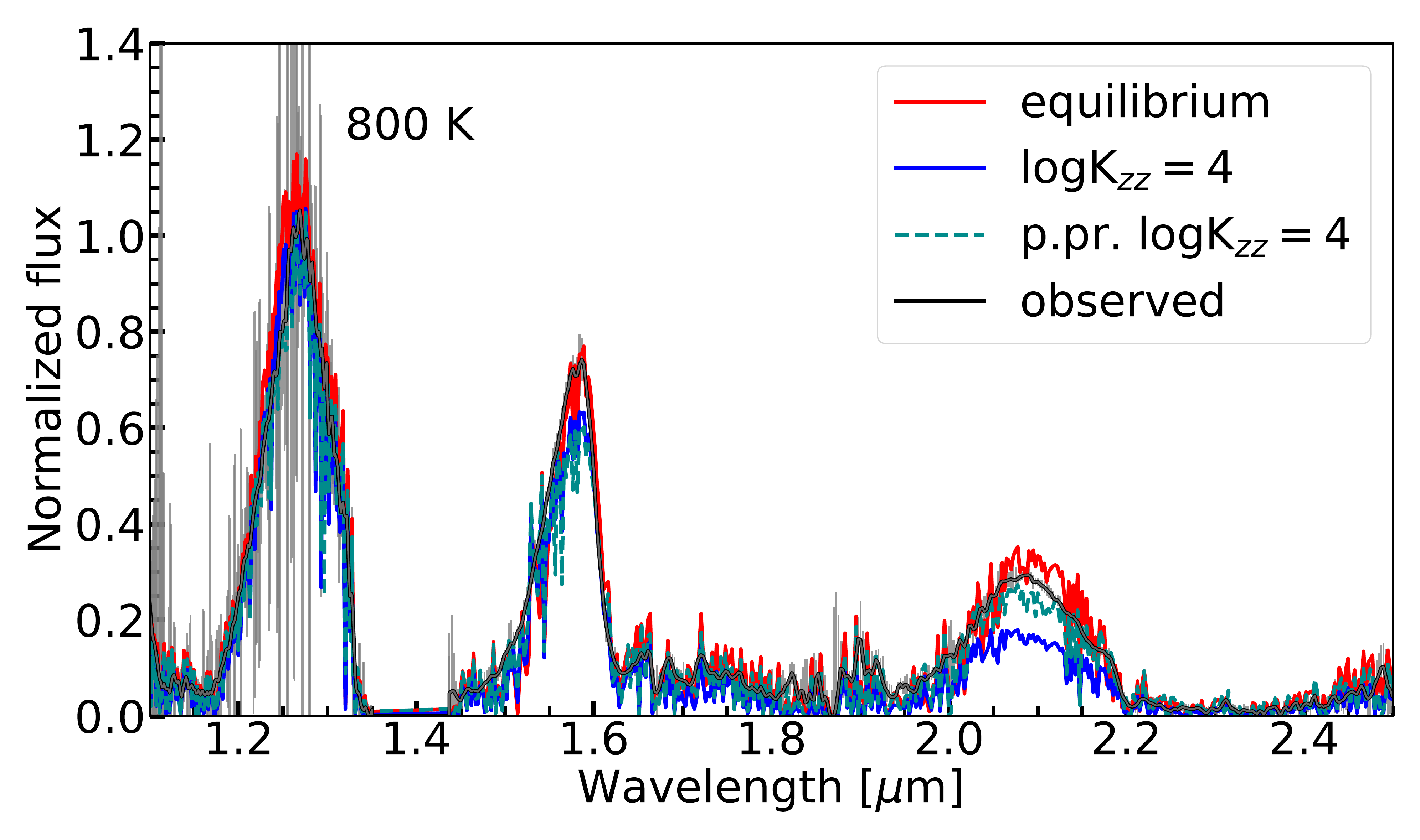}
\centering
\includegraphics[width=\linewidth]{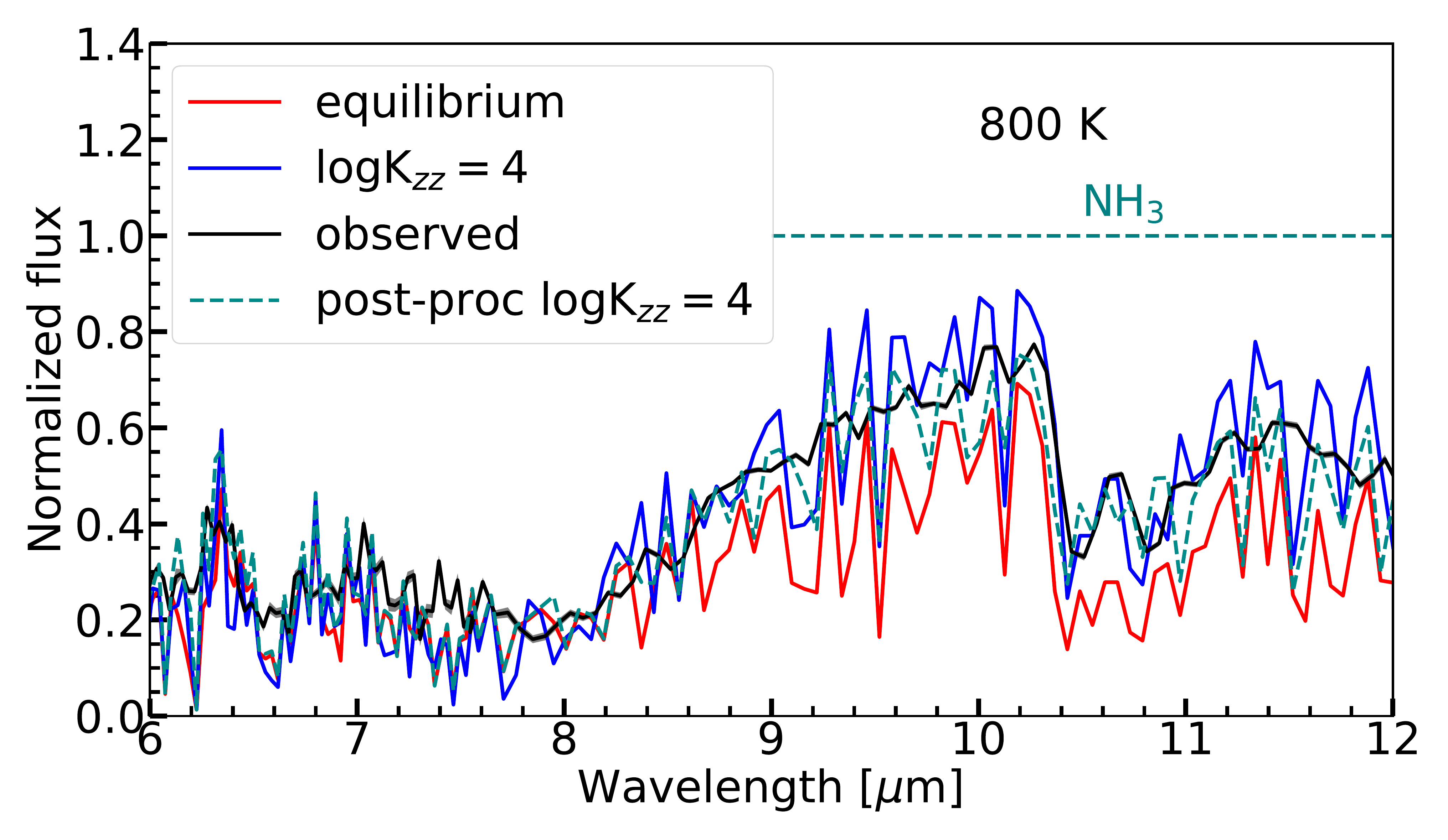}
\caption{Observed spectrum (black solid line) with error bars (gray lines) of Gl570D and 
disequilibrium models for an atmosphere with $T_\mathrm{eff}$=800~K, 
$\log g$=5 representative of the best-fit models of 
\citet[][]{saumon06} and \citet[][]{hubeny07}. The models are 
in equilibrium (red, solid line) or 
disequilibrium with $\log K_{zz}$= 4 (blue, solid line). We also show a post-processed disequilibrium spectrum with $\log K_{zz}$= 4 (cyan, dashed line) for comparison. Our model 
spectra were binned down to  
a variable resolution of 3,000 to 200 comparable to the observations for plot clarity. The error bars are overplotted in both spectra , but the error is too small to be visible in the NH$_3$ feature. }
\label{fig:gl570d_a}
\end{figure}

We fit the spectrum of Gl570D using models at 700~K, 750~K, 800~K and 850~K 
and log $g$=4.5, 4.75, 5.00 and 5.25. In Fig.~\ref{fig:gl570d_a}  
we show  our model spectra at ($T_\mathrm{eff}$, log $g$) = (800~K, 5.0) 
as representative of the best-fit cases of \citet[][]{saumon06} and 
\citet[][]{hubeny07}. 
The top panel shows the NIR Gl570D observations dataset 
(black line), as well as the (800~K, 5.0) 
equilibrium model (red model) and our $\log K_{zz}$=4 model (blue line), 
while the bottom panel shows the mid-IR Gl570D observations dataset. 
We also show a post-processed disequilibrium spectrum for the 
(800~K, 5.0) model atmosphere with $\log K_{zz}$=4 (cyan, dashed line) for comparison.
We binned down our model spectra to a variable resolution comparable to the 
observations for plot clarity.
In agreement with \citet[][]{saumon06} and \citet[][]{hubeny07} 
our (800~K, 5.0) disequilibrium models gave the best fit 
to the observations of Gl570D across the 1.1~$\mu$m - 14~$\mu$m spectrum (smaller $\chi^2$ by a factor of $\sim$3)
with the exception of the 2.0~$\mu$m - 2.2~$\mu$m window, where our disequilibrium models 
were underluminous in comparison to the observations ($\chi^2$ increased by a factor of $\sim$2.2). The (700~K, 5.0) models 
which are representative of the best-fit case of \citet[][]{line15} (not 
shown here) had a poor fit in the 10.0~$\mu$m - 12~$\mu$m window, where NH$_3$ dominates, and a worst overall $\chi^2$ fit (larger by a factor of 1.8) than the 800~K model. The underluminosity of our models in the K-band and a mismatch for wavelengths shortward of 1.1~$\mu$m is due to inaccuracies in our alkali opacities database and, potentially, our CIA and other opacities in the K-band. Addressing these issues is part of ongoing work. The post-processed spectrum provides a good fit to the observations of Gl570D, and is slightly underluminous in comparison to the observations in the 2.0~$\mu$m - 2.2~$\mu$m window, like our self-consistently calculated spectrum. However, the TP profile of the post-processed spectrum is 15~K to 100~K hotter than our self-consistent TP profile in the pressure range probed by the NIR Gl570D spectrum. This could lead to erroneous conclusions about the atmospheric structure when we characterize an atmosphere with post-processed spectra.

\begin{figure}[]
\centering
\includegraphics[width=\linewidth]{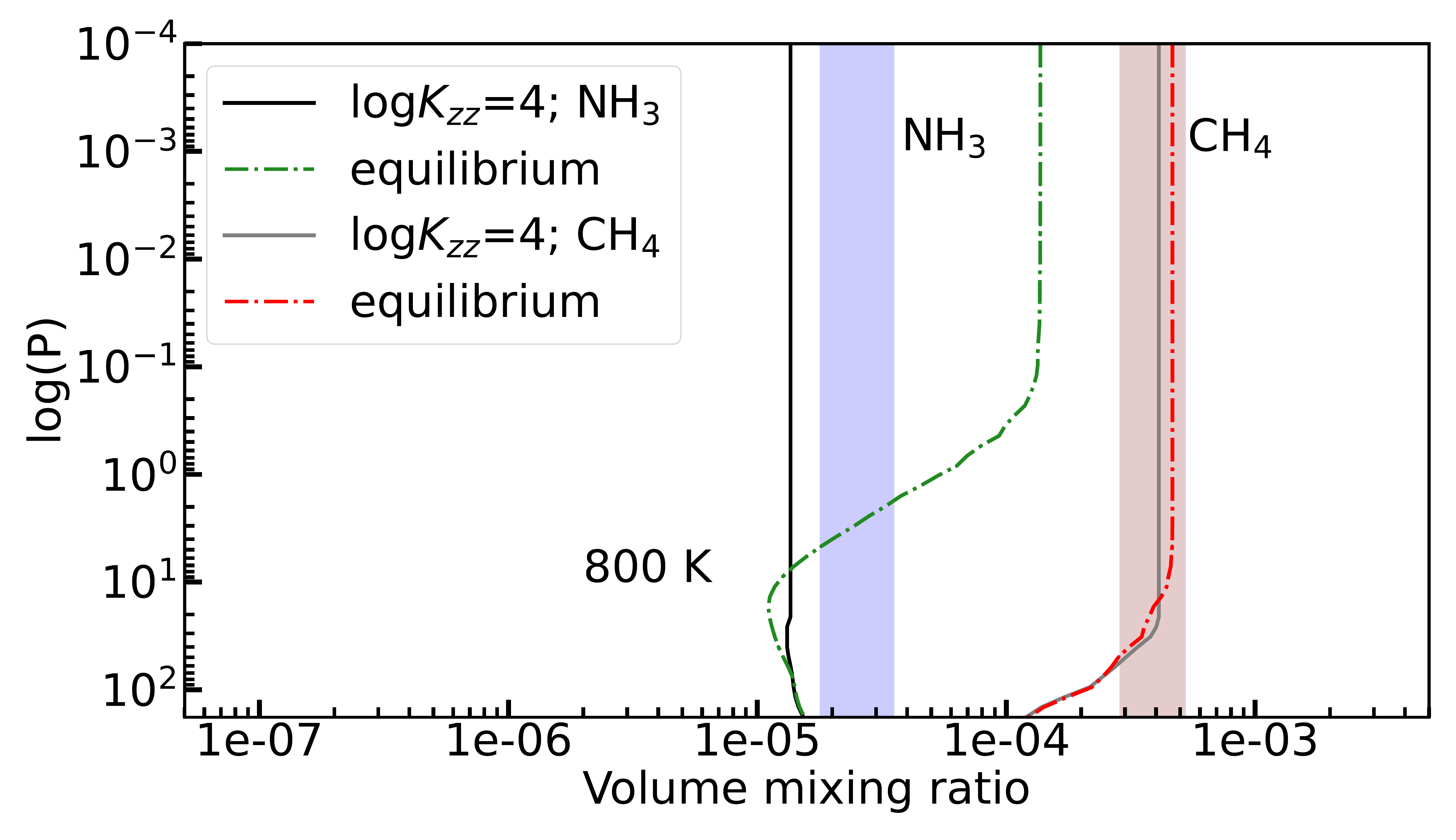}
\centering
\includegraphics[width=\linewidth]{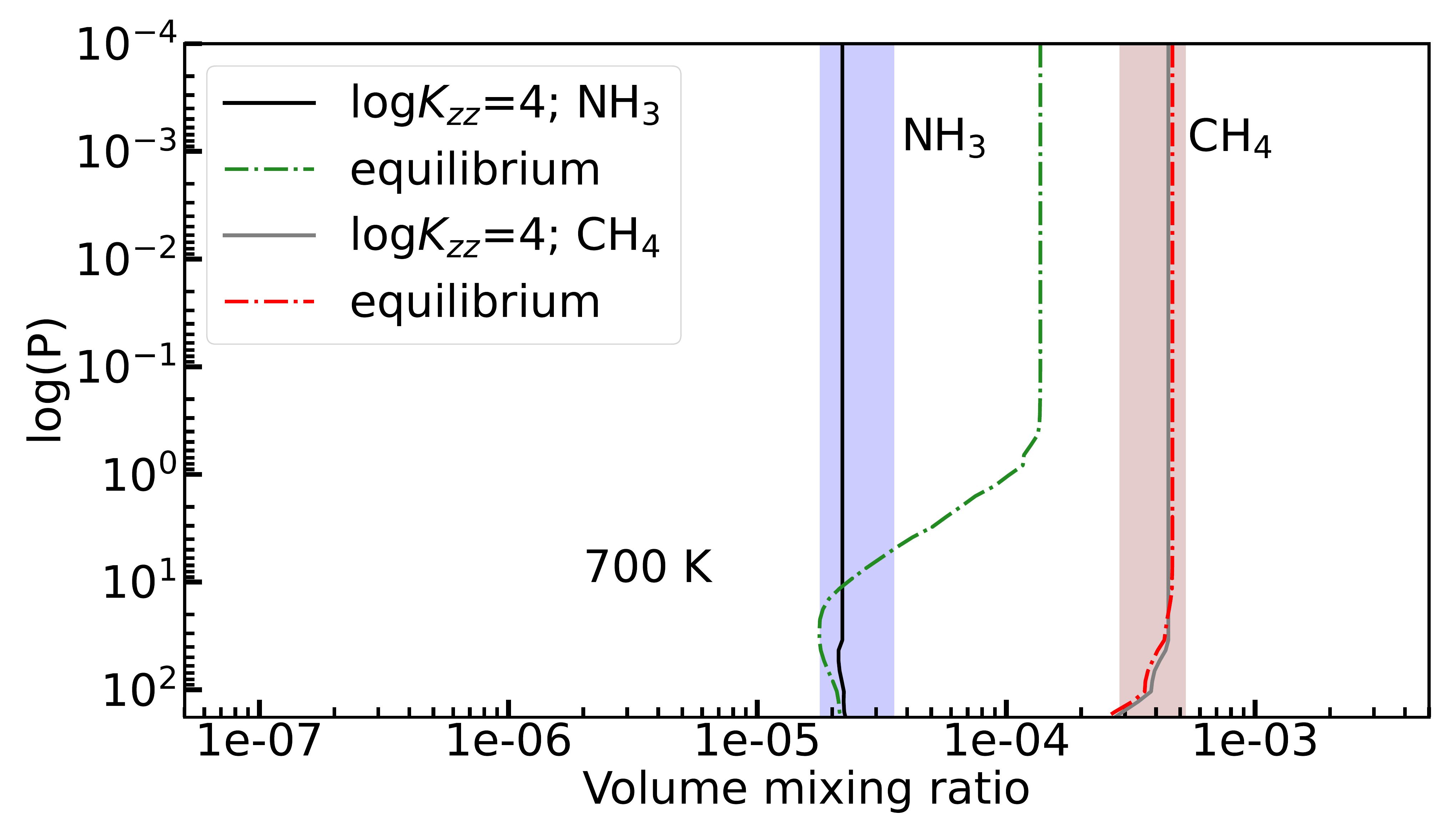}
\caption{Top panel: Volume mixing ratios of CH$_4$ and NH$_3$ 
for our best-fit model for Gl570D with ($T_\mathrm{eff}$, $\log g$) = (800~K, 5.0) (solid 
lines) against the corresponding equilibrium ratios (dashed-dotted lines) and
the retrieved ratios of \citet[][]{line15} (shaded areas). Bottom panel: 
Same as top panel but for the (700~K, 5.0) model, representative of 
the best-fit model of \citet[][]{line15} for Gl570D.}
\label{fig:gl570d_b}
\end{figure}

In Fig.~\ref{fig:gl570d_b} we plot the CH$_4$ and NH$_3$ 
volume mixing ratios of our quenched atmosphere models  
(solid lines) against the corresponding equilibrium ratios (dashed-dotted lines)
and the retrieved ratios of \citet[][]{line15} (shaded areas) for our best-fit 
disequilibrium chemistry model at 800~K (top panel) and the 700~K model (bottom 
panel) that is representative of the best-fit model of \citet[][]{line15}. 
For both models the CH$_4$ content of our model atmospheres was within the 
range retrieved by \citet[][]{line15}. For NH$_3$ however, our best-fit model 
has a lower volume mixing ratio than what \citet[][]{line15} retrieved. 
The NH$_3$ volume mixing ratio for the (700~K, 5.0) disequilibrium model was 
within the range retrieved by \citet[][]{line15}. We note, however, that 
the retrieval of \citet[][]{line15} took into account only the 
1.1-2.3\,$\mu$m spectral range, while our best-fit model was also  
driven by the fit to the strong NH$_3$ feature in the 
10.0-12\,$\mu$m window. This suggests that for retrievals in the 
\emph{JWST}  era the inclusion of the longer wavelength observations will be important to better constrain the NH$_3$ content of an atmosphere.

\section{Quenched Atmospheres Grid} \label{sect:thegrid}

To examine the effects of quenching on our model atmospheres as a function of 
temperature ($T_\mathrm{eff}$), gravity (log $g$) and eddy diffusion parameter 
($K_{zz}$) we have calculated a grid of models from $T_\mathrm{eff}$ 500~K to 1300~K (with steps of 50~K),
log $g$ ranging from 3.0 to 5.5 (with steps of 0.25; cgs) and log $K_{zz}$=2, 4 
and 7 (cgs). For a number of models we also run a case of $\log K_{zz}$=10.
Our model atmospheres are cloud--free and have solar metallicity. The
extension of the grid to cloudy atmospheres, and atmospheres of different
metallicities will be part of future work.

For every model we created an output file for the Sonora grid 
with the TP and composition profiles 
for the following species: H$_2$, He, CH$_4$, CO, CO$_2$, NH$_3$, N$_2$, H$_2$O, TiO,
VO, FeH, HCN, H, Na, K, PH$_3$ and H$_2$S. We then created high resolution 
emission spectra for these models using the radiative transfer code described in 
\citet[][]{morley15}.

In Sect.~\ref{sec:kzz_ts} we study the effect of quenching on 
the TP profiles of the atmospheres, in Sect.~\ref{sect:kzz_cmps} we study the 
effect of quenching on the composition profiles and in Sects.~\ref{sect:detect} 
and~\ref{sect:colors} we study the effect of quenching on the 
spectra and colors of our model atmospheres.

\subsection{TP profiles of quenched atmospheres} \label{sec:kzz_ts}

\begin{figure}[]
\centering
\includegraphics[width=\linewidth]{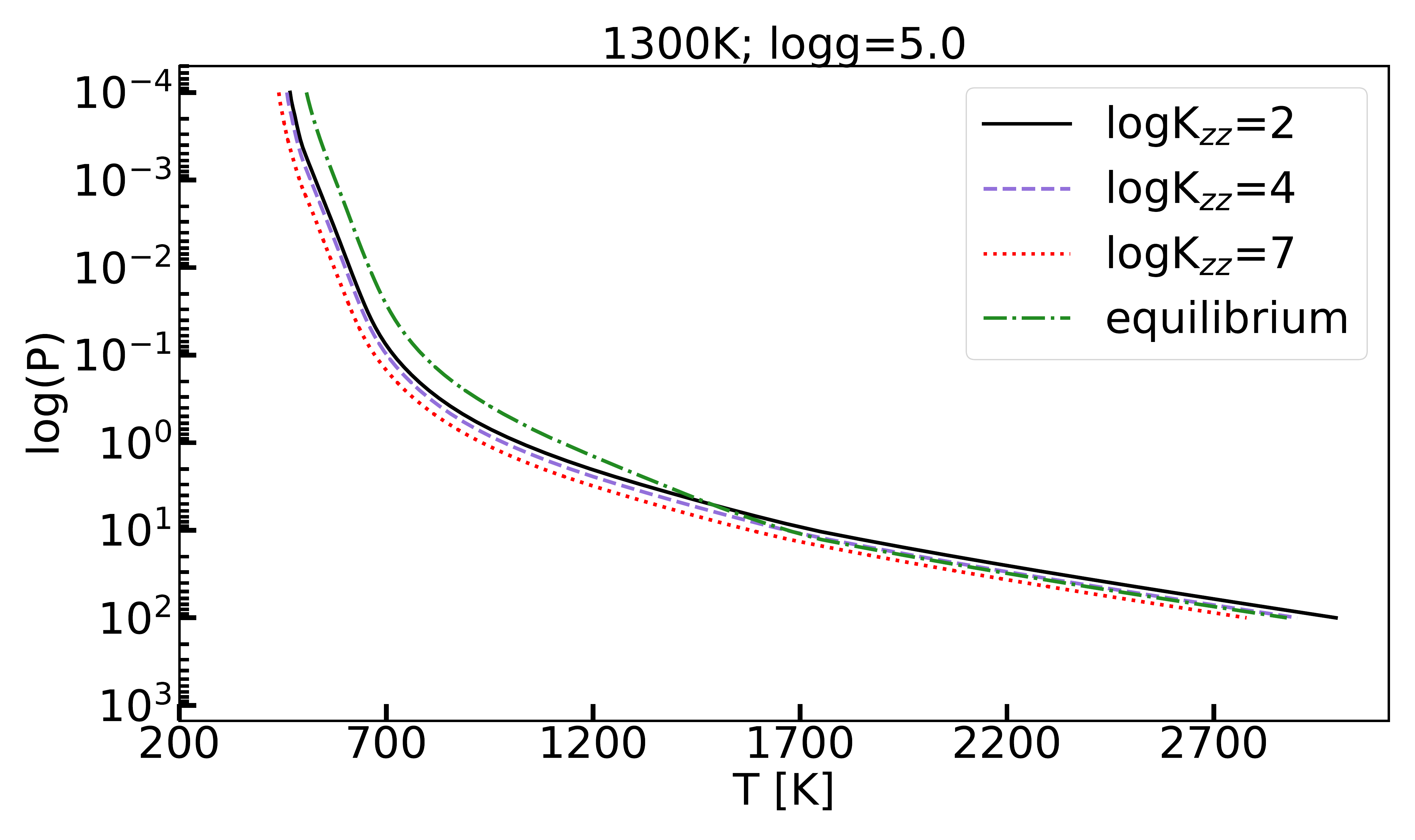}
\centering
\includegraphics[width=\linewidth]{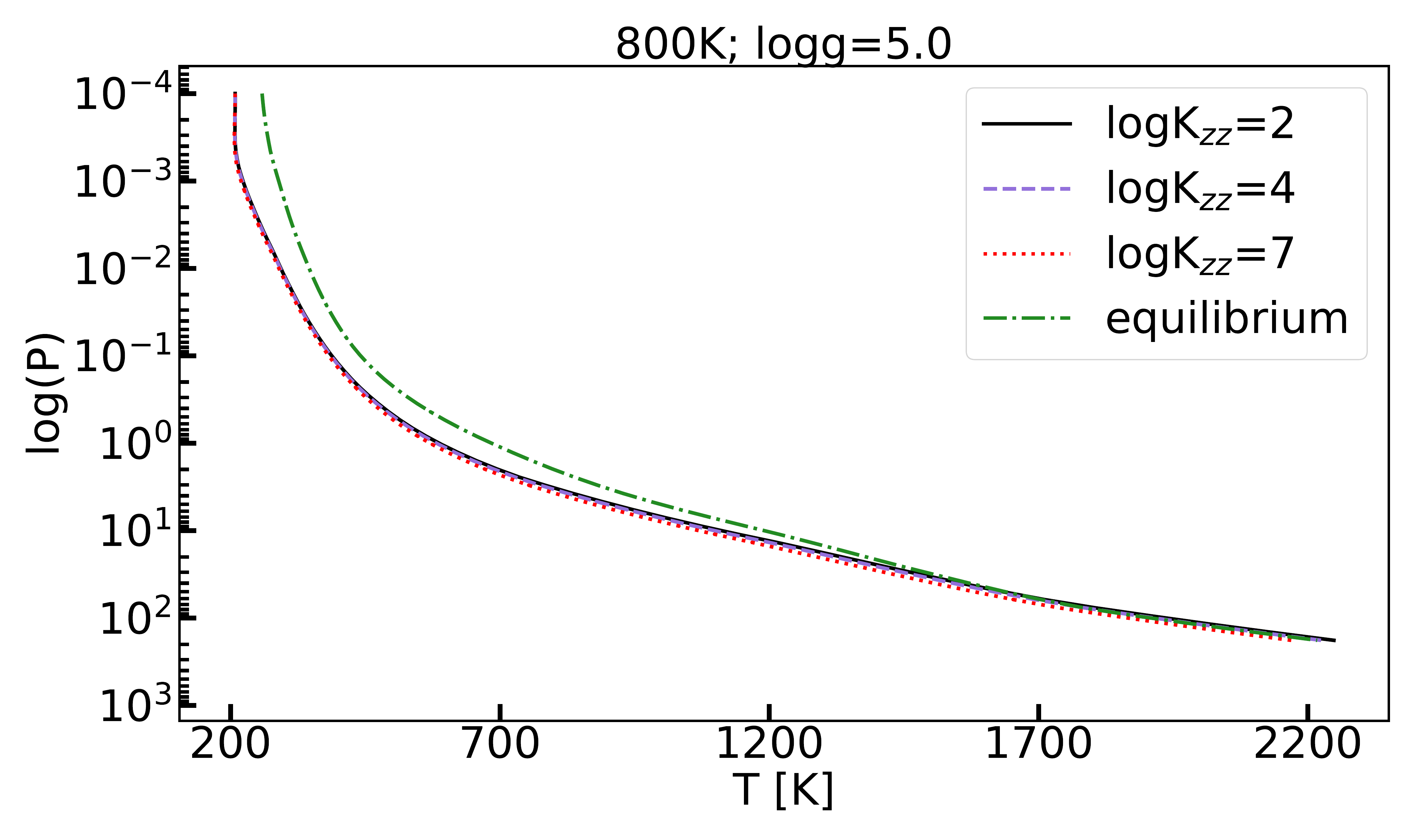}
\centering
\includegraphics[width=\linewidth]{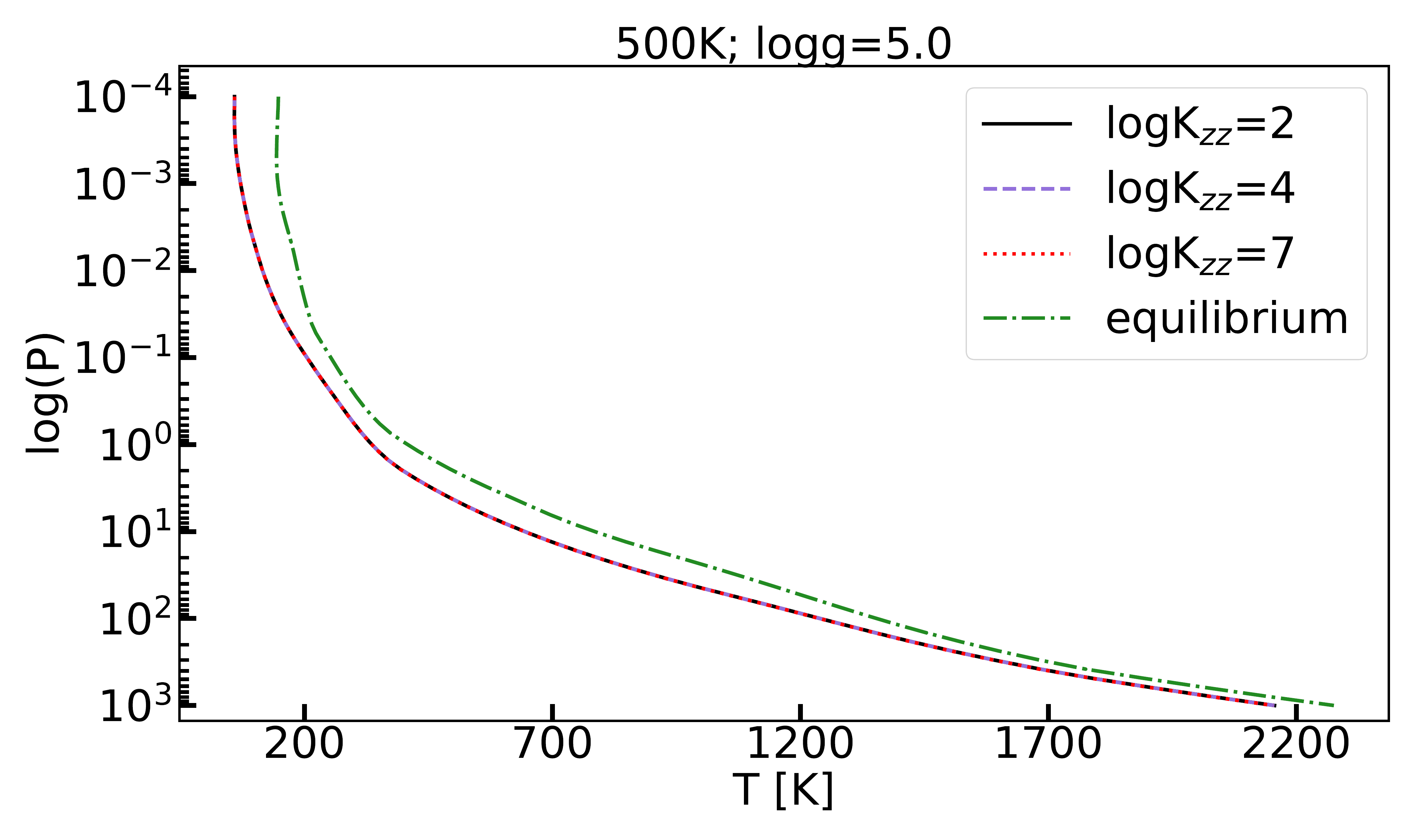}
\caption{Temperature as a function of pressure for models at 
[500~K, 5.00] (bottom panel) and for equilibrium chemistry (green, 
dashed--dotted lines), or disequilibrium with $\log K_{zz}$=2 (black lines), 
4 (purple, dashed lines) and 7 (red, dotted lines).}
\label{fig:tp_allk}
\end{figure}

In Fig.~\ref{fig:tp_allk} we show the TP profiles for 
atmospheres with $\log K_{zz}$=2 (black lines), 
4 (purple, dashed lines) and 7 (red, dotted lines) and 
for the corresponding equilibrium chemistry model atmospheres (green, 
dashed--dotted lines). 
The atmospheres have [$T_\mathrm{eff}$,log $g$] of [1300~K, 5.00] (top panel), 
[800~K, 5.00] (middle panel) and [650~K, 5.00] (bottom panel).

Both $\log g$ and $T_\mathrm{eff}$ 
of an atmosphere affect the influence of quenching to the 
TP profile of the atmosphere. 
For all models, directly above the quenching level of CH$_4$-H$_2$O-CO 
the atmospheres are colder by $\gtrsim$100~K due to the relative change 
in the volume mixing ratio of absorbers such as CH$_4$ and H$_2$O. 
At a pressure of $\sim$7 bar the relative change 
of temperatures were: 18.26\% ($\log K_{zz}$=2)--18.4\% ($\log K_{zz}$=7) 
for the [500~K, 5.00] model, 
7.9\% ($\log K_{zz}$=2)--11.4\% ($\log K_{zz}$=7) for the [800~K, 5.00] 
model, and -2.9\% ($\log K_{zz}$=2)--6.3\% ($\log K_{zz}$=7) for the 
[1300~K, 5.00] model. At a pressure of 0.5 bar the 
corresponding relative changes were: 14.5\%--14.6\%, 13.9\%--16.0\% 
and 10.0\%--17.3\%.

\begin{figure}[]
\centering
\includegraphics[width=\linewidth]{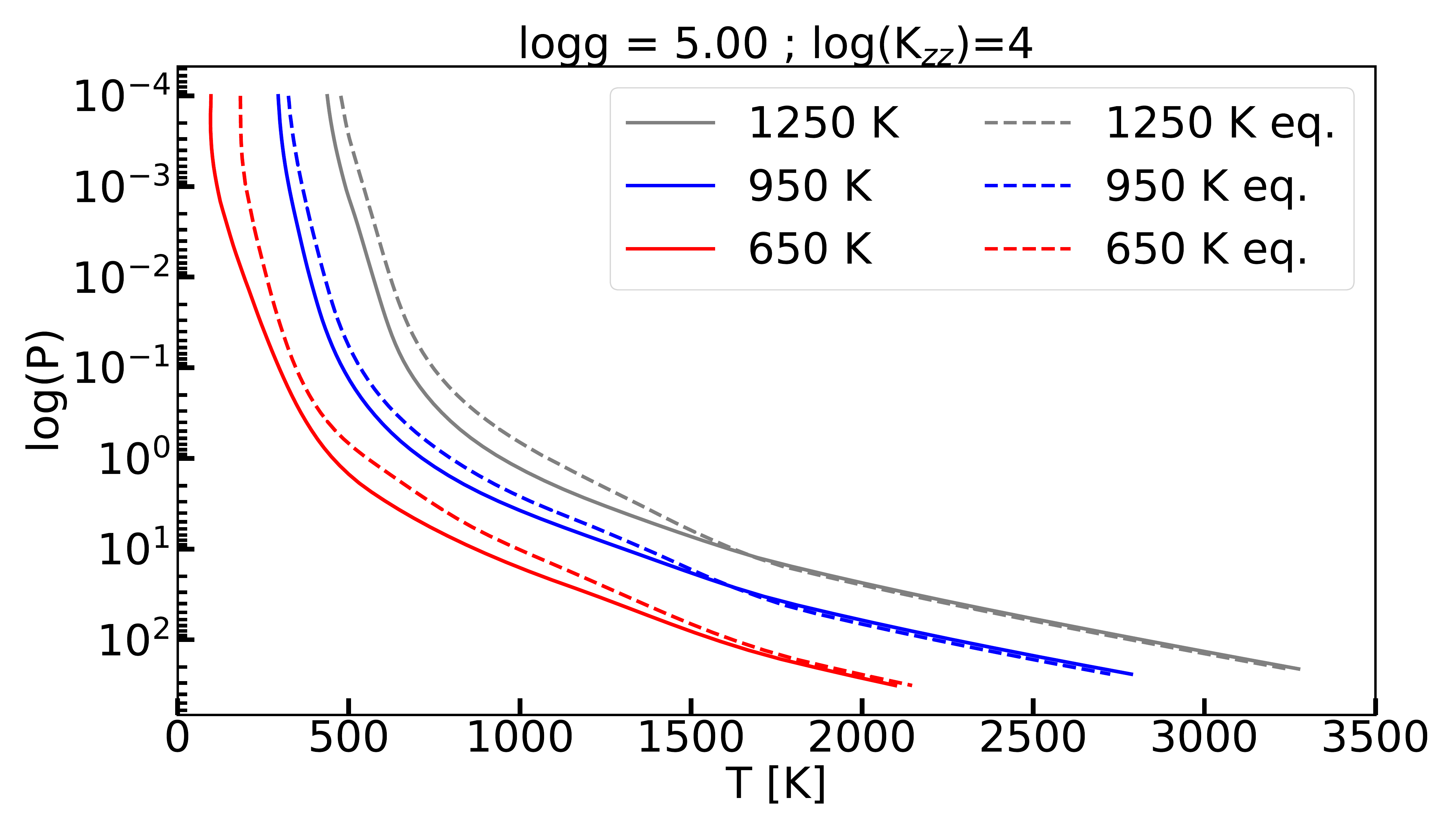}
\centering
\includegraphics[width=\linewidth]{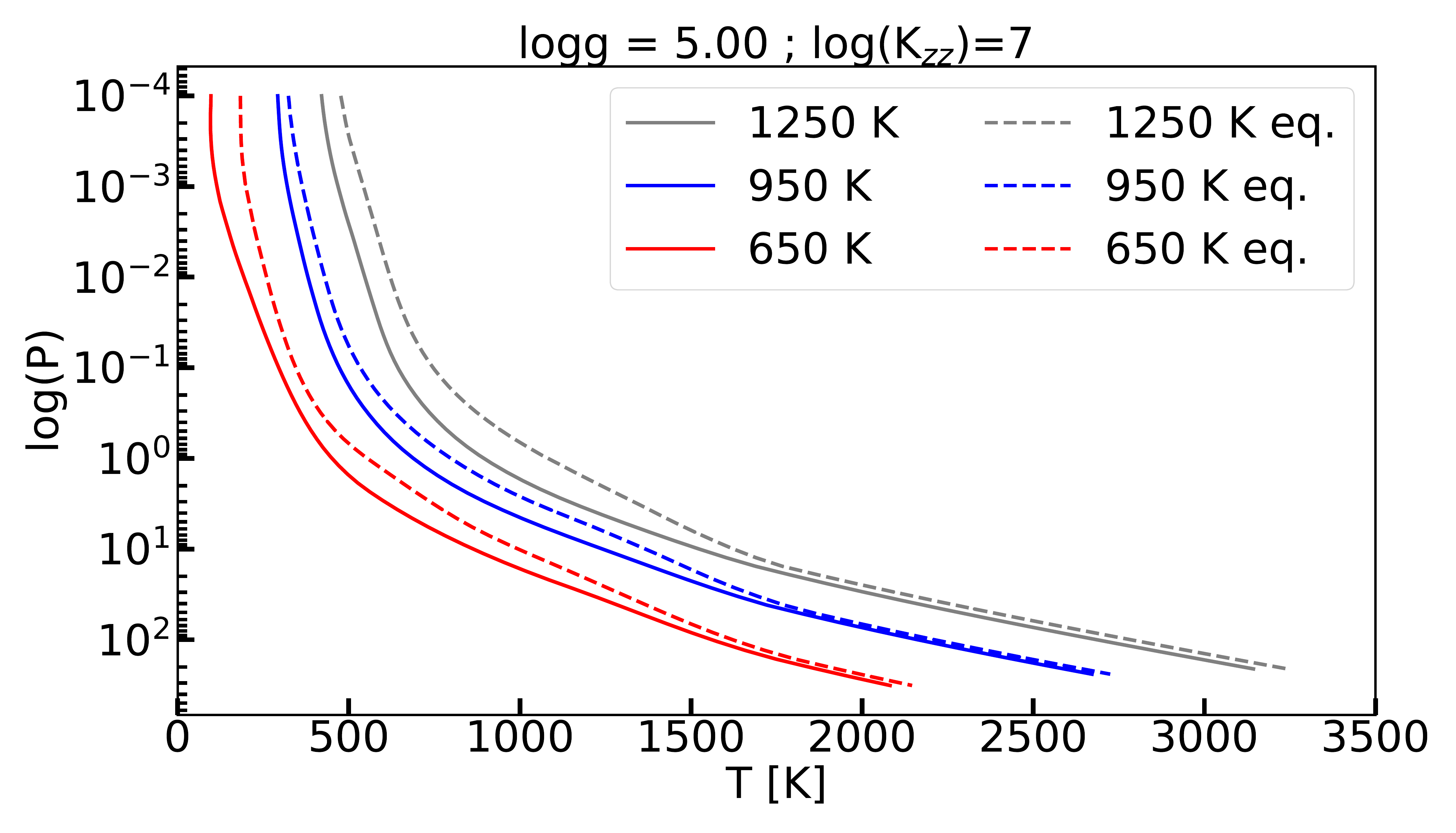}
\caption{TP profiles for quenched (solid lines) and equilibrium (dashed 
lines) models for atmospheres with $\log g$=5.0 
and $T_\mathrm{eff}$ of 1250~K (gray lines), 950~K (blue lines) 
and 650~K (red lines), and for 
$\log K_{zz}$= 4 (top panel) and 7 (bottom panel).}
\label{fig:tp_1000g_allt}
\end{figure}

\begin{figure}[]
\centering
\includegraphics[width=\linewidth]{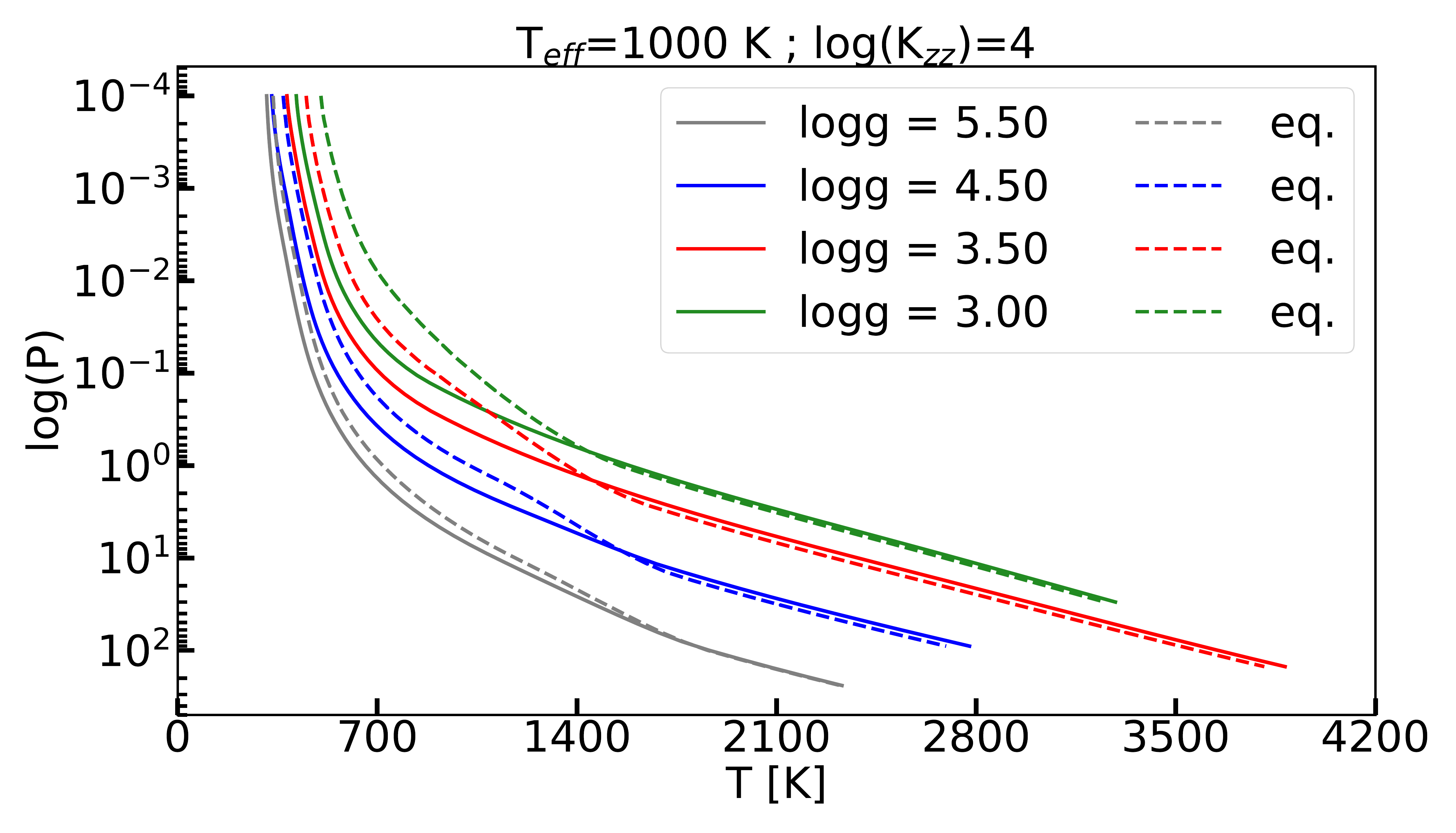}
\centering
\includegraphics[width=\linewidth]{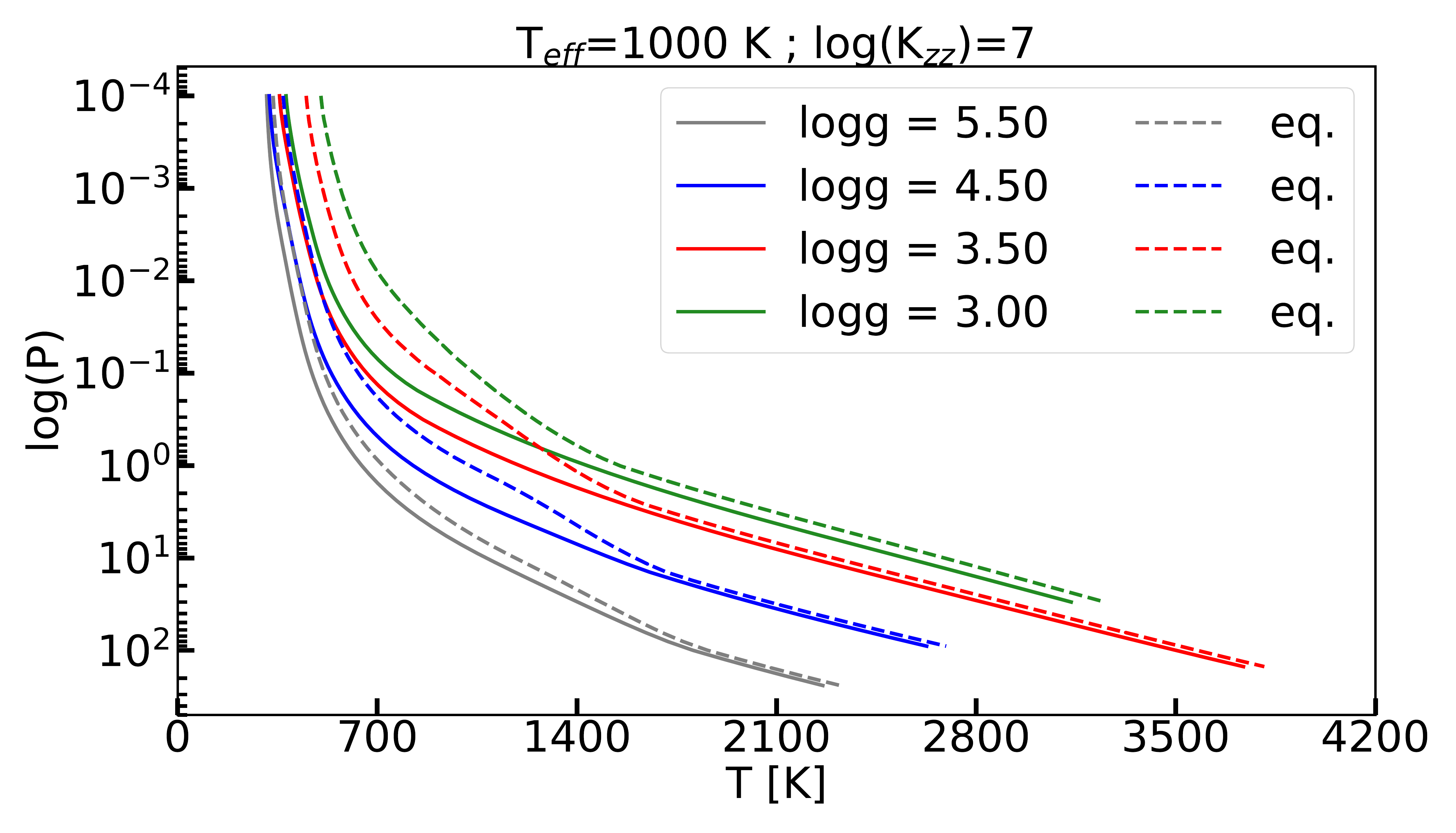}
\caption{TP profiles for atmospheres with $T_\mathrm{eff}$=1000~K and 
$\log g$=3.0 (green lines), 3.5 (red lines), 4.5 (blue lines) and 5.5 
(gray lines) and for $\log K_{zz}$= 4 (top panel) and 7 (bottom panel).}
\label{fig:tp_profs2}
\end{figure}

To study further the effect of $\log g$ and $T_\mathrm{eff}$ on the
TP profile of a model atmosphere, in 
Fig.~\ref{fig:tp_1000g_allt} we show the TP profiles for atmospheres 
with $\log g$=5.00, in disequilibrium (solid lines) with $\log K_{zz}$= 4 
(top panel) and 7 (bottom panel) and $T_\mathrm{eff}$ of 650~K 
(red lines), 950~K (blue lines) and 1250~K (gray lines) with their 
corresponding equilibrium profiles (dashed lines); 
and in Fig.~\ref{fig:tp_profs2} we show the TP profiles 
for atmospheres with $T_\mathrm{eff}$=1000~K,  $\log K_{zz}$= 4 
(top panel) and 7 (bottom panel) and $\log g$ ranging from 3.0 (green lines) to 
5.5 (gray lines). 

For a constant gravity $\log g$=5.0 (Fig.~\ref{fig:tp_1000g_allt}), 
the colder a model atmosphere was (i.e., the 
lower its $T_\mathrm{eff}$ was), the colder its upper atmosphere 
became in reference to the equilibrium model both for $\log K_{zz}$= 4 and 7. 
For the $\log K_{zz}$= 7 models, the upper atmosphere 
(P$\lesssim$0.1 bar) cooled by 3.7\%-20.9\% for the $T_\mathrm{eff}$ $\gtrsim$1250~K 
and by 3.9\%-24.8\% for the $T_\mathrm{eff}$ $\gtrsim$650~K models. 
The deeper atmosphere cooled down as well, but to a smaller degree.
For a constant temperature $T_\mathrm{eff}$=1000~K (Fig.~\ref{fig:tp_profs2}), 
the higher the surface gravity of the atmosphere was, the smaller $\delta$T was 
at all pressures for both $\log K_{zz}$= 4 and 7. 
This is due to the TP profile of the atmosphere shifting from deeper to lower 
pressures in the atmosphere and moving nearly vertical to the CO/CH4 equilibrium 
lines (see Fig.2 of \cite{zahnle14} for a reference CO/CH4 equilibrium 
line). This affects the opacities of our model atmospheres 
leading to the observed gravity dependence of $\delta$T.
%\textbf{This is due to the gravity dependence of the CH$_4$-CO quenching, with the higher (lower) gravity models shifting the TP profile at deeper (lower) pressures in the atmosphere. The TP profile moves parallel (vertical) to the CO/CH4 equilibrium lines (see Fig.2 of \cite{zahnle14}) at the pressures where quenching happens, which results in larger changes in the opacities of the atmosphere for the lower gravity models and thus, results in larger changes in the temperature of the upper atmosphere.}

\begin{figure}[]
\centering
\includegraphics[width=\linewidth]{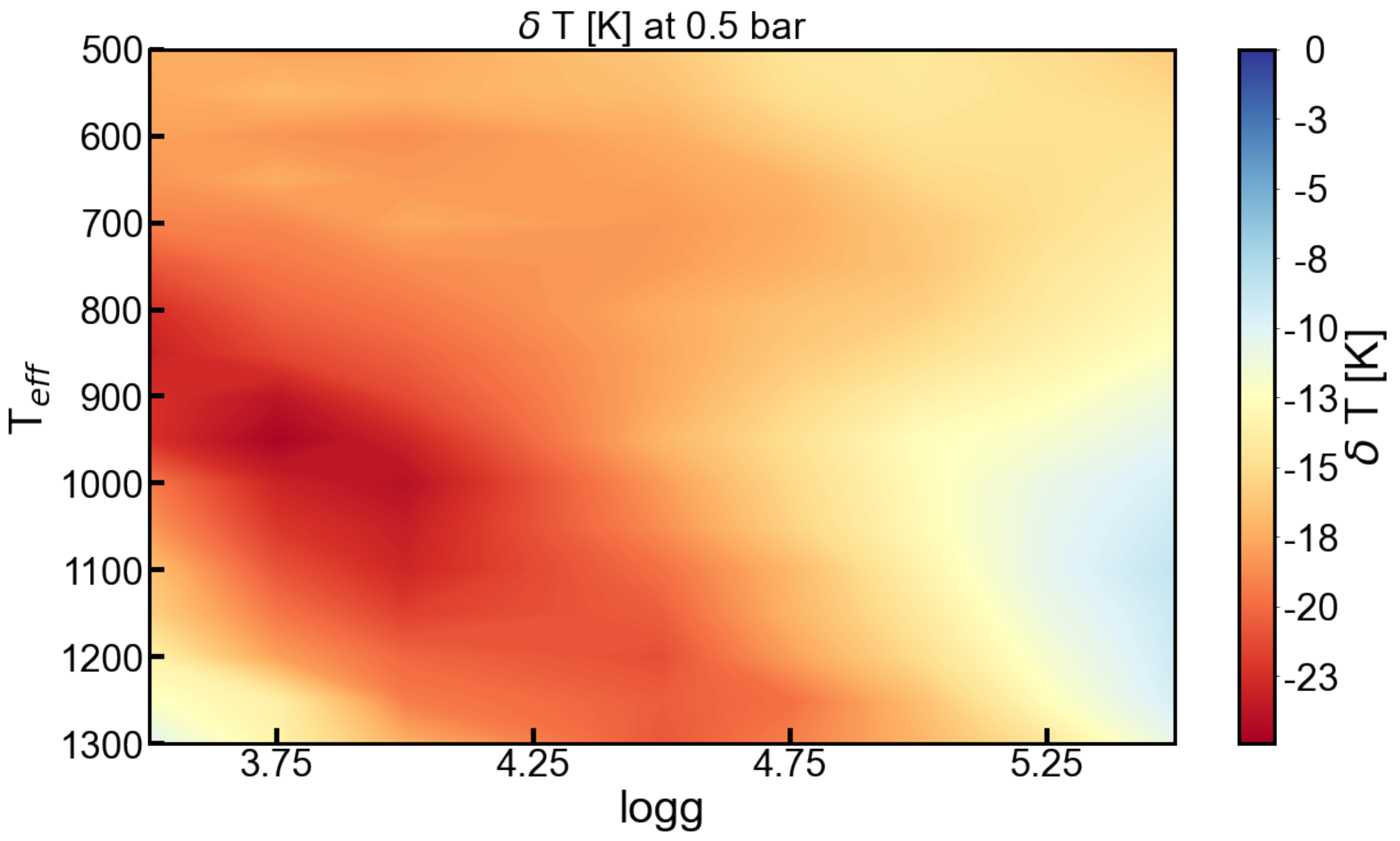}
\centering
\includegraphics[width=\linewidth]{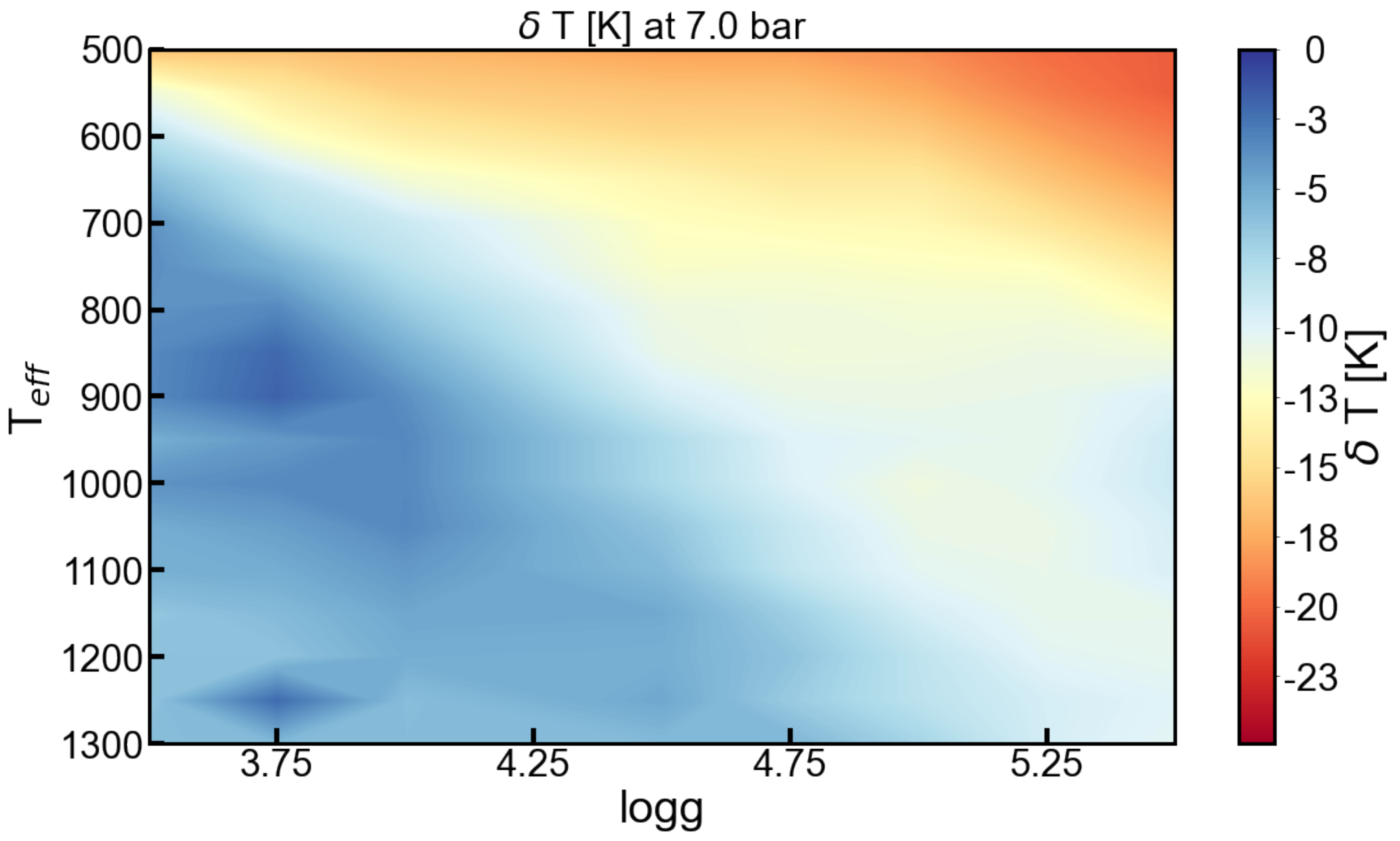}
\centering
\includegraphics[width=\linewidth]{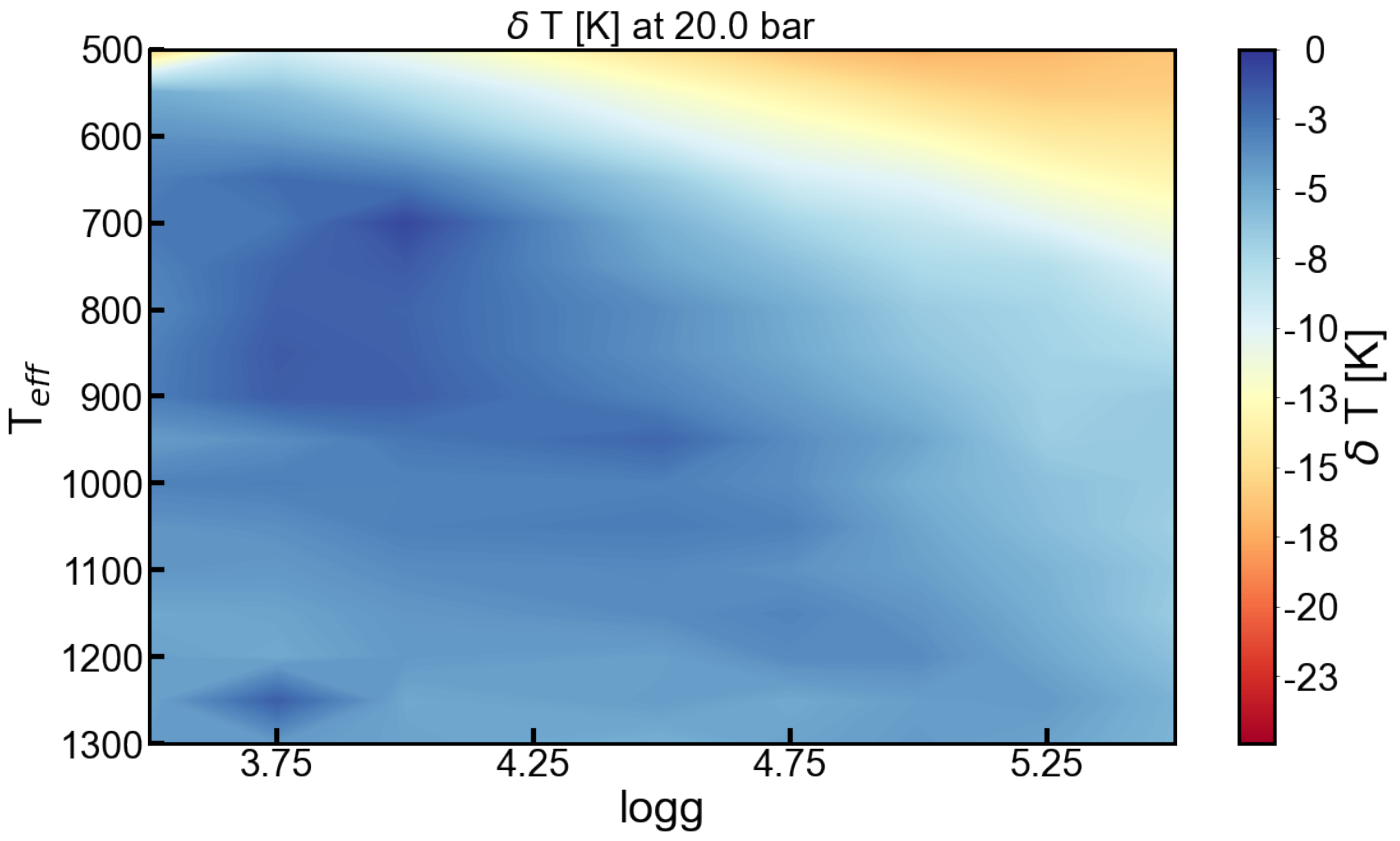}
\caption{Relative $\delta$T as a function of $\log g$ and $T_\mathrm{eff}$ 
for models with $\log K_{zz}$=7 at 0.5 bar (top panels), 7 bar (middle panels) 
and 20 bar (bottom panels). }
\label{fig:tp_allk_allt_kz2}
\end{figure}

To complete our picture of the effect of $T_\mathrm{eff}$ and gravity on $\delta$T
in Fig.~\ref{fig:tp_allk_allt_kz2} we show 
the absolute $\delta$T (=$T_\mathrm{eq}-T_\mathrm{deq}$) for all our grid models at 0.5 bar (top panels), 
7 bar (middle panels) and 20.0 bar (bottom panels), for $\log K_{zz}$=7. The 
pressures were chosen as representative of the upper, mid and deeper atmosphere. 
For $\log K_{zz}$= 7 our model atmospheres were cooler than the
equilibrium models. 
For $\log K_{zz}$= 2 (not shown here) our quenched model 
atmospheres were cooler than the 
equilibrium chemistry ones across most pressure layers, except at higher $T_\mathrm{eff}$ - low g models where at higher pressures the atmospheres heated up.

\subsection{Composition profiles of quenched atmospheres} \label{sect:kzz_cmps}

\begin{figure}[]
\centering
\includegraphics[width=\linewidth]{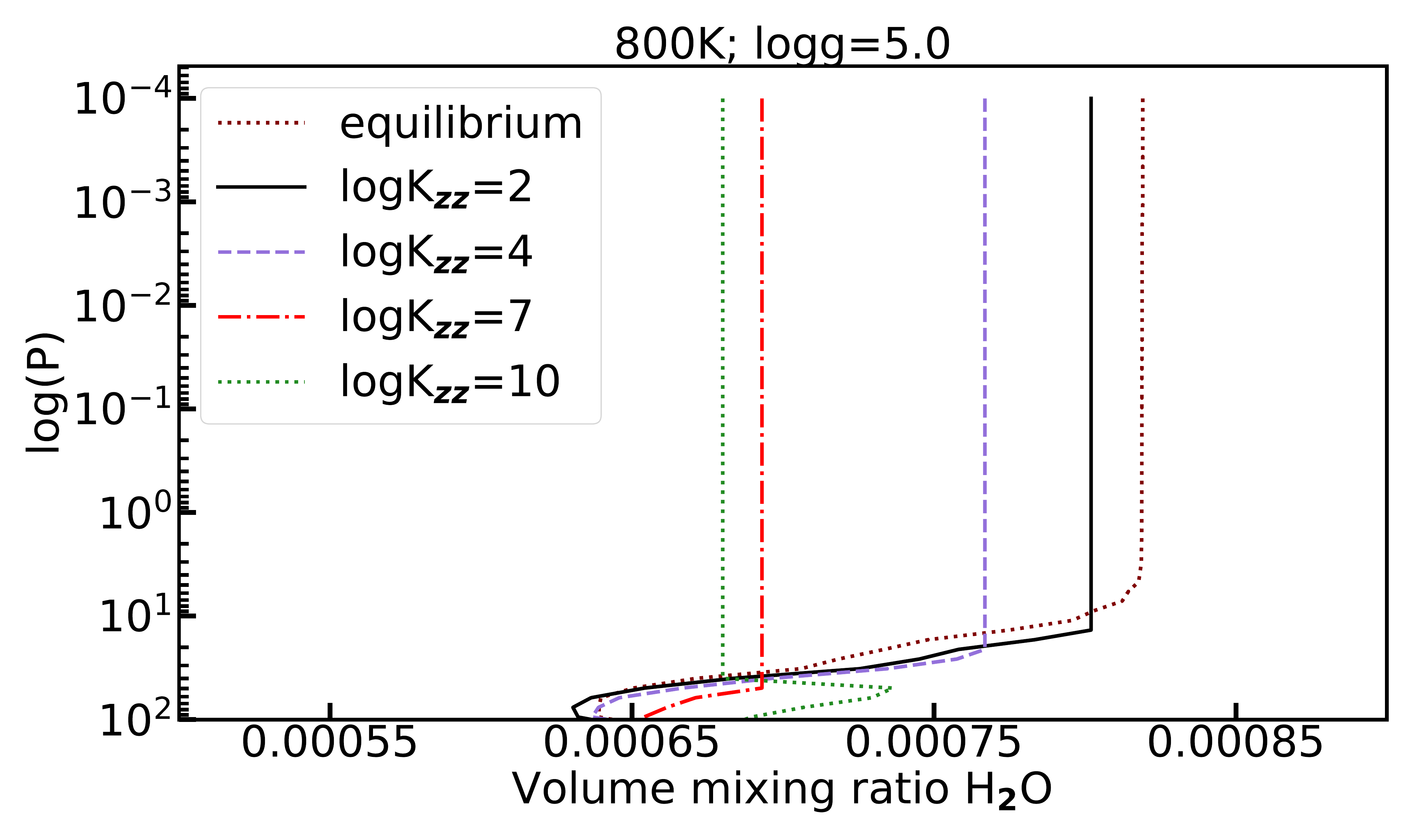}
\centering
\includegraphics[width=\linewidth]{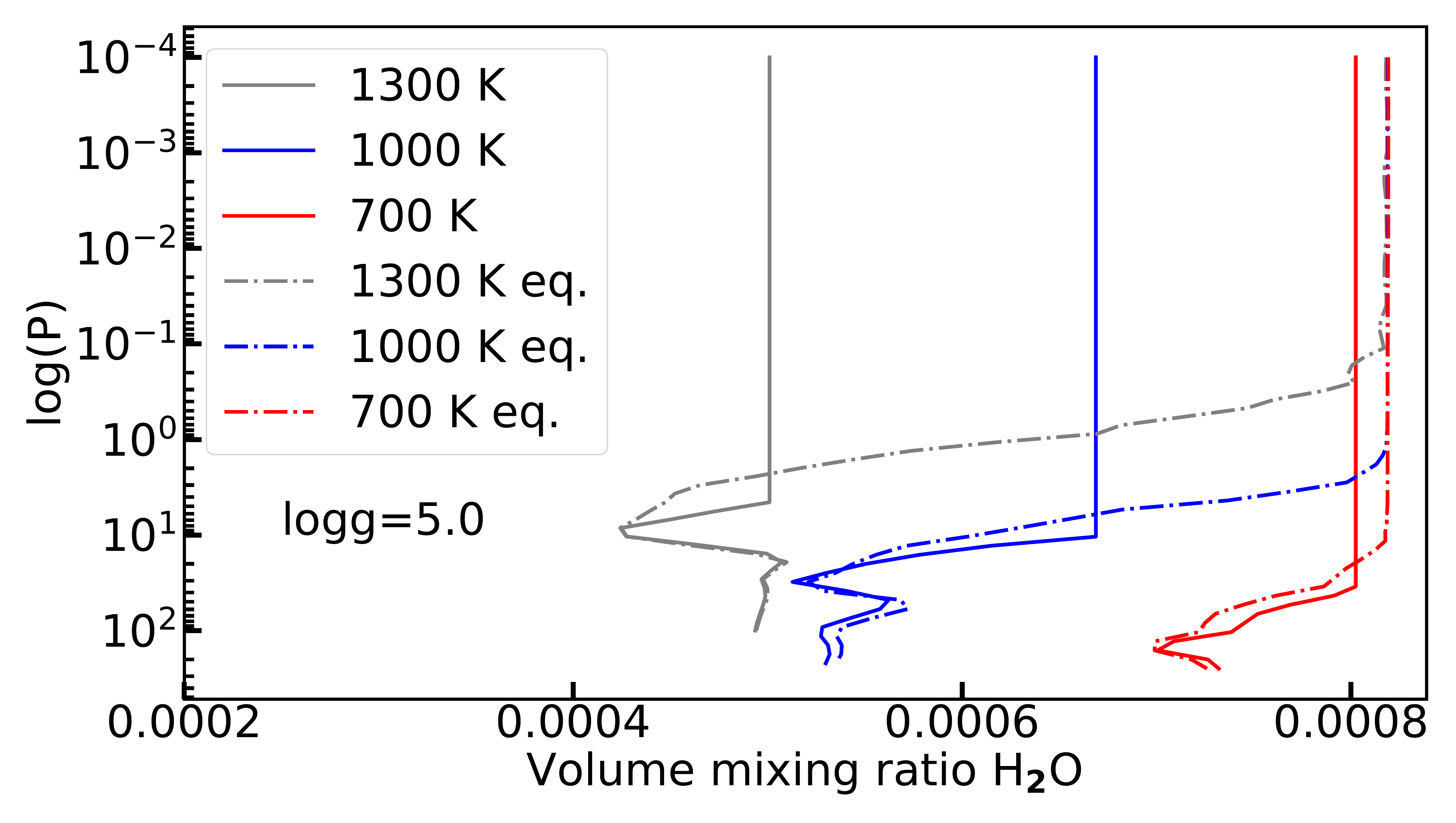}
\centering
\includegraphics[width=\linewidth]{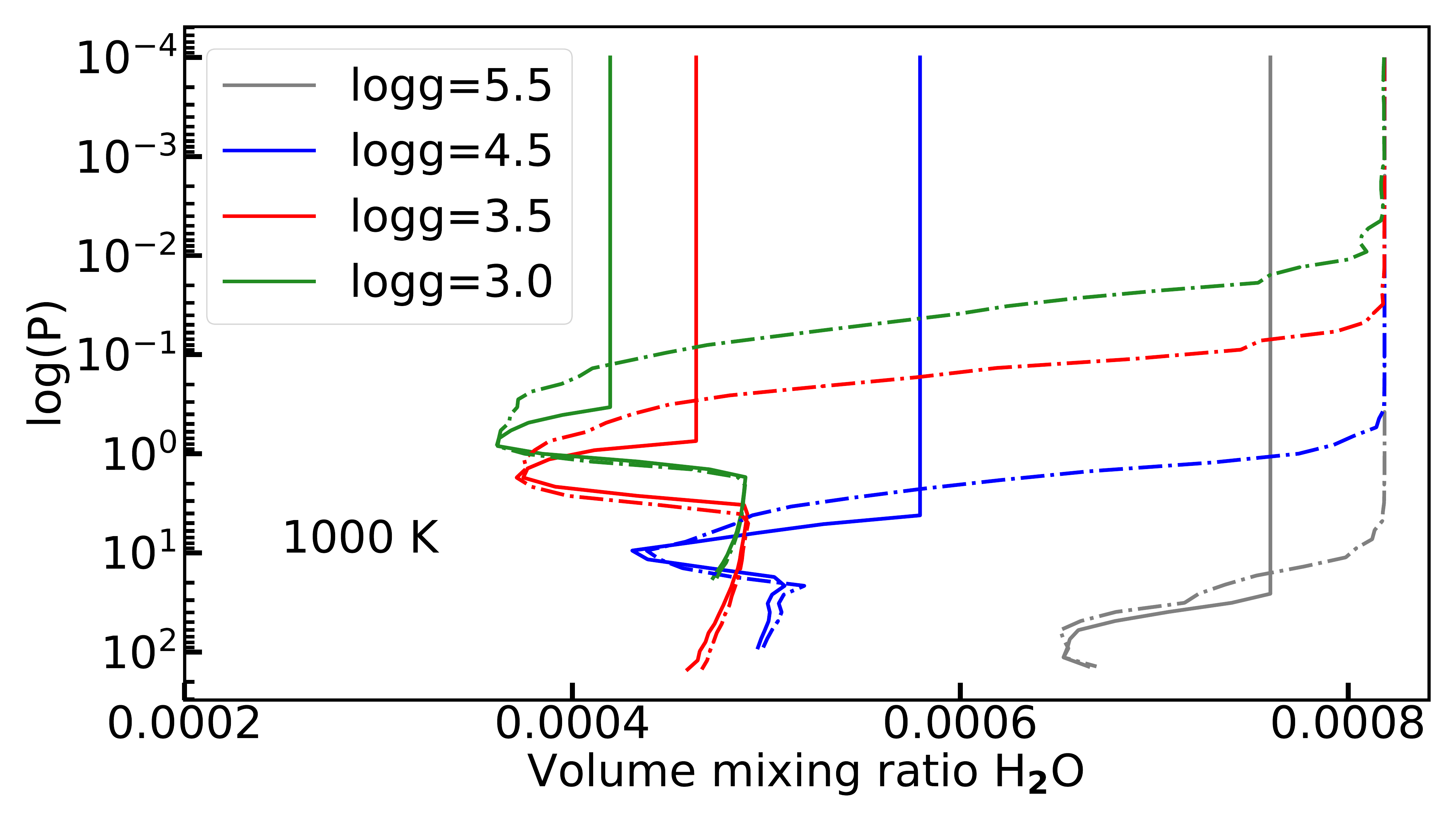}
\caption{Top panel: Volume mixing ratio of H$_2$O of our model atmospheres 
and that of the corresponding equilibrium atmosphere, for 
atmospheres with $T_\mathrm{eff}$ = 800~K, $\log g$=5.0 and 
$\log K_{zz}$= 2(black, solid line), 
4 (red, dashed--dotted line), 7 (red, dashed line) and 10 (green, dotted line). 
Middle panel: Volume mixing ratio of H$_2$O for our model atmospheres (solid lines)
in comparison to the corresponding equilibrium atmosphere profiles (dashed lines) for 
atmospheres with $\log g$=5.0, $\log K_{zz}$=4 and $T_\mathrm{eff}$ of 
1300~K (gray lines), 1000~K (blue lines) and 700~K (red lines). 
Bottom panel: Volume mixing ratio of H$_2$O for our model atmospheres (solid lines)
in comparison to the corresponding equilibrium atmosphere profiles (dashed lines) for 
atmospheres with $T_\mathrm{eff}$ = 1000~K, $\log K_{zz}$=4 
and $\log g$ of 3.0 (green lines),3.5 (red lines), 4.5 (blue lines) and 
5.5 (gray lines).}
\label{fig:h2o_allk_allkzz}
\end{figure}

In Fig.~\ref{fig:h2o_allk_allkzz} (top panel) we show the 
volume mixing ratio profile of H$_2$O for the quenched 
and equilibrium model atmospheres for a model atmosphere with 
$T_\mathrm{eff}$ = 800~K, $\log g$=5.0 and $\log K_{zz}$= 2 (black, solid line), 
4 (purple, dashed line), 7 (red, dashed--dotted line) and 10 (green, dotted line). We
also show (middle panel) the volume mixing ratio of 
H$_2$O for model atmospheres with $\log g$=5.0,
$\log K_{zz}$= 4 and $T_\mathrm{eff}$ of 700~K (red lines), 1000~K (blue lines) 
and 1300~K (gray lines); and (bottom panel) the volume mixing ratio of 
H$_2$O for model atmospheres with $\log K_{zz}$ = 4, $T_\mathrm{eff}$ = 1000~K 
and $\log g$ of 3.0 (green lines), 3.5 (red lines), 4.5 (blue lines) and 5.5 (gray lines).

\begin{figure}[]
\centering
\includegraphics[width=.95\linewidth]{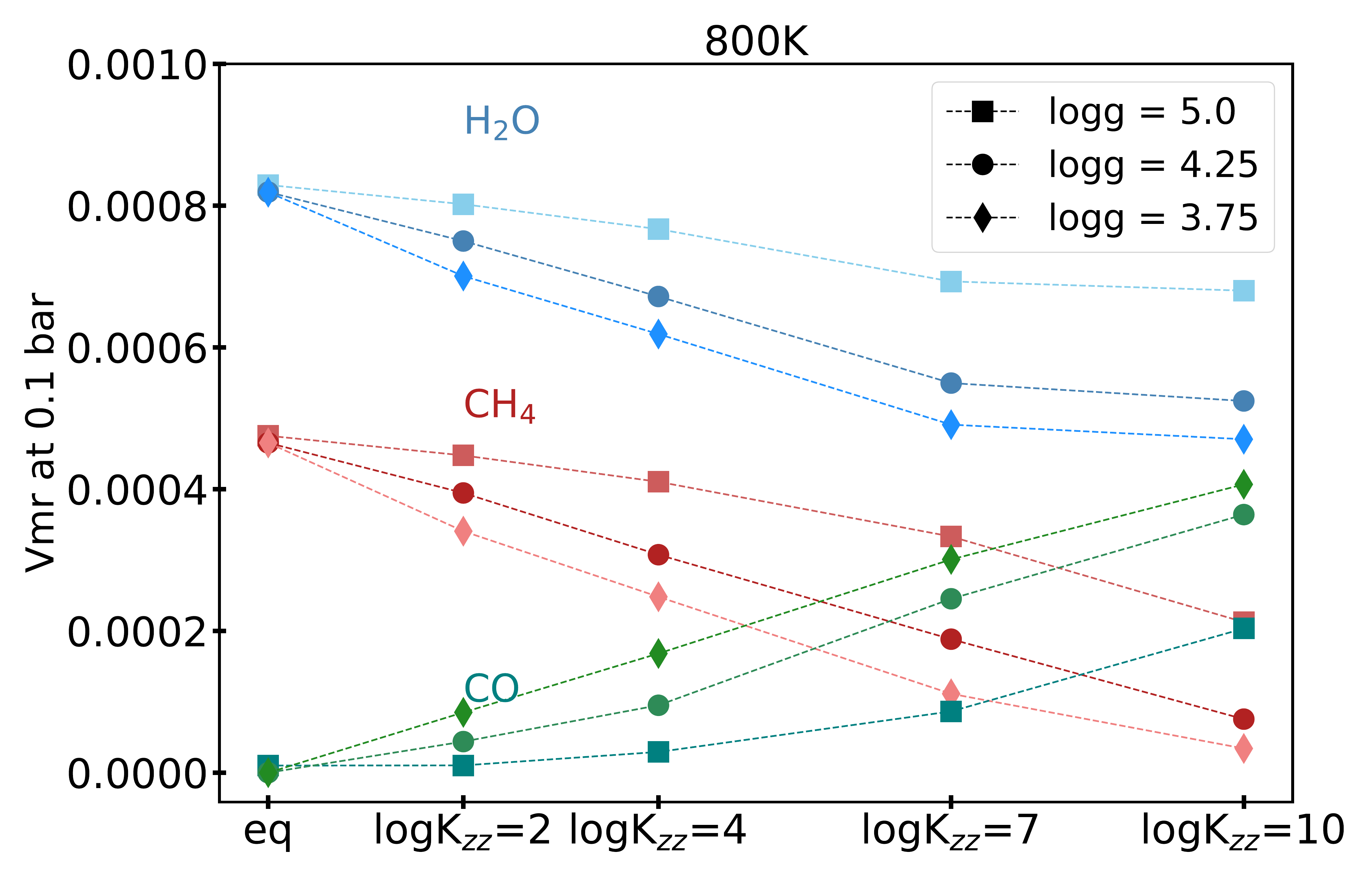}
\caption{Volume mixing ratio of H$_2$O (blue), CH$_4$ (red) and CO (green) at 0.1 bar for different $\log K_{zz}$ values and for a model atmosphere with $T_\mathrm{eff}$ = 800~K and $\log g$=5.0 (squares),  4.25 (circles) and 3.75 (diamonds) . }
\label{fig:h2o_allk_onet_difg}
\end{figure}

As expected, with increasing $\log K_{zz}$ our model atmospheres depart 
further from equilibrium chemistry, with the H$_2$O volume mixing ratio relatively 
reduced by 2\% for $\log K_{zz}$= 2 to 17\% for $\log K_{zz}$= 10. 
For a constant $\log g$ the departure from equilibrium chemistry depends on the 
temperature of our model atmospheres (Fig.~\ref{fig:h2o_allk_allkzz}, middle panel). In 
the upper atmosphere, for $\log K_{zz}$=4 the H$_2$O volume mixing ratio of our atmosphere 
was reduced by 2\% at 700~K to 38.8\% at 1300~K.
Finally, as expected \citep[see also][]{zahnle14}, for a constant temperature the departure 
from equilibrium chemistry depends \emph{strongly} on the gravity of our model 
atmosphere (bottom panel of Fig.~\ref{fig:h2o_allk_allkzz} and Fig.~\ref{fig:h2o_allk_onet_difg}). 
The smaller $\log g$ is, the larger the depletion of H$_2$O higher up
in the atmosphere is. Due to quenching happening higher up in the atmosphere, the H$_2$O 
profile of the atmosphere in deeper layers coincides with the equilibrium chemistry model
profile. On the other hand, the larger $\log g$ is the larger the depletion of 
H$_2$O deeper in the atmosphere is. As an indication of the changes in the volume mixing ratio of other species with $\log g$ and $\log K_{zz}$, Fig.~\ref{fig:h2o_allk_onet_difg} shows the volume mixing ratio at 0.1 bar for H$_2$O (blue markers), CH$_4$ (red markers) and CO (green markers). Our model atmospheres have $T_\mathrm{eff}$ = 800~K,  $\log g$= 5.0 (squares), 4.25 (circles) and 3.75 (diamonds) and different $\log K_{zz}$ values. The volume mixing ratio of all three species changes with $\log K_{zz}$ for both gravities. Similar to H$_2$O (bottom panel of Fig.~\ref{fig:h2o_allk_allkzz}), the smaller $\log g$ is, the larger the change in CH$_4$ and CO is with $\log K_{zz}$. As noted in \citet[][]{zahnle14} $\log g$ affects the $\log K_{zz}$ for which CO dominates the atmosphere.

\begin{figure}[]
\centering
\includegraphics[width=\linewidth]{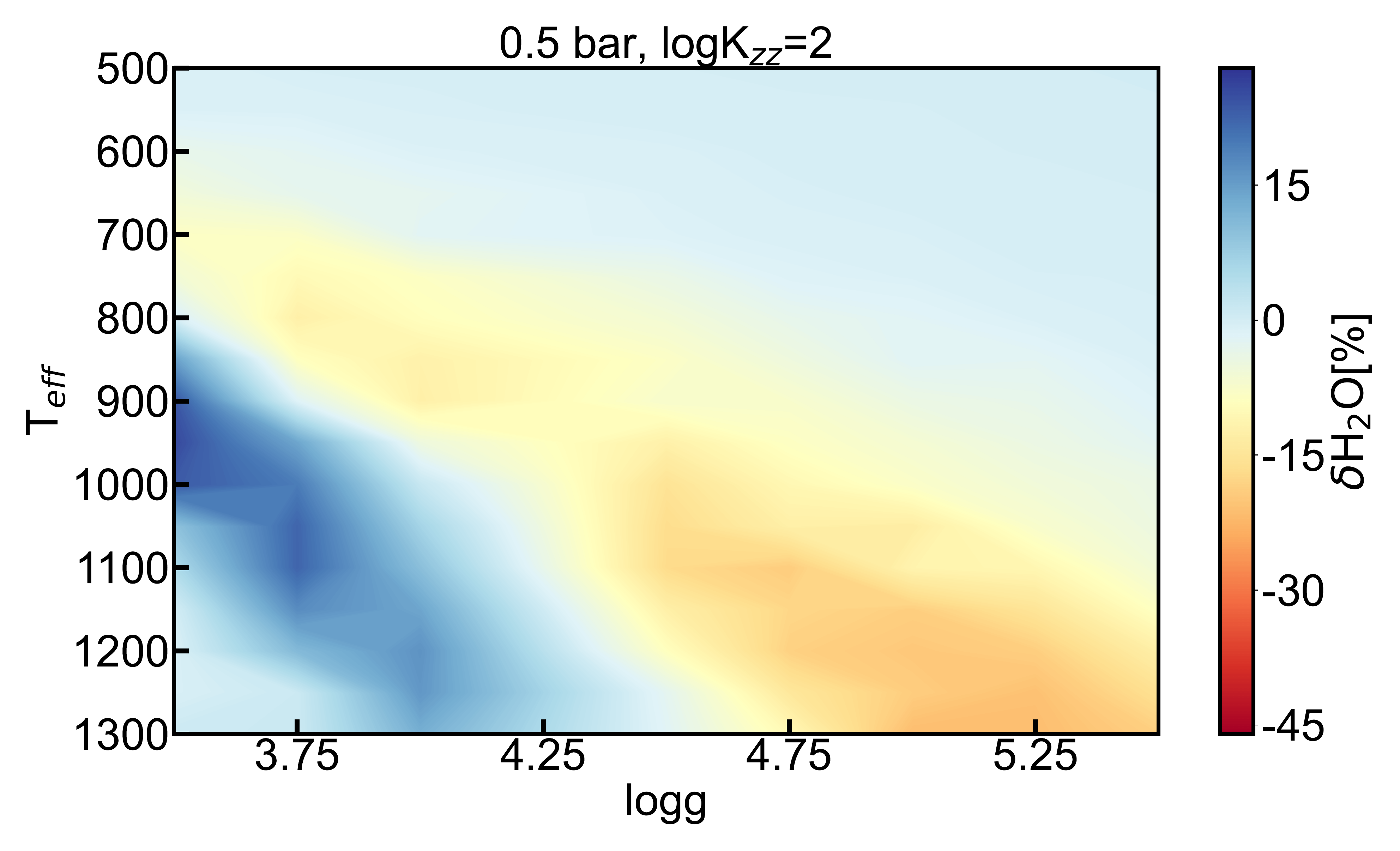}
\centering
\includegraphics[width=\linewidth]{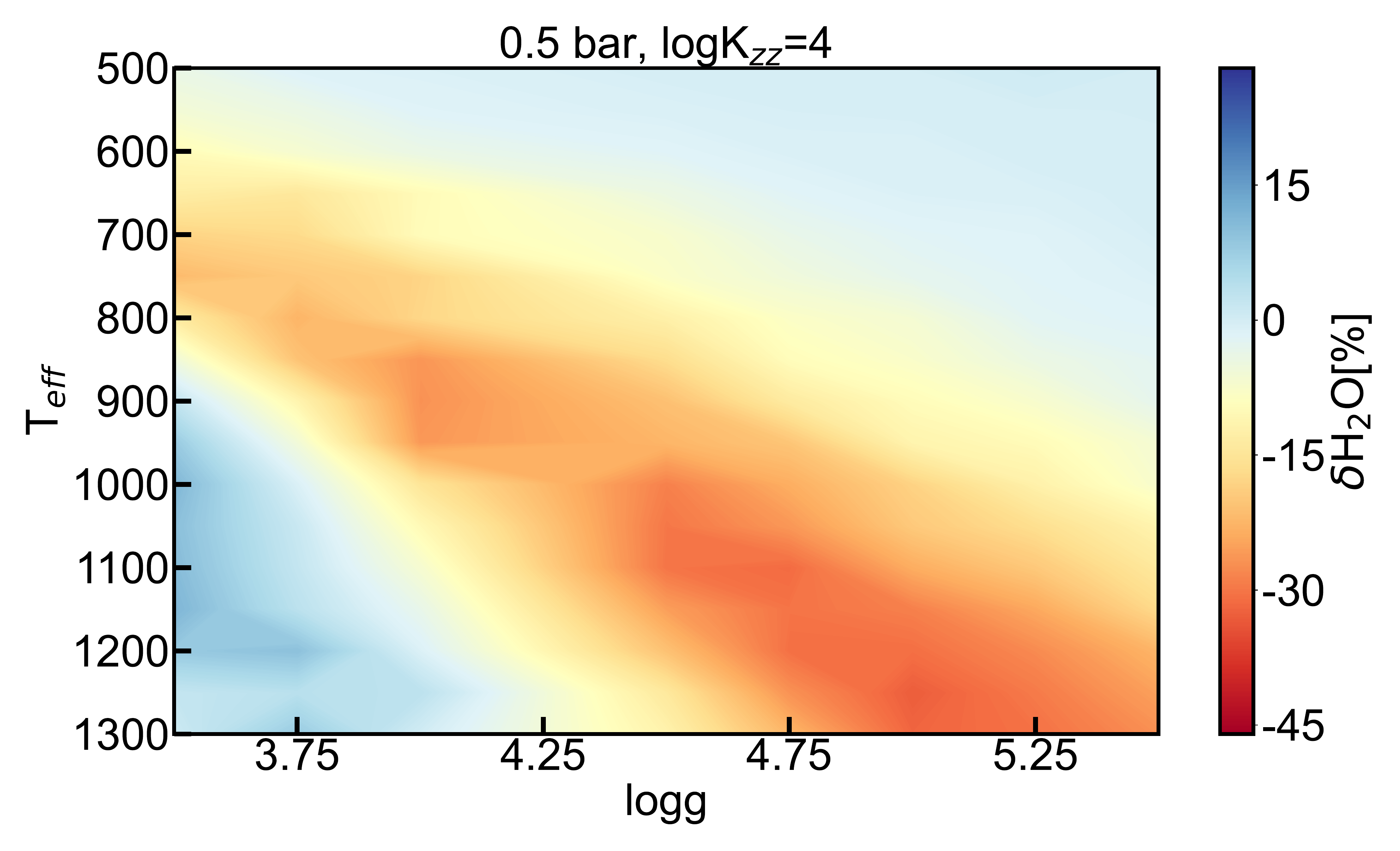}
\caption{Relative change in H$_2$O at a pressure of 0.5 bar between a model with  
$\log K_{zz}$= 2 (top panel) or 4 (bottom panel) 
and the equilibrium chemistry model, as a function of gravity 
and effective temperature. }
\label{fig:h2o_0p5bar}
\end{figure}

\begin{figure}[]
\centering
\includegraphics[width=\linewidth]{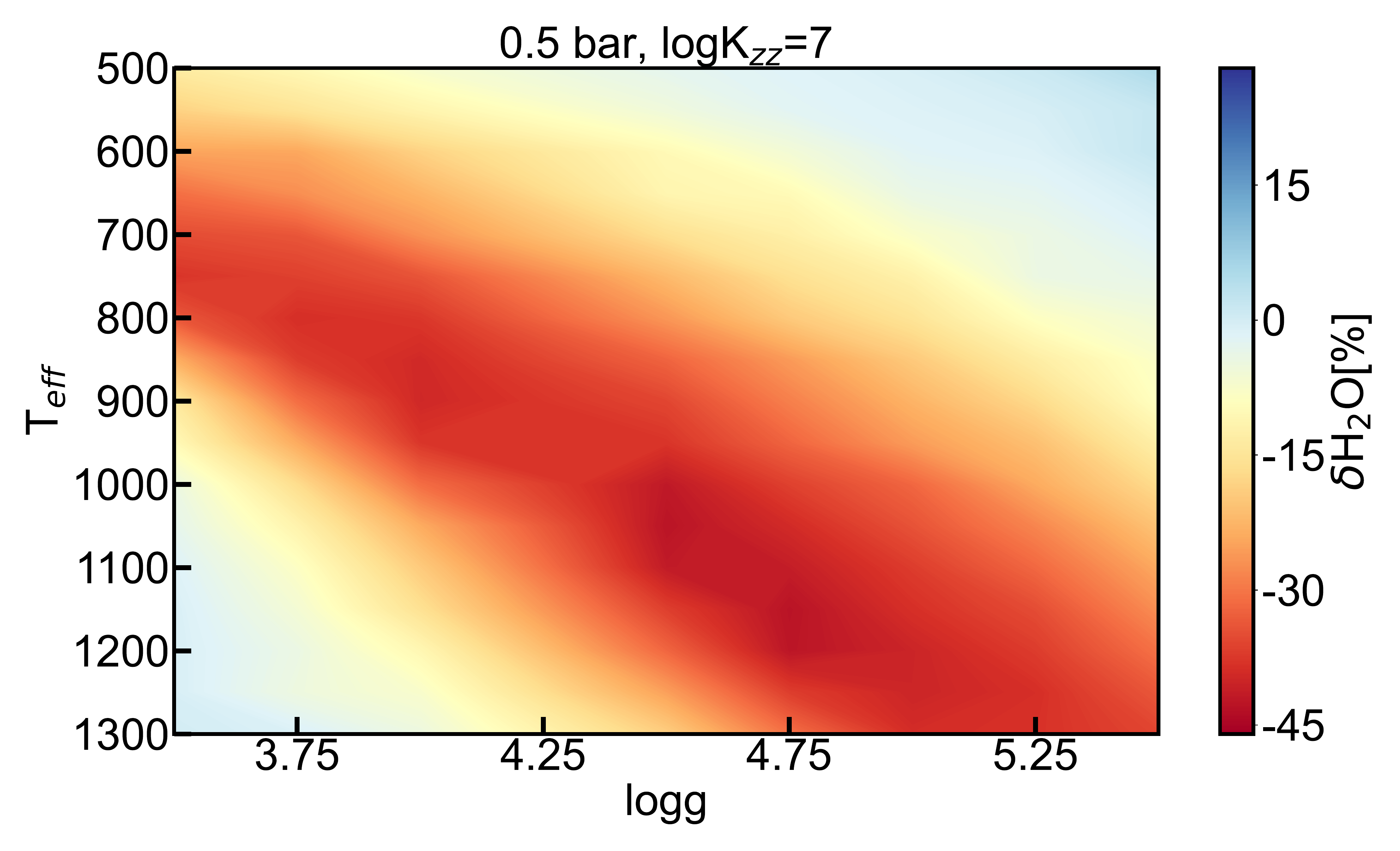}
\centering
\includegraphics[width=\linewidth]{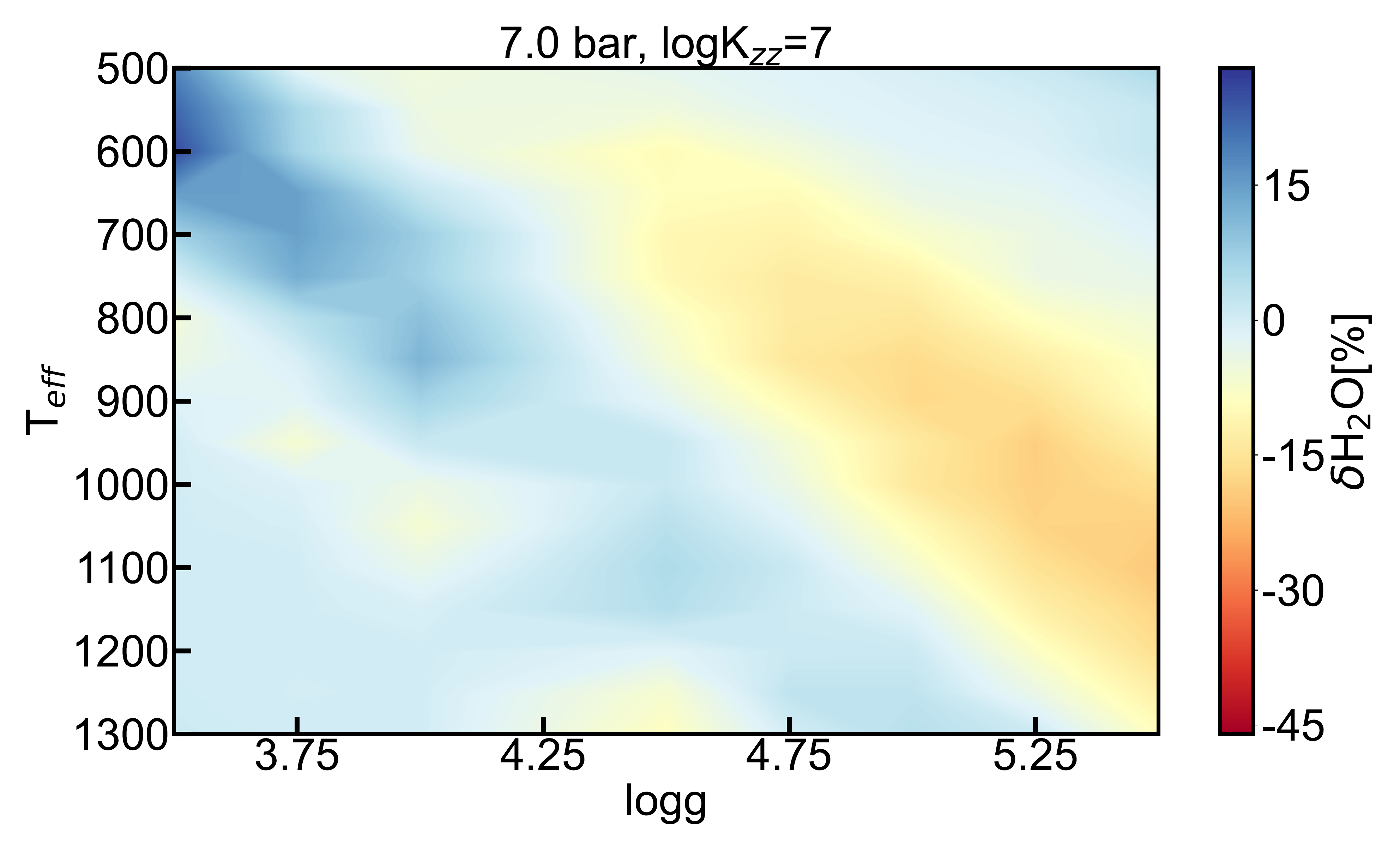}
\centering
\includegraphics[width=\linewidth]{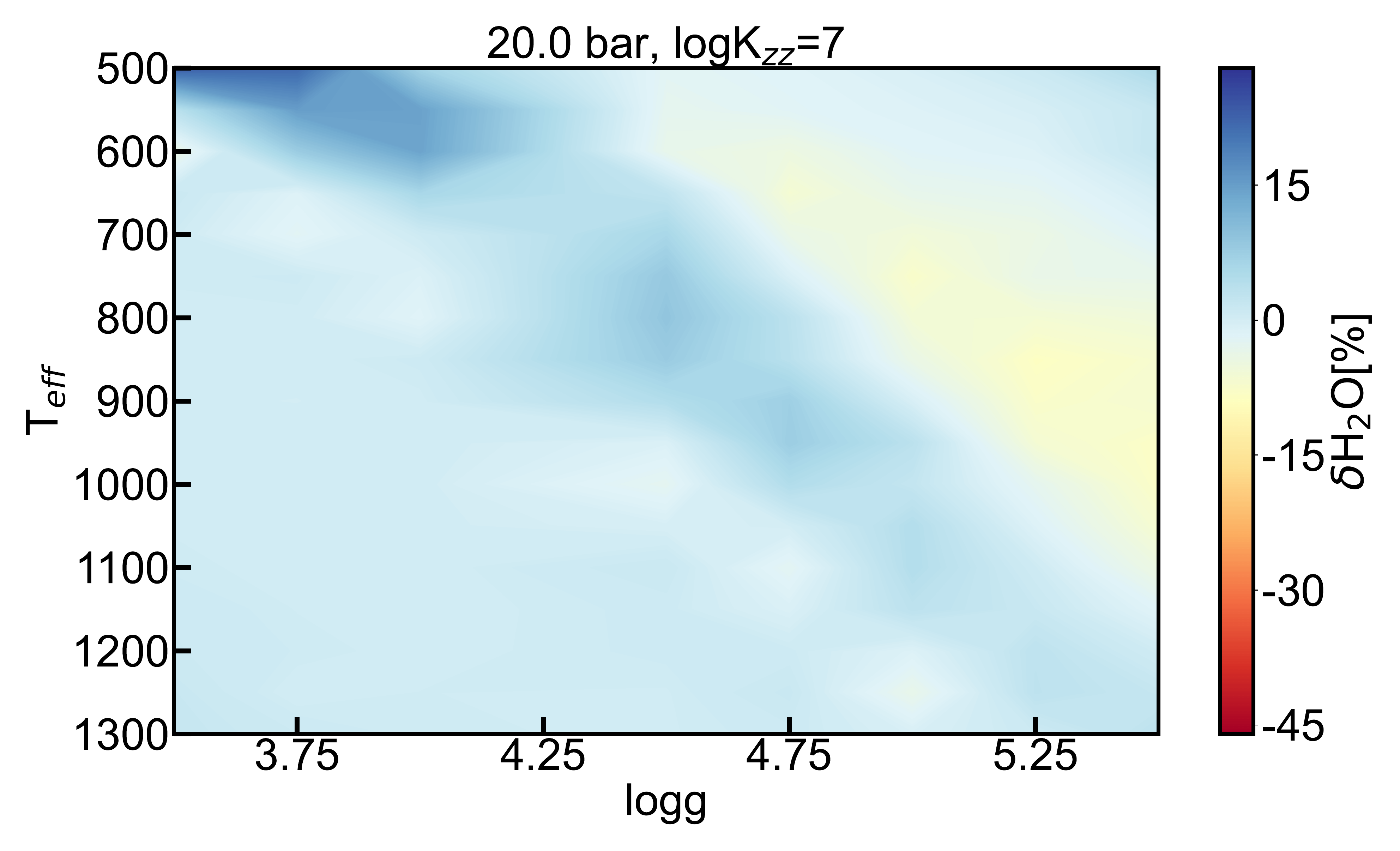}
\caption{Relative change in H$_2$O at 0.5 bar (top panel), 7 bar (middle panel) and 
20 bar (bottom panel) between a model with $\log K_{zz}$= 7 and the 
equilibrium chemistry model.}
\label{fig:h2o_dif_bar_kzz7}
\end{figure}

Finally, Figs.~\ref{fig:h2o_0p5bar} and~\ref{fig:h2o_dif_bar_kzz7}  
show the 
relative changes in the H$_2$O %and CH$_4$ %and NH$_3$ 
content of our model atmospheres as a 
function of $T_\mathrm{eff}$ and $\log g$, and for various $\log K_{zz}$ and 
pressures. These figures are indicative of the trends in the relative change of 
different species' volume mixing ratios, and do not intend to show a complete picture 
of the grid. 

\begin{figure*}
\centering
\includegraphics[width=.9\textwidth]{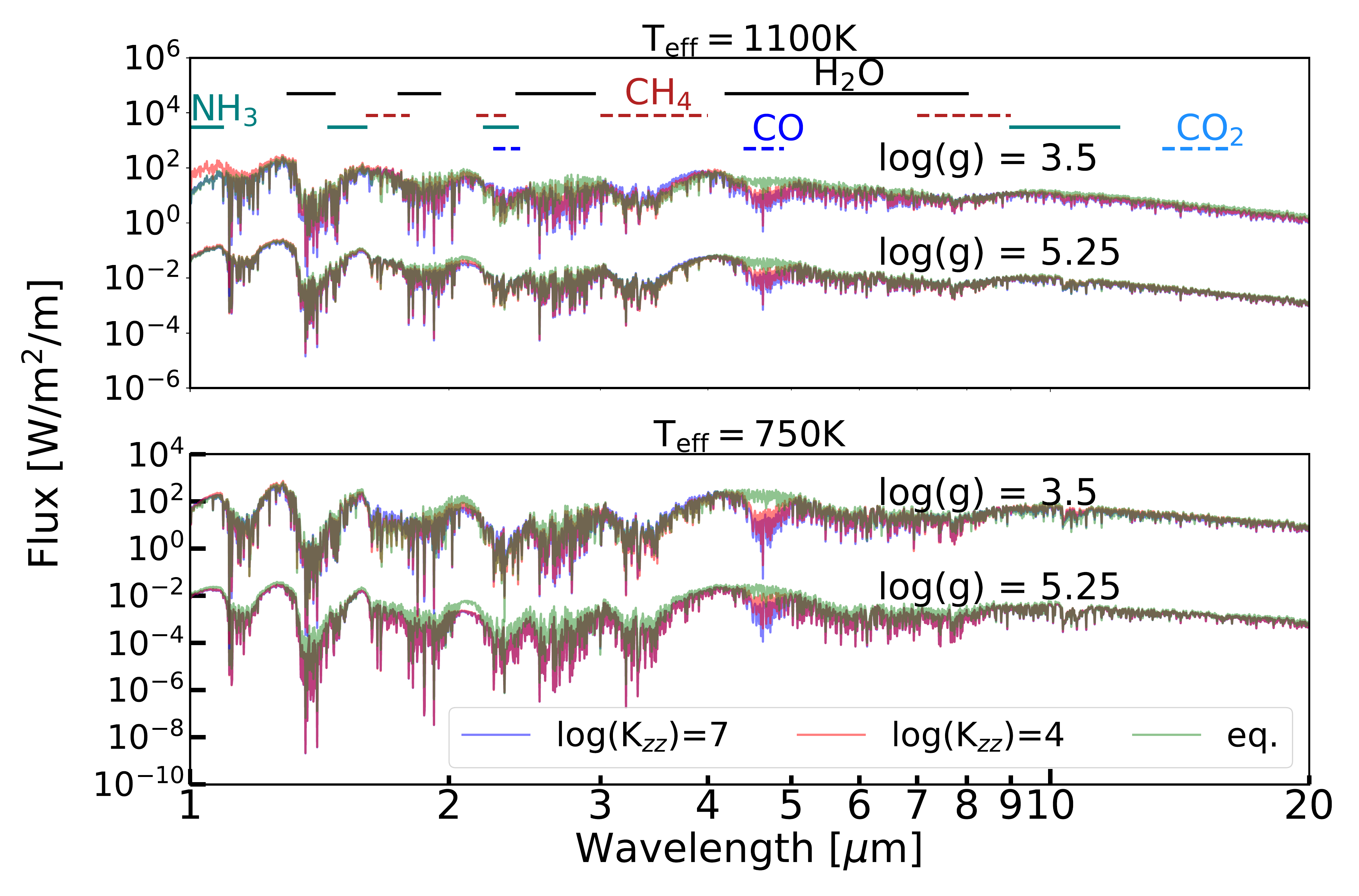}
%3162g_dif_k_log_log_flux.pdf}
\caption{Quenching changes the composition and thus the spectrum of our model 
atmospheres. Here shown are spectra of a selection of our model atmospheres 
(top panel 1100~K and bottom panel 750~K)
with log $g$=3.5 and 5.25, and log $K_{zz}$=4 (red line) or 7 (blue line) 
and the corresponding equilibrium model spectrum (green line). The spectra are shifted 
by an arbitrary amount for plot clarity and downsampled to a resolution of 800. Also shown are some of the major 
absorption bands for CH$_4$, CO, H$_2$O and NH$_3$, and CO$_2$.}
\label{fig:3162_flx_loglog}
\end{figure*}

\subsection{Detection of CH4, CO and NH3 in quenched atmospheres} \label{sect:detect}

In Fig.~\ref{fig:3162_flx_loglog} we plot spectra of our model atmospheres 
for $\log g$=3.5 and 5.25, 
$T_\mathrm{eff}$ of 750~K (top two spectra) or 1100~K (bottom two spectra) and 
$\log K_{zz}$=4 (red line) or 7 (blue line). We also plot the 
corresponding spectra of model atmospheres with equilibrium chemistry 
(green line). For plotting clarity we shifted our spectra by 
arbitrary amounts and binned our spectra down to a resolution of 500.

Quenching of CH$_4$, CO, H$_2$O and NH$_3$ affected the composition profiles 
of these species in our model atmospheres 
and thus the detectability of these species in the atmosphere. 
For $\log g$=5.25 and 
$T_\mathrm{eff} \gtrsim$ 1100~K the changes in the 
composition of our model atmospheres in reference to the equilibrium model 
were small for CH$_4$, H$_2$O and NH$_3$ resulting in non-detectable 
changes in the model spectra in the major absorption bands of these species. 
Notably, quenching affected the strong CO band around 4.7~$\mu$m 
resulting in a large difference ($\delta$F$\sim$40\%-50\% for 
$\log K_{zz}$=4 or 7 respectively at 1100~K, to $\delta$F$\sim$50\%-60\% 
at 750~K) for all atmospheres. In this Section we test the detectability 
of CH$_4$, CO, H$_2$O and NH$_3$ for our model atmospheres with JWST.

\begin{figure}
\centering
\includegraphics[width=\linewidth]{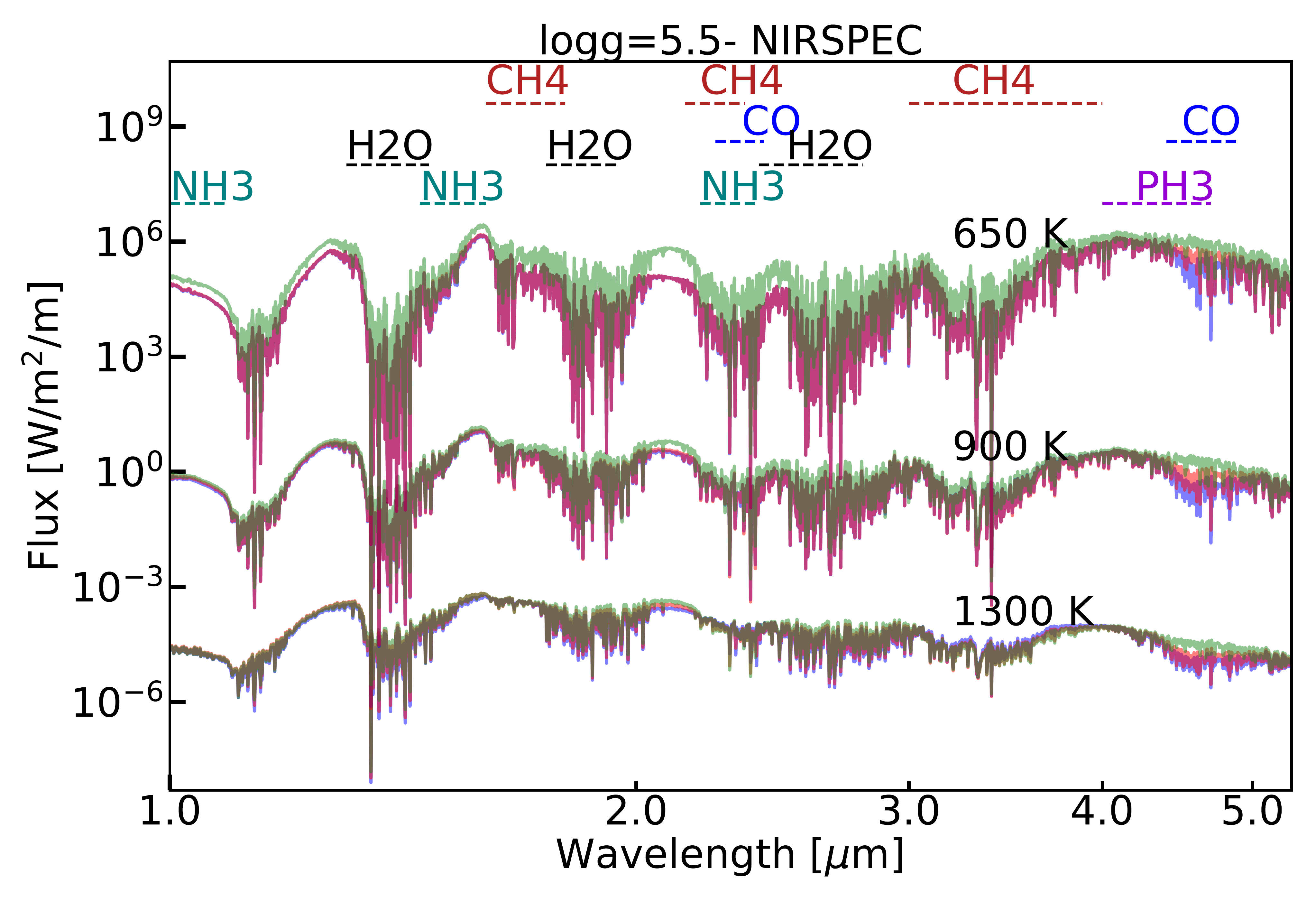}
\centering
 \includegraphics[width=\linewidth]{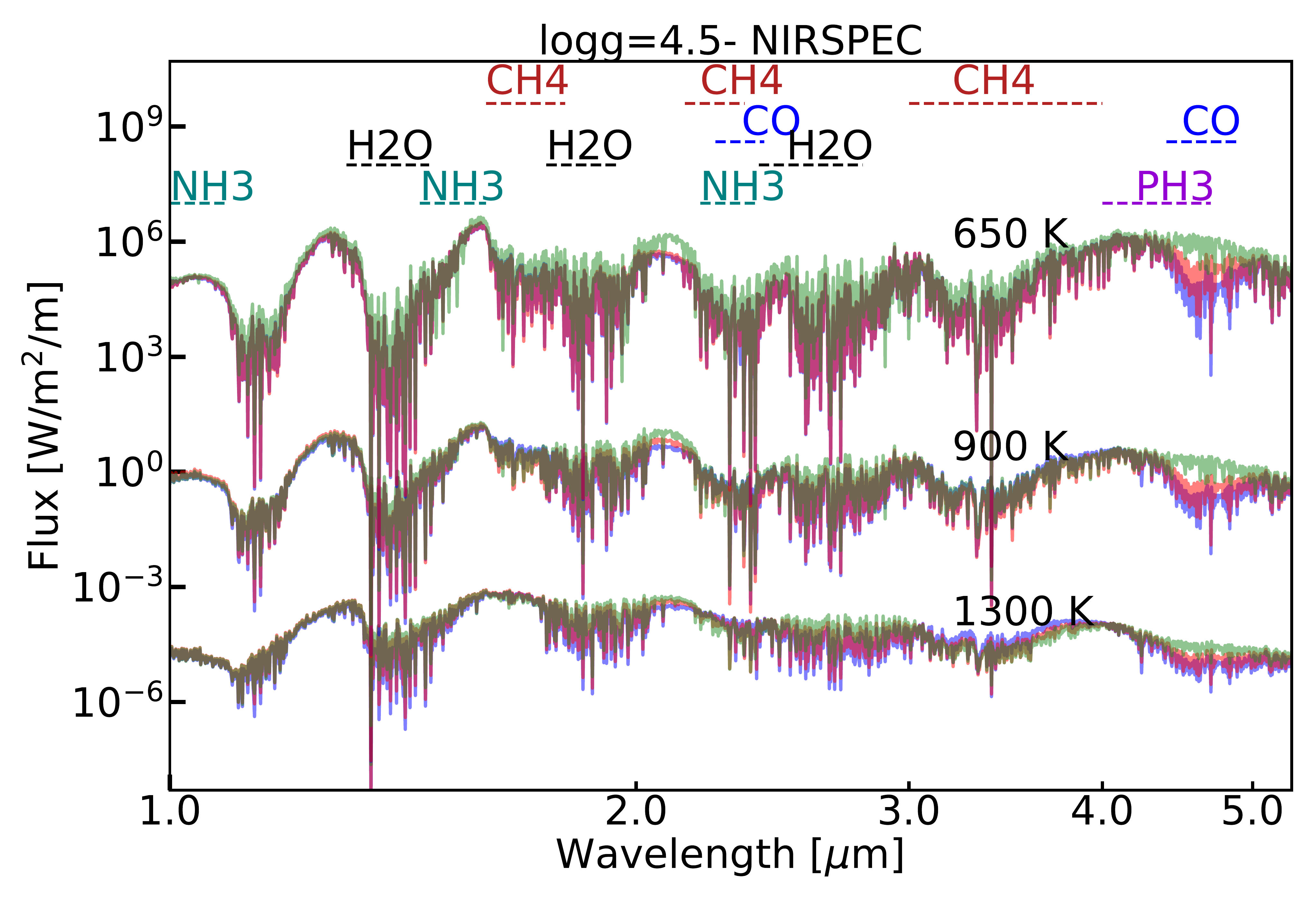}
 \centering
 \includegraphics[width=\linewidth]{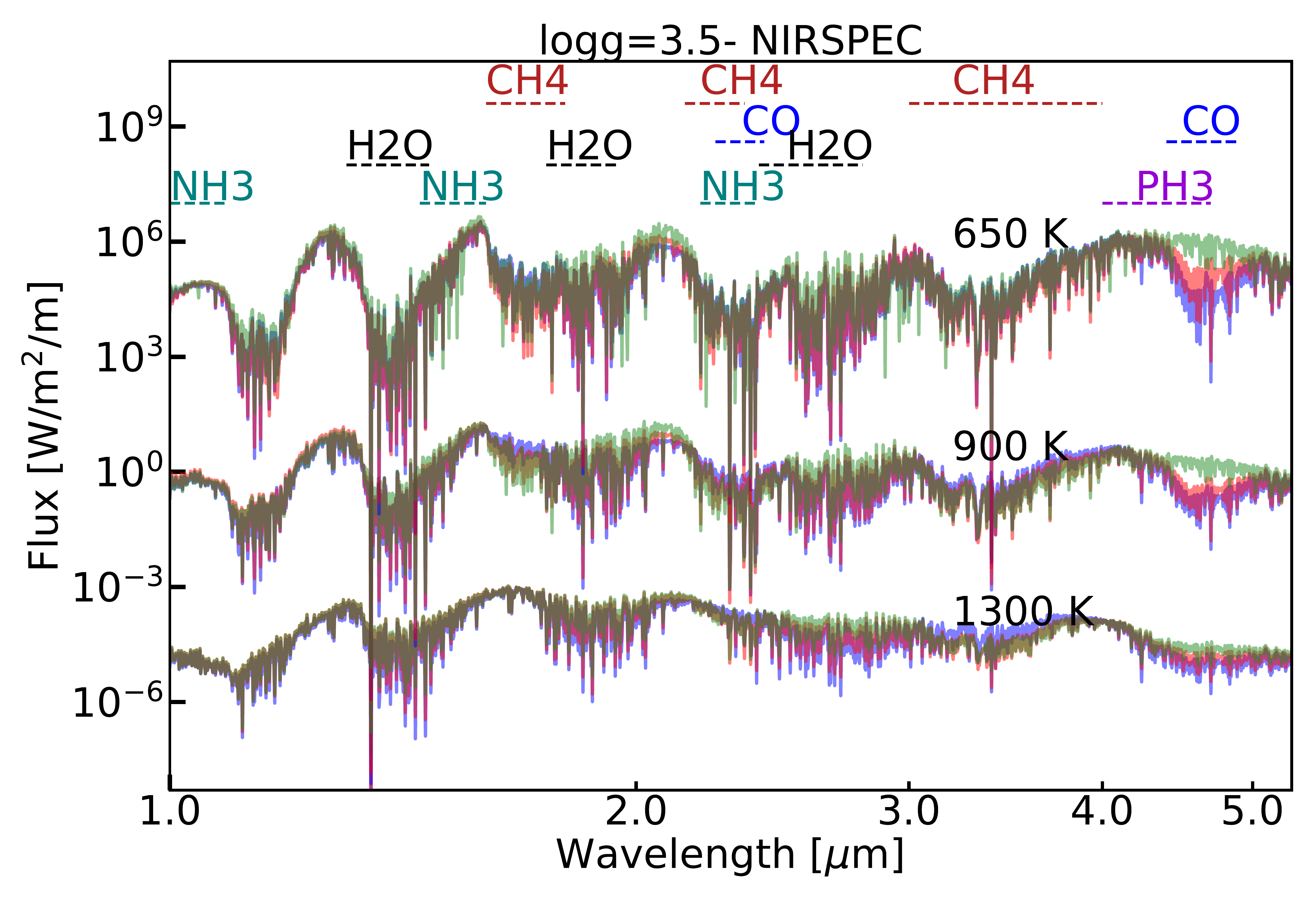}
\caption{Spectra of a selection of our model atmospheres as they would be observed by \emph{JWST} with NIRSPEC at high resolution (G140H + G235H + G395H) . 
Top panel: atmospheres with $\log g$=5.5
   and $T_\mathrm{eff}$ ranging from 650~K to 1300~K. Green lines are atmospheres with 
   equilibrium chemistry, while red and blue lines are quenched atmospheres with 
   $\log K_{zz}$=4 or 7 respectively. Medium and lower panels: Same as 
   top panel but for $\log g$=4.5 and 3.5 respectively. }
\label{fig:jwst_nirspec_g395h}
\end{figure}

In Fig.~\ref{fig:jwst_nirspec_g395h} (top panel) we plot our model atmosphere spectra 
for atmospheres with $T_\mathrm{eff}$ ranging from 650~K to 1300~K and 
$\log g$=5.5. Red lines are used for the $\log K_{zz}$=4 models, blue lines for the 
$\log K_{zz}$=7 models and green lines for the equilibrium chemistry models. 
The spectra are plotted as they would be observed with \emph{JWST} 
using NIRSPEC (covering the 0.97 $\mu$m - 5.14 $\mu$m range). Note that these spectra also cover the 
wavelength range observed with 
the NIRISS Single-Object Slitless Spectroscopy (NIRISS SOSS; covering the 0.6 $\mu$m - 2.8 $\mu$m range), but at a 
higher resolution, so we omit showing the NIRISS SOSS spectra here. 
We also plot the same models but for $\log g$=4.5 (middle panel) and 
3.5 (bottom panel).

For high- and intermediate-gravity atmospheres 
($\log g$=5.5 and 4.5) 
\emph{JWST} NIRISS-SOSS could detect departures from equilibrium 
in the H$_2$O and CH$_4$ content 
of an atmosphere for the cooler atmospheres ($\lesssim$600~K), a 
small change in the NH$_3$ content might be detectable around  
 $\sim$1.6$\mu$m, 
but no changes were observable for CO. This is due to the fact that the single CO absorption band in the NIRISS-SOSS wavelength range overlaps with absorption bands of other major species like CH$_4$ and H$_2$O. For the low
gravities ($\log g$=3.5) \emph{JWST} NIRISS-SOSS could be 
able to separate easier the different $\log K_{zz}$ cases in intermediate to cooler 
atmospheres ($T_\mathrm{eff} \lesssim$ 900~K). 
The inclusion of the 4-5~$\mu$m window with NIRSPec, allowed to 
detect deviations from equilibrium for the H$_2$O,
CH$_4$ and possibly NH$_3$ content of our model atmospheres like NIRISS-SOSS did, 
but also allowed the detection of variation in the CO content of our 
model atmospheres around the 4.7~$\mu$m CO-absorption band, especially 
for the cooler atmospheres ($T_\mathrm{eff} \lesssim$ 900~K). 
Finally, a small change was potentially observable around 4.14~$\mu$m 
due to changes in the PH$_3$ content of the colder atmospheres. We note 
though, that the overlap with a H$_2$O and a CO absorption band may hinder 
this detection at low resolutions. The PH$_3$ feature could be easier detected 
for higher metallicity objects \citep[][]{visscher2006,miles20}. Exploring the effect of disequilibrium chemistry on atmospheres with non-solar metallicities will be part of future work.

Finally, we note that the use of MIRI spectroscopy would allow the detectability of 
NH$_3$ in disequilibrium through changes in the 10.5$\mu$m NH$_3$ 
feature. In particular, our 500 K 
and 550 K models at a medium resolution (not shown here) showed a decrease in the flux in the 
10.5~$\mu$m NH$_3$ absorption feature by $\sim$40\% and  $\sim$30\% in comparison to the 
equilibrium flux, allowing the detection of NH$_3$ in disequilibrium for these cooler model 
atmospheres.

\subsection{Colors of quenched atmospheres} \label{sect:colors}

Quenching changed the composition of our model atmospheres and thus their 
spectra in comparison to the equilibrium models (see also
Sect.~\ref{sect:detect}). In this section we study the effect of 
quenching in the color of our model atmospheres. In particular, we 
study the effect of quenching on the color that \emph{JWST} NIRCam would observe 
for our model atmospheres. We used NIRCam's F115W filter (J-band), 
F162M (H), F210M (K) and F356W (L). 

In Fig.~\ref{fig:cmd_all} we show the Color Magnitude Diagram (CMD) for our model 
atmospheres in J-H (top panel), H-K (upper middle panel) and J-K (bottom panel) for 
$\mathrm{log}g$ = 3.5 , 4.25 and 5.25. 
We plot the equilibrium model CMDs (red-orange dots) and our 
disequilibrium model CMDs for $\log K_{zz}$= 7 
(purple-magenta diamonds). Quenching clearly affected the colors of 
our model atmospheres. 

In  $H-K$ and $J-K$ all disequilibrium atmospheres are bluer than the equivalent equilibrium 
chemistry models for all $\log K_{zz}$. 
In $J-H$ the temperature and gravity of the atmosphere influenced the color change 
in reference to the equilibrium models. The colder atmospheres 
($\lesssim$700~K)  turned redder for all models and all $\log K_{zz}$. At intermediate 
temperatures (700~K$\lesssim T_\mathrm{eff} \lesssim$1050~K)  
atmospheres shifted bluer for $\log K_{zz}= 2$ and 4, while they turned redder for $\log K_{zz}= 7$.
Finally the hotter ($\gtrsim$1100~K) atmospheres turned redder for $\log K_{zz}$= 7, while 
for $\log K_{zz}$= 2 and 4 they turned bluer, or had approximately the same color as the 
equilibrium models.

To explain the color differences between our equilibrium and disequilibrium chemistry 
models we focus on the $\log g$=5.25 models. For the hotter models 
($\gtrsim$1,000~K) most pressure layers were depleted in H$_2$O, CH$_4$ and NH$_3$.  
Some layers appeared to have an overabundance of these species due the disequilibrium 
atmosphere following a different TP profile than the equilibrium one. 
For example, at 1300K the disequilibrium model had more 
H$_2$O than the equilibrium model deeper in the atmosphere ($\gtrsim$10~bar) and in a 
narrow pressure layer around $\sim$0.5 and $\sim$2~bar, 
but it was depleted in H$_2$O in all other layers. 
CH$_4$ was depleted at all pressures $\lesssim$ 1 bar and was overabundant for deeper 
layers. NH$_3$ was overabundant at all pressures.

On average, the J and H band probed the same pressure layers in our 1300~K model atmospheres.
The H band (at a high resolution) though, probed a wider range of pressure across 
the band, varying from $\sim$1 bar at the edges, down to $\sim$15 bar in the 
center of the band (Fig.~\ref{fig:representative_cf}). The J band on the other hand, probed the $\sim$15-20 bar region 
throughout the band, with the exception of some narrow lines where lower pressures were 
probed and the wings of the band where pressures around $\sim$3.5 
bar were probed. The pressures probed by both J and H bands were depleted in H$_2$O, 
had slightly overabundant NH$_3$, while CH$_4$ was slightly overabundant for J and depleted 
in pressures probed by the H band. This resulted in the J band being 
dimmer for the disequilibrium model than the equilibrium one ($\delta F \sim$14\%), 
while the H band had 
comparable brightness for the disequilibrium and equilibrium models. 
This resulted in redder J-H colors for the disequilibrium model.

The disequilibrium model K band got dimmer than the equilibrium model K band ($\delta F \sim$33\%), 
resulting in overall bluer J-K colors than the equilibrium chemistry models. The 
reason for the dimmer K band was that its longer wavelengths are dominated by a  
CH$_4$ and a NH$_3$ band, and its shorter ones by an H$_2$O band. The K band 
probed pressures between $\sim$0.5 and $\sim$4.5 bar
were NH$_3$ was overabundant by $\sim$50\%-150\% (relative variation). 
In these pressures H$_2$O was slightly overabundant as well, while CH$_4$ was slightly 
underabundant. 
This resulted in a dimmer K-band for $\lambda\lesssim$ 
2.15~$\mu$m. The longer wavelength part of the K band was slightly 
dimmer or comparable to the equilibrium model one, due to including an NH$_3$ and a CH$_4$ 
window. Finally, errors in our CIA opacity database could also affect our K band and result in further dimming as for Gl570D (see Sect.~\ref{sect:gl570d}). 
Overall this resulted in a dimmer K-band (relative change of $\sim$34\% vs  $\sim$13\% for the 
J-band for the $\log K_{zz}$= 7 model) and bluer J-K and H-K colors.

For intermediate models at $\sim$900~K and $\log g$=5.25, H$_2$O and CH$_4$ were depleted
for most pressures except the deeper atmosphere (deeper than $\sim$10bar for $\log K_{zz}$= 4 
to $\sim$30bar for $\log K_{zz}$= 7).  Finally NH$_3$ was
depleted for all pressures $\lesssim$ 10 bar. The pressure ranges probed by the J, H 
and K bands in our 900K model differ from those at 1300K. The J band for the 900K model probed pressures 
around 50 bar (see Fig.~\ref{fig:representative_cf}), which resulted in the J band being dimmer for the disequilibrium than the 
equilibrium models (relative $\delta F \lesssim$18\%), since the pressure range probed 
covers areas where H$_2$O is overabundant by a few percent ($\sim$4\%; 
all percentages are relative variations).
The H band probed pressures around 15 bar. In the pressures probed by the H band, 
CH$_4$ was depleted by $\sim$20-30\% and NH$_3$ was overabundant by $\sim$10\%-30\% 
(for $\log K_{zz}$= 4 and 7), which resulted in an overall dimmer H band (average 
relative $\delta F \sim$ 20\%) and a slightly bluer J-H color for our disequilibrium atmosphere 
than the equilibrium atmosphere. Finally, the K band of our disequilibrium model 
became dimmer than the equilibrium one (average 
relative $\delta F \sim$ 46\%) due to an overabundance of NH$_3$ and 
CH$_4$ in the pressures probed by the band. This resulted in bluer J-K colors for the 
disequilibrium atmospheres. 

Finally, for the even colder models at $\sim$650~K, the pressures probed by the J, H and K bands 
are $\gtrsim$25 bar, which are overabundant in CH$_4$, NH$_3$ and H$_2$O, resulting in a 
dimmer J band (relative $\delta F \sim$ 48\%), a less dim H band 
(relative $\delta F \sim$ 35\%) and a dimmer K band  (relative $\delta F \sim$ 81\%) 
resulting in redder J-H colors and bluer J-K and H-K colors.

In Fig.~\ref{fig:cmd_jh_hk_dupuy} we plot our M$_\mathrm{J}$ vs J-H  and J-K 
CMD against the observations of \citet[][]{dupuy12} (top and middle panel), %\textcolor{red}{
and the Spitzer IRAC M$_\mathrm{Ch1}$ vs Ch1-Ch2 against the ensemble of T and Y dwarfs presented in \citet[][]{kirkpatrick2019}. %}. 
Note that 
for this plot we used the MKO and Spitzer IRAC filters on 
our model spectra, to match the data from \citet[][]{dupuy12}  and \citet[][]{kirkpatrick2019}.  
We also note that we don't intend this plot as a characterization 
effort for any of the \citet[][]{dupuy12} or the \citet[][]{kirkpatrick2019} targets since both our disequilibrium 
and equilibrium models are cloud-free, solar metallicity models 
while a number of these targets are expected to be (at least partially) 
cloudy and could potentially have non-solar metallicities. 
Finally, to keep our plot consistent with the data we only 
plot models with $T_\mathrm{eff} \lesssim$ 1000 K for the Spitzer IRAC dataset.

The disequilibrium chemistry models turned the J-H colors of 
our atmospheres redder than the equilibrium chemistry models for the 
later T-type atmospheres. This is in agreement with observations. 
However, comparing our models with the 
\citet[][]{dupuy12} observations in the T-dwarf regime it can be seen that our 
models are still bluer than the data for intermediate and low gravities 
(3.5 and 4.25 here). All our model atmospheres are cloud-free. 
The introduction of clouds and hazes in an atmosphere is known to turn the colors 
of atmospheres redder \citep{morley12}. The color discrepancy between our model atmospheres and observations 
necessitates extending the disequilibrium chemistry model grid to cloudy atmospheres. 
This will be part of a future paper.
Based on their J-H and H-K (not shown here) colors a 
number of the observed early T dwarfs reside in both the equilibrium and
disequilibrium chemistry space. When J-H, J-K and H-K  is taken into account, our 
disequilibrium chemistry models fit better the mid T and later dwarfs than the equilibrium models. 
Additionally, the disequilibrium chemistry models give 
a better fit to the Spitzer 3.6 $\mu$m and 4.5 $\mu$m (Ch1 
and Ch12; bottom panel of Fig.~\ref{fig:cmd_jh_hk_dupuy}) 
for most of the ensemble of T dwarfs presented in \citet[][]{kirkpatrick2019}. 
For the latest T ($\gtrsim$T8.5) and Y dwarfs disequilibrium chemistry alone cannot provide a good match to the observed Spitzer colors. However, clouds are expected to play a crucial role in these atmospheres \citep{morley12} and affect their colors. 
In a future paper we will extend the disequilibrium chemistry model grid to cloudy atmospheres and revisit this plot. %}

Our findings support the importance of 
disequilibrium chemistry for T type dwarfs, which was already 
suggested by \citet[][]{saumon06,saumon07} based on observations of Gl570D, 
2MASSJ04151954-0935066 and 2MASSJ12171110-0311131.  A number of the T dwarfs in our sample like 2MASSJ11145133-2618235 \citep[][]{legget07}, 
ULASJ141623.94+134836.3 \citep[][]{burgasser10}, 
2MASSJ09393548-2448279 \citep[][]{burgasser08} and 2MASSJ12373919+6526148 \citep[][]{liebert07} have already been suggested 
to be in disequilibrium. For example, 
\citet[][]{burgasser08} showed that the spectrum of 2M0939 (T8)
was best fit with a $\log K_{zz}$=4 model. ULAS J1416+1348 
(T7.5) was also found to be best-fit by a $\log K_{zz}$=4 model by 
\citet[][]{burgasser10}. On the other hand, 2M1237 (T7) and 2M1114 (T7.5) were 
observed to have faint K-bands which \citet[][]{liebert07} and \citet[][]{legget07}
attributed to a subsolar metallicity ([m/H]$\sim$-0.3 for 2M1114) or 
high gravity. These authors did not explore the possibility of 
disequilibrium chemistry for these targets. Long wavelength coverage spectra 
that help retrieve the abundances of multiple species would 
allow us to disentangle the effect of metallicity and 
disequilibrium chemistry for these targets. %}

\section{Discussion and Conclusions}\label{sect:discussion}

\emph{JWST} will enable the imaged exoplanet and brown dwarf community to study in more detail 
a range of exoplanet and brown dwarf atmospheres and perform comparative studies of 
their properties 
as a function of atmospheric properties ($T_\mathrm{eff}$, $\log g$, 
metallicity etc).  
The long-wavelength-coverage, high quality spectra of exoplanet and brown dwarf atmospheres 
that \emph{JWST} will acquire will allow us to probe \textit{simultaneously} 
a wider range of pressures than ever before and constrain 
chemistry and cloud changes in atmospheres as a function of pressure. 
\emph{JWST} will also provide us with time-resolved observations of 
imaged exoplanets and cooler brown dwarfs of comparable 
quality to what \emph{HST} does for L/T transition brown dwarfs today\citep{kostov13}.  
This will allow us to constrain the time-variability of chemistry and clouds 
in these atmospheres. However, to do that accurately we will need models that 
properly account for vertical mixing in the atmosphere. 

The departure of an atmosphere from equilibrium chemistry, i.e., how 
strong its vertical mixing is, 
depends on the atmosphere's properties and the eddy diffusion coefficient 
($\log K_{zz}$ in this paper) of an
atmosphere. The latter is also an important parameter for constraining the 
cloud formation in the atmosphere \citep[e.g.,][]{marley13}, thus 
observational constrains of $\log K_{zz}$ are of high 
importance to the community. 
Recently, e.g., \citet{miles20} presented low resolution ground-based observations of 
seven late T to Y brown dwarfs and compared their observations against models 
of atmospheres with disequilibrium chemistry 
to constrain the $\log K_{zz}$ of these atmospheres. \citet{miles20} 
showed that $\log K_{zz}$ spans a range of 
values in these atmospheres from 4 to 8.5, and discussed how comparing 
these values against the maximum $\log K_{zz}$ predicted from theory
can help constrain the existence of detached convective zones in warmer atmospheres.
In the coming decade \emph{JWST} will allow for the first time to 
constrain changes in $\log K_{zz}$ as a function of pressure in
atmospheres and potential trends with atmospheric properties such as
$T_\mathrm{eff}$ and $\log g$. Such observations will allow us to constrain 
in unprecedented detail the vertical structure of atmospheres, 
including the potential existence of detached convective zones predicted by theory 
\citep[see, e.g.,][]{marley15}. The accuracy of our 
results though, will depend on the accuracy of our models. For this reason  
sets of models that use a self-consistent scheme to study 
the effect of quenching on atmospheres are necessary for the \emph{JWST} era.

In this paper we presented an extension of the atmospheric structure code of M.~S.~Marley and collaborators (e.g., \citealt[][]{marley19} to calculate the TP and composition profiles of atmospheres in disequilibrium, in a self-consistent way (see Sect.\ref{sect:code}). We validated the new code against its well tested equilibrium chemistry version, as well as previously published results (Sect.~\ref{sect:validation}), and 
tested the fit of our new non-equilibrium models against
observations of Gl570D (Sect.~\ref{sect:gl570d}), which is considered the archetypal cloud-free 
brown dwarf atmosphere with signs of disequilibrium chemistry.  
A number of differences between our models and the observed spectra were noted 
that are related to inaccuracies in our alkali opacity database for wavelengths
shortward of 1.1~$\mu$m and, potentially, our CIA opacity in the K-band. Addressing these issues is part of ongoing work.

The extension of our code presented here, opens up the possibility 
to model more complex atmospheres with variable $\log K_{zz}$ 
profiles in the future. 
\citet{flasar1978} and later \citet{dong2014}, e.g., 
predicted that Jupiter and Saturn will show latitudinal variation of 
$\log K_{zz}$ due to the planet's rotation. \citet{dong2014}, 
e.g., showed that the $\log K_{zz} (P)$ profiles in Jupiter 
should change with latitude, with the value of 
$\log K_{zz}$ changing by more than an order of magnitude for a 
given temperature layer between $0^\circ$ and $80^\circ$ latitude. The 
changes in Saturn were about two orders of magnitude. Similar 
variable profiles with pressure and latitude could exist in imaged 
atmospheres and they would affect the chemical profiles and 
cloud formation in these atmospheres. Additionally, the existence 
of detached convective zones in some atmospheres would locally 
change the $\log K_{zz}$ profile, making the use of a simple 
$\log K_{zz}$ profile as used in this paper inaccurate for the 
characterization of the three-dimensional structure of these atmospheres.
Thus, updated models with variable $\log K_{zz}$ may be needed 
in the future.

\begin{figure}[]
\centering
\includegraphics[width=\linewidth]{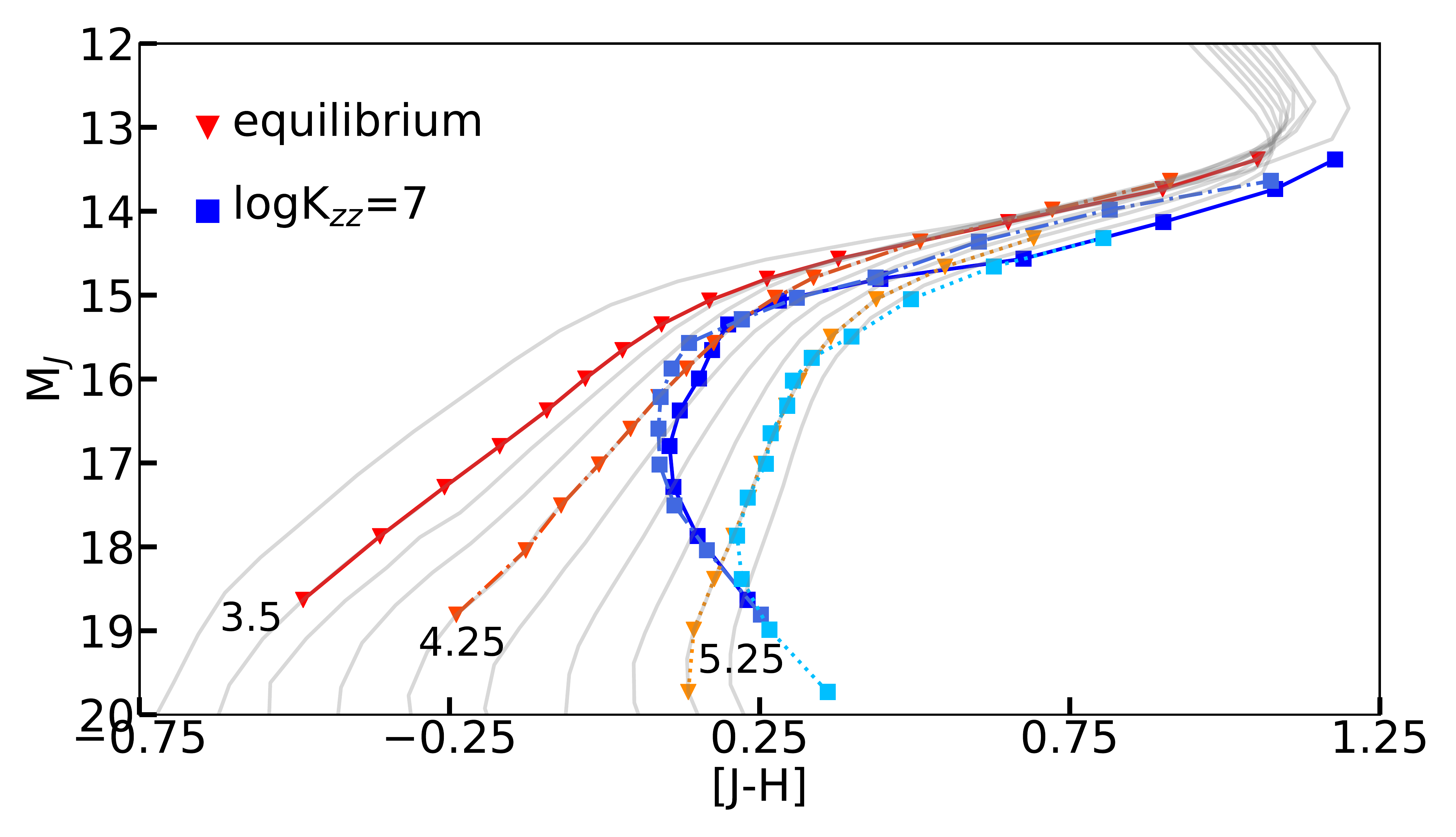}
\centering
\includegraphics[width=\linewidth]{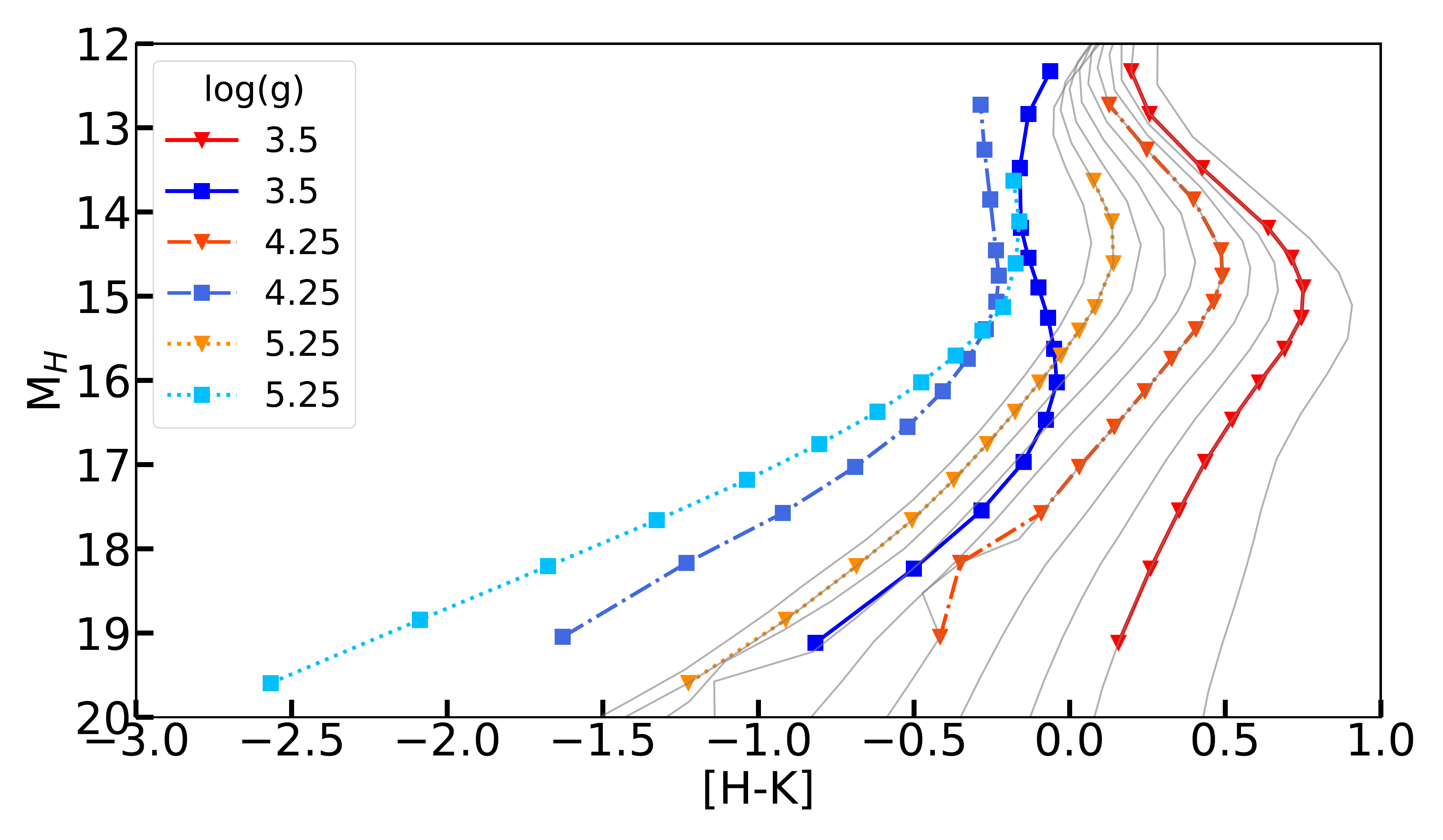}
\centering
\includegraphics[width=\linewidth]{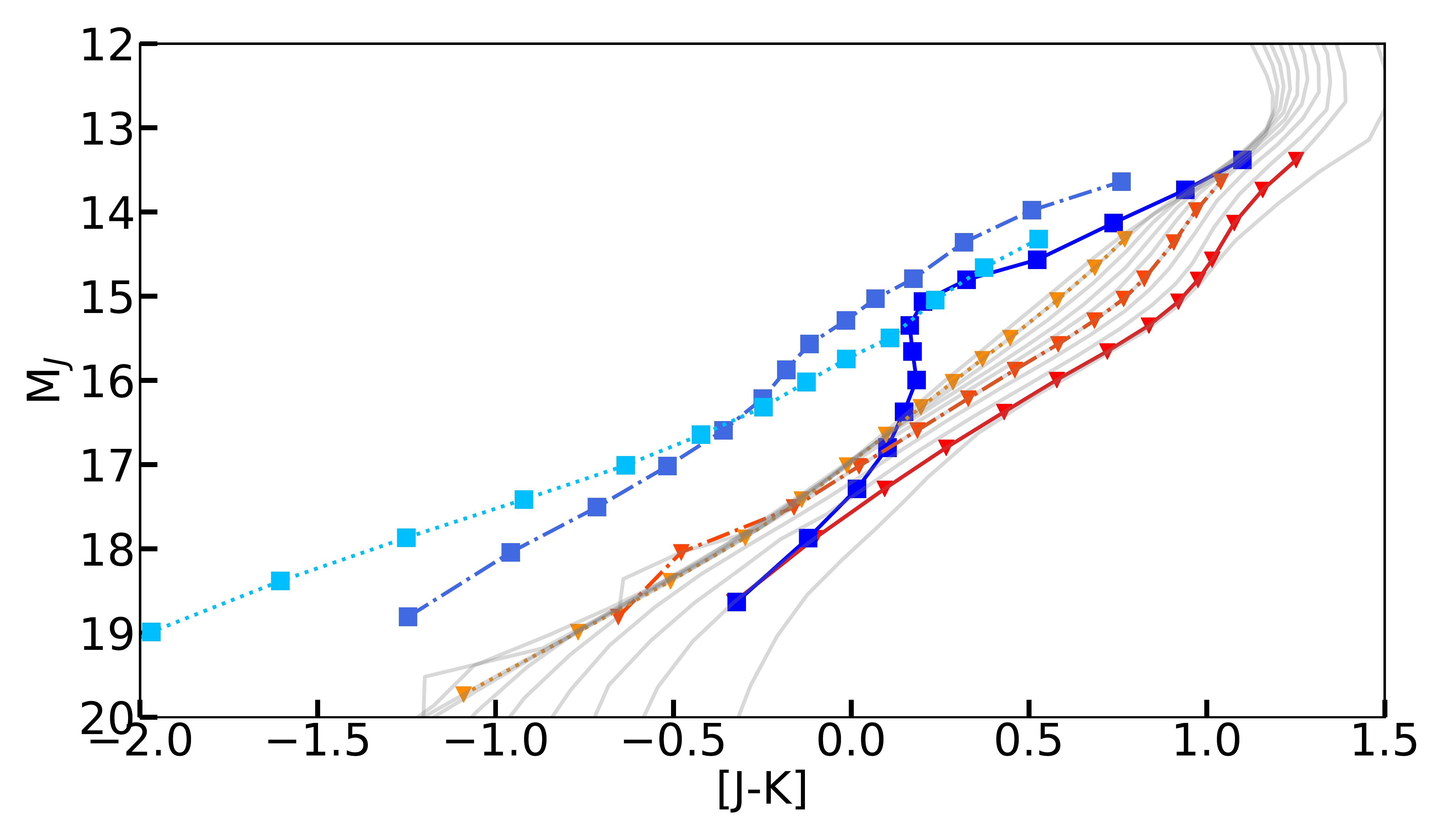}
\caption{Color Magnitude Diagrams (CMDs) for three gravities ($\log g=3.5$ (solid lines), 
4.25 (dashed-dotted lines) and 5.25 (dotted lines)) of our equilibrium 
atmospheres (red-orange triangles) and disequilibrium atmospheres with 
$\log K_{zz}$=7 (blue squares). Radii computed from our equilibrium model evolution (Marley et al. 2021). Our model atmospheres range from 500K to 
1000K with a step of 50K, and 1000K to 1300K with a step of 100K. Our $J$, $H$ and $K$  correspond to {\em JWST} F115W, F162M, and F210M respectively. Overplotted with gray lines are the equilibrium tracks from the Sonora Bobcat models \citet{marley19}. The gravities of the overplotted Bobcat models range from $\log g = 3.0$ to $5.5$ with a step of 0.25. }
\label{fig:cmd_all}
\end{figure}

\begin{figure}[]
\centering
\includegraphics[width=\linewidth]{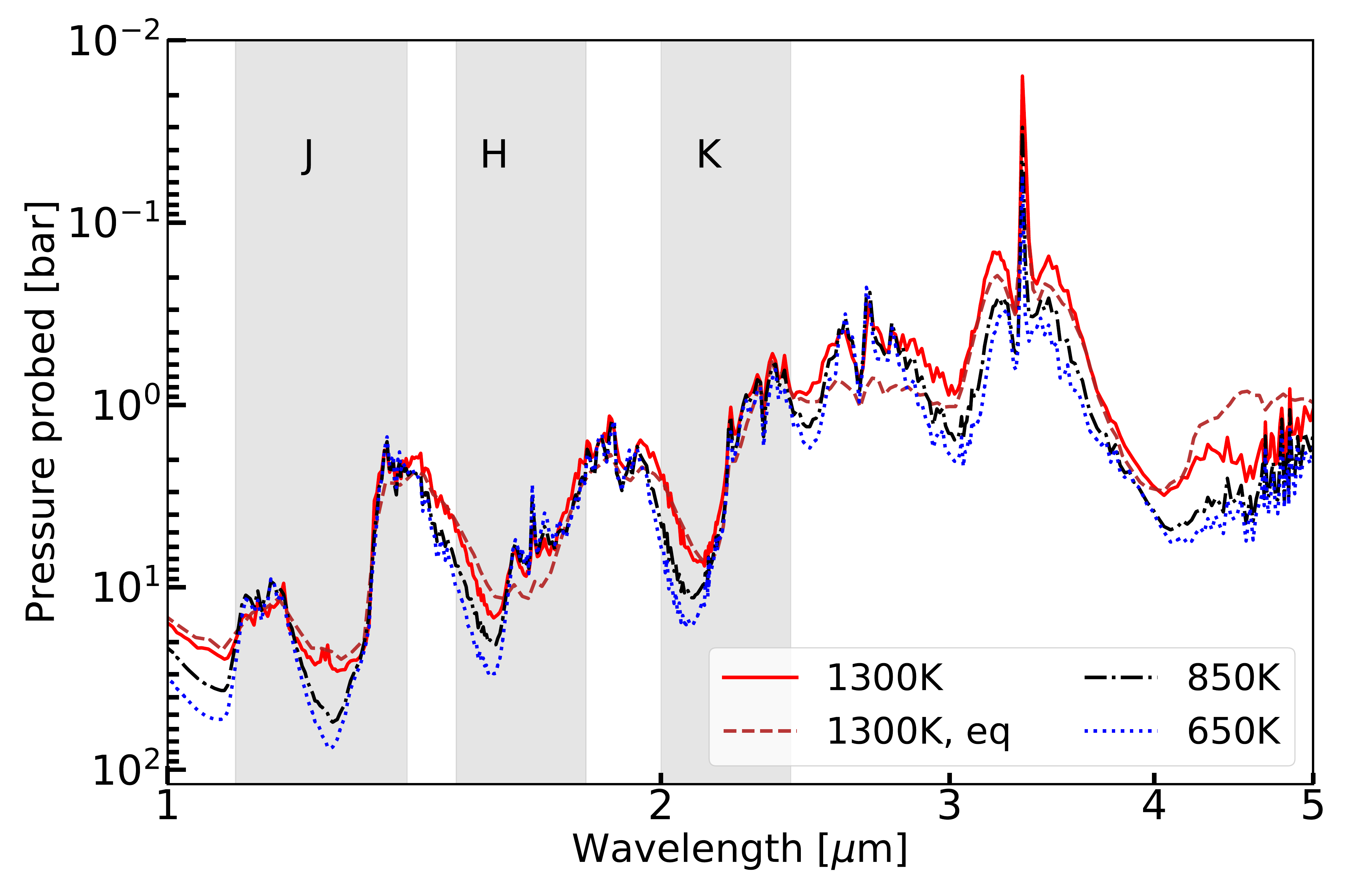}
\caption{ Different bands probe different pressures in an atmosphere.  
Here shown are the pressures probed at different wavelengths 
for our $\log K_{zz}$=4 models with 
$T_\mathrm{eff}=$ 1300 K (red line), 850 K (black line) and 650 K (blue 
line) and $\log g=5.25$. Also shown is the equilibrium 1300 K model 
(brown, dashed line) for comparison. The J, H and K band passes are also 
shown for convenience. The abundance profiles of the major absorbents in the pressures probed by the different bands differ between our equilibrium and 
disequilibrium model atmospheres. These differences lead to different 
amounts of absorption and thus colors for the disequilibrium 
and the equilibrium atmospheres. }
\label{fig:representative_cf}
\end{figure}

\begin{figure}[]
\centering
\includegraphics[width=\linewidth]{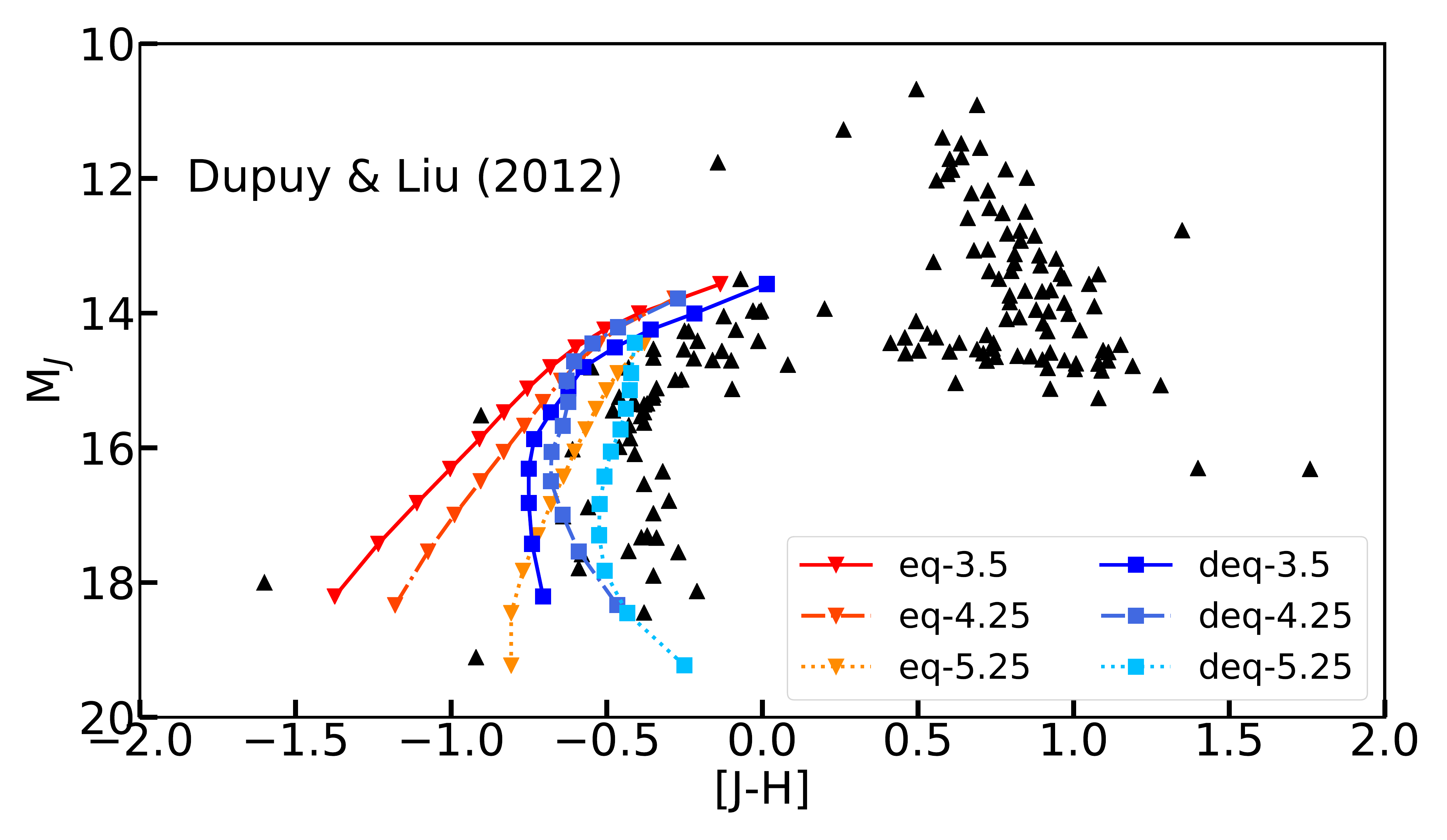}
\centering
\includegraphics[width=\linewidth]{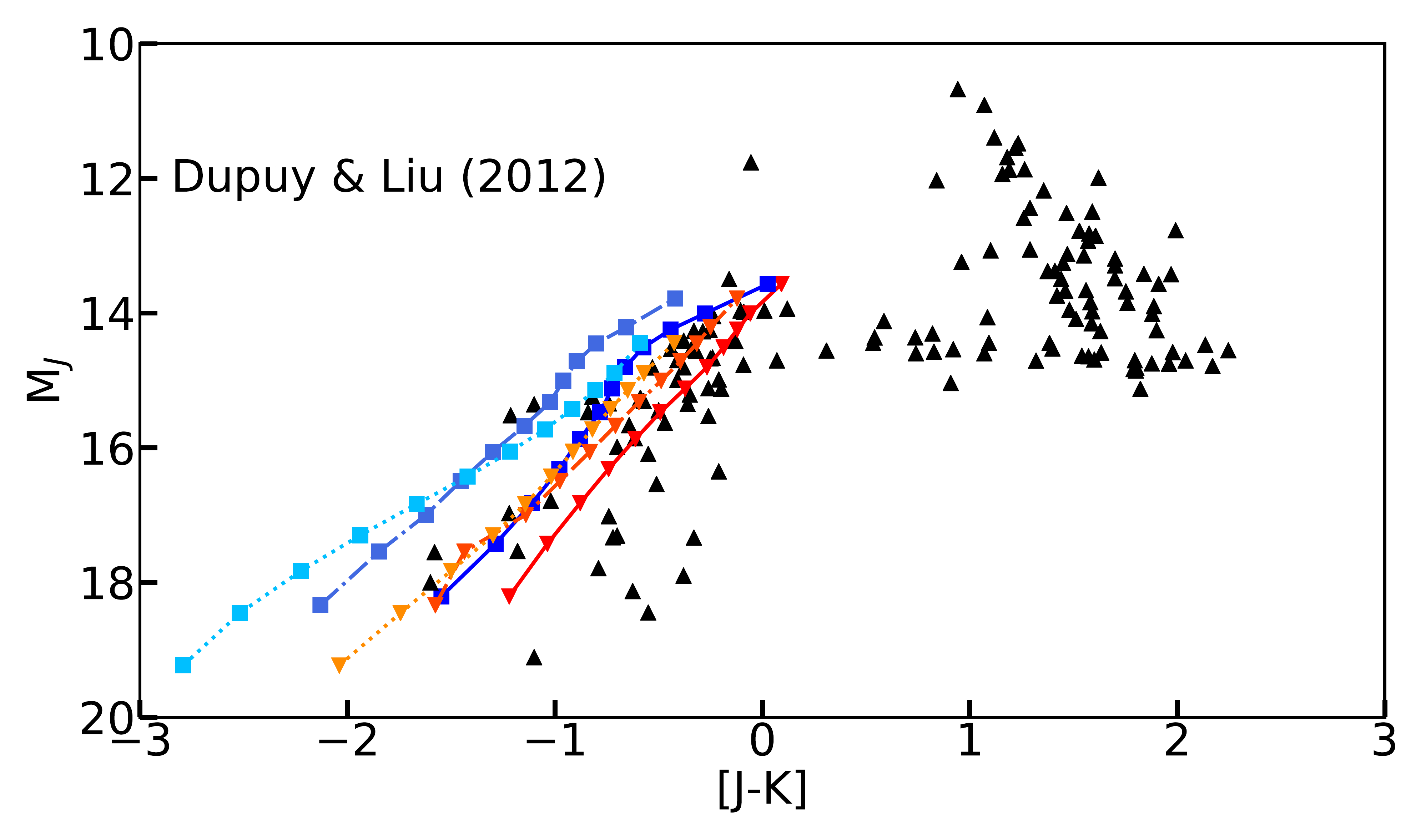}
\includegraphics[width=\linewidth]{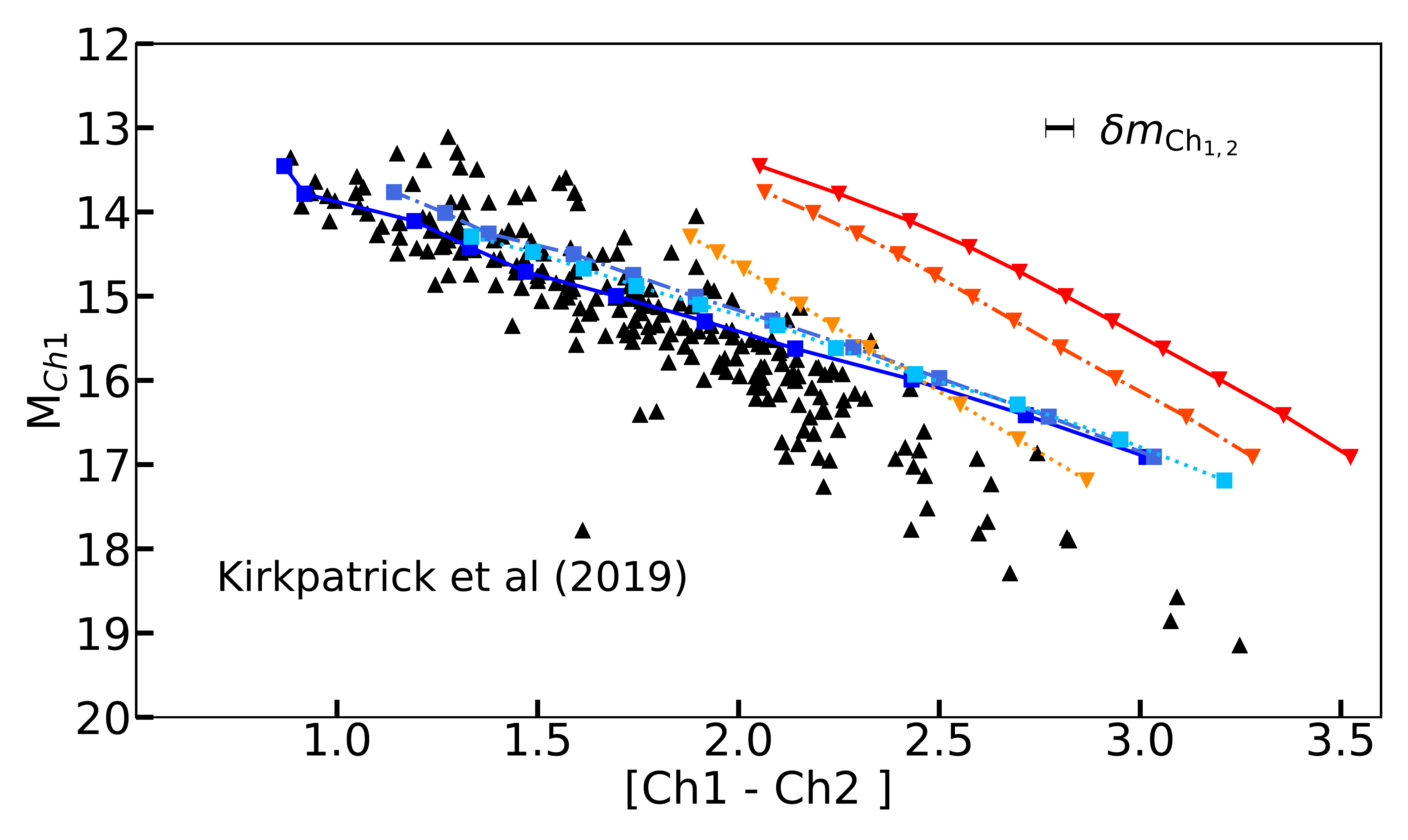}
\caption{CMDs of our equilibrium and 
disequilibrium atmospheres with $\log K_{zz}$=7, and 
the MKO observations of \citet[][]{dupuy12} (top and middle panel) and Spitzer data from
\citet[][]{kirkpatrick2019} (bottom panel) . Color and line coding are similar 
to Fig.~\ref{fig:cmd_all}. Note that for this plot we used the MKO (top and middle panel) 
and Spitzer IRAC filters (bottom panel) on our model spectra, to match the observations. 
For the IRAC data we also plot the average error bar ($\delta m$) in the photometry for reference. }
\label{fig:cmd_jh_hk_dupuy}
\end{figure}

In the \emph{JWST}  era the Direct Imaging community will have 
access to high resolution observations of imaged atmospheres in the 
near and mid infrared. In Sect.~\ref{sect:detect} we showed that using NIRISS 
and NIRSpec observations we should be able to distinguish between 
(at least the cloud-free) cooler atmospheres with disequilibrium or 
equilibrium chemistry. In particular, both NIRISS and NIRSpec should allow 
the detection of disequilibrium in H$_2$O, CH$_4$ and NH$_3$, while the longer
wavelength observations of 
NIRSpec should also allow the detection of CO in disequilibrium. 
In Sect.~\ref{sect:gl570d} when we compared our best-fit model 
volume mixing ratio of NH$_3$ against the retrieved ratio 
of \citet[][]{line15}, we showed that omitting the 
10~$\mu$m -12~$\mu$m observations could have affected their best-fit model. 
This suggests that in the \emph{JWST} era MIRI MRS observations will also 
be important for accurate NH$_3$ retrievals.

\begin{figure}[]
\centering

\includegraphics[width=\linewidth]{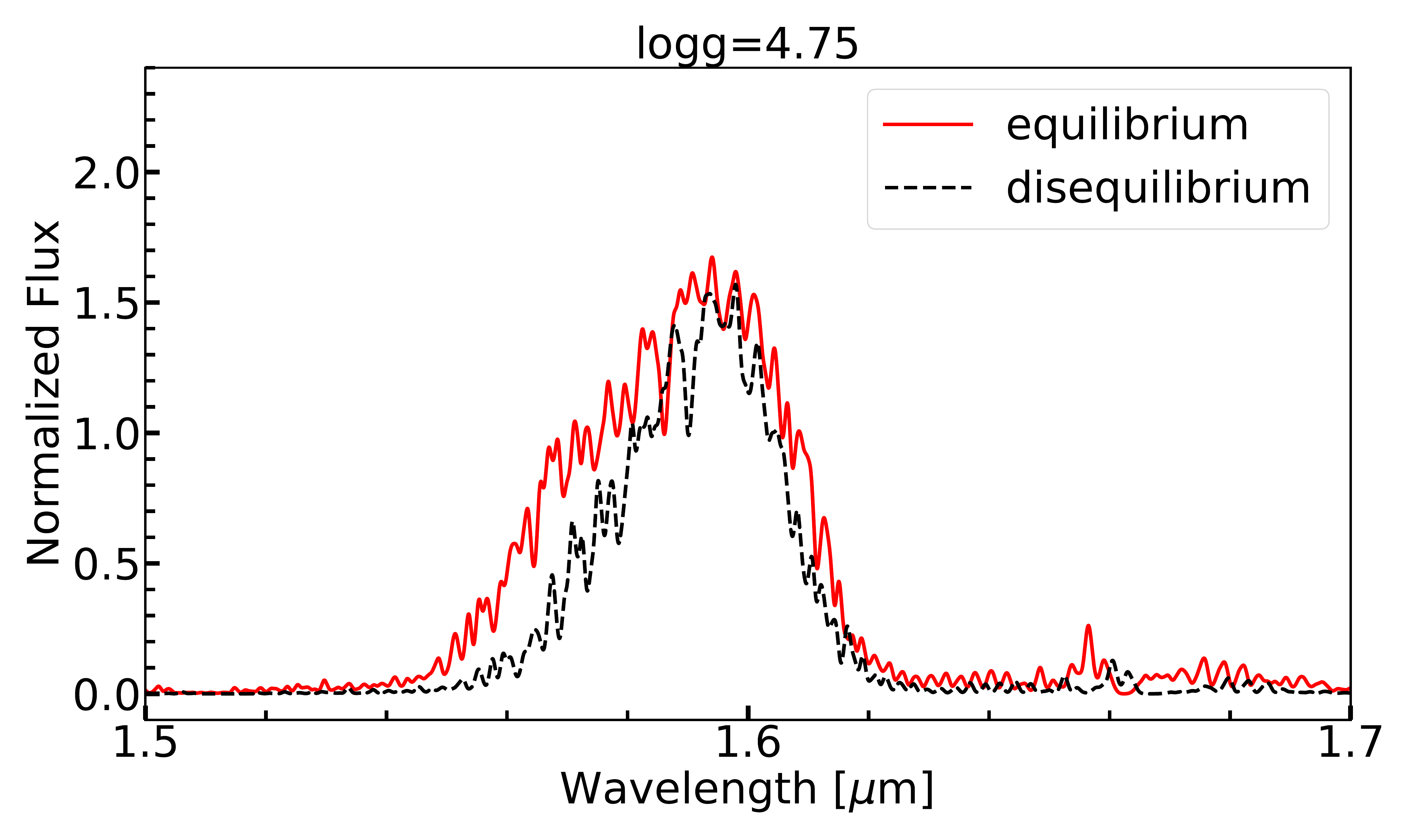}
\caption{Normalized H-band for model atmospheres with $\log g$=4.75 
at 350K. Solid line is an atmospheres with equilibrium 
chemistry and dashed line with $\log K_{zz}$=7. The models are binned down to a resolution of 1800 for plot clarity. The model in disequilibrium shows a 
comparable NH$_3$ enhancement to the tentative detection on WISEP J1738+2732  \citep[][]{cushing11}.}
\label{fig:hband_350k}
\end{figure}

NH$_3$ has been detected in disequilibrium in some cooler T dwarfs
\citep[][]{canty15} and Y dwarfs \citep[][]{cushing11}. 
Overall our coolest disequilibrium model at 500~K, for 
$\log g$= 5.0  was depleted in NH$_3$ (relative difference of 
-50\%) as expected. This was detectable in the major NH$_3$ feature around
10.5~$\mu$m, which showed a lack of NH$_3$ in the quenched model atmosphere. 
However, in the deeper pressures probed in the 
1.0-1.3~$\mu$m and 1.5-1.6~$\mu$m  windows, which include 
NH$_3$ absorption windows, the cooling in the TP profile  
of our atmosphere (at those pressures) resulted in an overabundance 
of NH$_3$ which should be detectable for some atmospheres. We are 
currently extending our grid  
to cooler atmospheres (down to 200~K), in the realm of Y-dwarfs, 
where \citet[][]{cushing11} reported a possible detection of NH$_3$ in the atmosphere 
of WISEP-J1738 (350~K - 400~K) in the 1.5-1.6~$\mu$m  window.  
Fig.~\ref{fig:hband_350k} shows an example of how NH$_3$ excess in 
our cooler atmospheres changed the H-band in a comparable way to the 
tentative detection on WISEP J1738+2732 by \citealp[][]{cushing11}.

A similar dependence on gravity appeared in our detection of CO, with the lower gravity models showing detectable disequilibrium CO absorption around 4.7~$\mu$m for the cooler atmospheres. Finally, for most of our model atmospheres the detection of CH$_4$ is easier at higher gravities than at lower gravities, (see Fig.~\ref{fig:jwst_nirspec_g395h}) in agreement with observations that suggest that low gravity planetary mass objects are depleted in CH$_4$ in comparison to their brown dwarf analogues (that have a higher surface gravity).

In this paper we showed that, at least for cloud-free atmospheres, disequilibrium 
chemistry results in redder J-H colors for our model atmospheres
(Sect.~\ref{sect:colors}). For 
smaller values of $\log K_{zz}$ only the cooler quenched 
atmospheres ($\lesssim$ 700~K) turned red at all gravities. At 
$\log K_{zz}$ = 7 disequilibrium chemistry resulted in redder colors 
even for atmospheres as hot as 1200~K (Fig.~\ref{fig:cmd_all} and 
Sect.~\ref{sect:colors}). 
A number of brown dwarfs and planetary mass objects have been 
detected that are redder in J-H than their standard field counterparts, like WISEP 
J004701.06+680352.1 \citep[W0047;][]{gizis12}, 2MASS J12073346-3932539, the planets 
of the HR8799 system \citep[2M1207b][]{oppenheimer13,currie11,skemer14} and 
others. 
Most of these atmospheres are expected 
to be cloudy. For example, both W0047 and 2M1207b show rotational variability which 
is related to cloud patchiness \citep[][]{lew16,zhou16}, and planetary mass 
companion observations are best fit by cloudy models 
\citep[e.g.,][]{skemer14}. Clouds will affect the colors of the atmosphere, 
but a number of these atmospheres are also CH$_4$ depleted \citep[][]{biller18} 
hinting to disequilibrium chemistry in their atmospheres 
\citep[see, e.g.,][]{barman15}. Our cloud-free model atmospheres at the 
temperatures and gravities that are representative of the planets of the 
HR8799 system (900~K- 1100~K and $\log g$ $\lesssim$4) and 
for $\log K_{zz}$ = 7  
\citep[][]{barman15} became redder in J-H with $\delta$[J-H] = 0.12 to 0.26. 
Observations of the HR8799 system planets found a color difference from 
field brown dwarfs that is comparable or larger 
than what our cloud-free atmospheres showed. 
Clouds and disequilibrium chemistry are expected to interplay in the 
atmosphere and affect its color.
Clouds are expected to turn atmospheres redder 
and deplete some of the available chemical 
species changing the mixing ratio of species in the atmosphere.  This hints to 
the importance of extending the Sonora grid to cloudy atmospheres with 
disequilibrium chemistry, which will be part of future work.

Finally, we compared the Color Magnitude Diagrams (CMD) 
of our disequilibrium and equilibrium models 
against MKO observations of brown dwarfs by \citet[][]{dupuy12}  and Spitzer data from \citet{kirkpatrick2019} (Sect.~\ref{sect:colors}).
We noted that our disequilibrium chemistry models give a better fit 
to photometry of mid to late T type brown dwarfs, supporting the importance of disequilibrium chemistry for T type dwarfs \citep[][]{saumon06,saumon07}. In particular, only 
disequilibrium models could fit the Spitzer colors of \citet[][]{kirkpatrick2019}. The 
equilibrium models have more CH$_4$ and less CO than the observed atmospheres so they 
appear redder than the latter. On the other hand, disequilibrium chemistry increases the CO content 
of the atmosphere and reduces its CH$_4$ content (Fig.~\ref{fig:jwst_nirspec_g395h}), which 
results to bluer colors in agreement with the observations. %}
A number of these target atmospheres have already been suggested 
to be in disequilibrium. Previous fitting of spectra for some of these 
targets suggested that a subsolar metallicity or high gravity, may be responsible 
for their different colors, but the authors did not explore the possibility of 
disequilibrium chemistry for these targets. Refitting these observations with 
equilibrium and disequilibrium chemistry models at different metallicities would be 
of interest. Extending the Sonora grid to atmospheres of different metallicities 
with disequilibrium chemistry will be part of future work.

Following the Sonora bobcat models, our models will be archived in Zenodo under \url{https://zenodo.org/record/4450269}.

\bigskip
\emph{Acknowledgements.} 
TK acknowledges the University of Central Florida Advanced Research Computing Center high-performance computing resources made available for conducting the research reported in this paper (\url{https://arcc.ist.ucf.edu}). Part of this work was performed under the auspices of the U.S. Department of Energy under Contract No.
89233218CNA000001. JJF, MSM, CVM, RL, and RSF acknowledge the support of NASA Exoplanets Research Program grant 80NSSC19K0446. 
%mention Stokes as well

 \newcommand{\noop}[1]{}


\begin{thebibliography}{66}
\expandafter\ifx\csname natexlab\endcsname\relax\def\natexlab#1{#1}\fi

\bibitem[{{Allard}(2014)}]{allard14}
{Allard}, F. 2014, in IAU Symposium, Vol. 299, Exploring the Formation and
  Evolution of Planetary Systems, ed. M.~{Booth}, B.~C. {Matthews}, \& J.~R.
  {Graham}, 271--272

\bibitem[{{Allard} {et~al.}(2012){Allard}, {Homeier}, \& {Freytag}}]{allard12}
{Allard}, F., {Homeier}, D., \& {Freytag}, B. 2012, Philosophical Transactions
  of the Royal Society of London Series A, 370, 2765

\bibitem[{{Amundsen} {et~al.}(2017){Amundsen}, {Tremblin}, {Manners},
  {Baraffe}, \& {Mayne}}]{amundsen17}
{Amundsen}, D.~S., {Tremblin}, P., {Manners}, J., {Baraffe}, I., \& {Mayne},
  N.~J. 2017, \aap, 598, A97

\bibitem[{{Barman} {et~al.}(2015){Barman}, {Konopacky}, {Macintosh}, \&
  {Marois}}]{barman15}
{Barman}, T.~S., {Konopacky}, Q.~M., {Macintosh}, B., \& {Marois}, C. 2015,
  ApJ, 804, 61

\bibitem[{{Biller} \& {Bonnefoy}(2018)}]{biller18}
{Biller}, B.~A. \& {Bonnefoy}, M. {Exoplanet Atmosphere Measurements from
  Direct Imaging}, 101

\bibitem[{{Buenzli} {et~al.}(2015){Buenzli}, {Saumon}, {Marley}, {Apai},
  {Radigan}, {Bedin}, {Reid}, \& {Morley}}]{buenzli15}
{Buenzli}, E., {Saumon}, D., {Marley}, M.~S., {Apai}, D., {Radigan}, J.,
  {Bedin}, L.~R., {Reid}, I.~N., \& {Morley}, C.~V. 2015, ApJ, 798, 127

\bibitem[{{Burgasser} {et~al.}(2010){Burgasser}, {Cruz}, {Cushing}, {Gelino},
  {Looper}, {Faherty}, {Kirkpatrick}, \& {Reid}}]{burgasser10}
{Burgasser}, A.~J., {Cruz}, K.~L., {Cushing}, M., {Gelino}, C.~R., {Looper},
  D.~L., {Faherty}, J.~K., {Kirkpatrick}, J.~D., \& {Reid}, I.~N. 2010, ApJ,
  710, 1142

\bibitem[{{Burgasser} {et~al.}(2006){Burgasser}, {Geballe}, {Leggett},
  {Kirkpatrick}, \& {Golimowski}}]{burgasser06}
{Burgasser}, A.~J., {Geballe}, T.~R., {Leggett}, S.~K., {Kirkpatrick}, J.~D.,
  \& {Golimowski}, D.~A. 2006, ApJ, 637, 1067

\bibitem[{{Burgasser} {et~al.}(2004){Burgasser}, {McElwain}, {Kirkpatrick},
  {Cruz}, {Tinney}, \& {Reid}}]{burgasser04}
{Burgasser}, A.~J., {McElwain}, M.~W., {Kirkpatrick}, J.~D., {Cruz}, K.~L.,
  {Tinney}, C.~G., \& {Reid}, I.~N. 2004, \aj, 127, 2856

\bibitem[{Burgasser {et~al.}(2008)Burgasser, Tinney, Cushing, Saumon, Marley,
  Bennett, \& Kirkpatrick}]{burgasser08}
Burgasser, A.~J., Tinney, C.~G., Cushing, M.~C., Saumon, D., Marley, M.~S.,
  Bennett, C.~S., \& Kirkpatrick, J.~D. 2008, \apj, 689, L53

\bibitem[{{Burningham} {et~al.}(2011){Burningham}, {Leggett}, {Homeier},
  {Saumon}, {Lucas}, {Pinfield}, {Tinney}, {Allard}, {Marley}, {Jones},
  {Murray}, {Ishii}, {Day-Jones}, {Gomes}, \& {Zhang}}]{burningham11}
{Burningham}, B., {Leggett}, S.~K., {Homeier}, D., {Saumon}, D., {Lucas},
  P.~W., {Pinfield}, D.~J., {Tinney}, C.~G., {Allard}, F., {Marley}, M.~S.,
  {Jones}, H.~R.~A., {Murray}, D.~N., {Ishii}, M., {Day-Jones}, A., {Gomes},
  J., \& {Zhang}, Z.~H. 2011, \mnras, 414, 3590

\bibitem[{{Burningham} {et~al.}(2017){Burningham}, {Marley}, {Line}, {Lupu},
  {Visscher}, {Morley}, {Saumon}, \& {Freedman}}]{burningham17}
{Burningham}, B., {Marley}, M.~S., {Line}, M.~R., {Lupu}, R., {Visscher}, C.,
  {Morley}, C.~V., {Saumon}, D., \& {Freedman}, R. 2017, MNRAS, 470, 1177

\bibitem[{{Canty} {et~al.}(2015){Canty}, {Lucas}, {Yurchenko}, {Tennyson},
  {Leggett}, {Tinney}, {Jones}, {Burningham}, {Pinfield}, \& {Smart}}]{canty15}
{Canty}, J.~I., {Lucas}, P.~W., {Yurchenko}, S.~N., {Tennyson}, J., {Leggett},
  S.~K., {Tinney}, C.~G., {Jones}, H.~R.~A., {Burningham}, B., {Pinfield},
  D.~J., \& {Smart}, R.~L. 2015, \mnras, 450, 454

\bibitem[{{Currie} {et~al.}(2011){Currie}, {Burrows}, {Itoh}, {Matsumura},
  {Fukagawa}, {Apai}, {Madhusudhan}, {Hinz}, {Rodigas}, {Kasper}, {Pyo}, \&
  {Ogino}}]{currie11}
{Currie}, T., {Burrows}, A., {Itoh}, Y., {Matsumura}, S., {Fukagawa}, M.,
  {Apai}, D., {Madhusudhan}, N., {Hinz}, P.~M., {Rodigas}, T.~J., {Kasper}, M.,
  {Pyo}, T.-S., \& {Ogino}, S. 2011, ApJ, 729, 128

\bibitem[{{Cushing} {et~al.}(2011){Cushing}, {Kirkpatrick}, {Gelino},
  {Griffith}, {Skrutskie}, {Mainzer}, {Marsh}, {Beichman}, {Burgasser},
  {Prato}, {Simcoe}, {Marley}, {Saumon}, {Freedman}, {Eisenhardt}, \&
  {Wright}}]{cushing11}
{Cushing}, M.~C., {Kirkpatrick}, J.~D., {Gelino}, C.~R., {Griffith}, R.~L.,
  {Skrutskie}, M.~F., {Mainzer}, A., {Marsh}, K.~A., {Beichman}, C.~A.,
  {Burgasser}, A.~J., {Prato}, L.~A., {Simcoe}, R.~A., {Marley}, M.~S.,
  {Saumon}, D., {Freedman}, R.~S., {Eisenhardt}, P.~R., \& {Wright}, E.~L.
  2011, ApJ, 743, 50

\bibitem[{{Cushing} {et~al.}(2006){Cushing}, {Roellig}, {Marley}, {Saumon},
  {Leggett}, {Kirkpatrick}, {Wilson}, {Sloan}, {Mainzer}, {Van Cleve}, \&
  {Houck}}]{cushing06}
{Cushing}, M.~C., {Roellig}, T.~L., {Marley}, M.~S., {Saumon}, D., {Leggett},
  S.~K., {Kirkpatrick}, J.~D., {Wilson}, J.~C., {Sloan}, G.~C., {Mainzer},
  A.~K., {Van Cleve}, J.~E., \& {Houck}, J.~R. 2006, \apj, 648, 614

\bibitem[{{Drummond} {et~al.}(2016){Drummond}, {Tremblin}, {Baraffe},
  {Amundsen}, {Mayne}, {Venot}, \& {Goyal}}]{drummond2016}
{Drummond}, B., {Tremblin}, P., {Baraffe}, I., {Amundsen}, D.~S., {Mayne},
  N.~J., {Venot}, O., \& {Goyal}, J. 2016, \aap, 594, A69

\bibitem[{{Dupuy} \& {Liu}(2012)}]{dupuy12}
{Dupuy}, T.~J. \& {Liu}, M.~C. 2012, \apjs, 201, 19

\bibitem[{{Fegley} \& {Lodders}(1996)}]{fegley96}
{Fegley}, Bruce, J. \& {Lodders}, K. 1996, \apjl, 472, L37

\bibitem[{{Flasar} \& {Gierasch}(1978)}]{flasar1978}
{Flasar}, F.~M. \& {Gierasch}, P.~J. 1978, Geophysical and Astrophysical Fluid
  Dynamics, 10, 175

\bibitem[{{Fortney} {et~al.}(2008){Fortney}, {Lodders}, {Marley}, \&
  {Freedman}}]{fortney08}
{Fortney}, J.~J., {Lodders}, K., {Marley}, M.~S., \& {Freedman}, R.~S. 2008,
  \apj, 678, 1419

\bibitem[{{Fortney} {et~al.}(2005){Fortney}, {Marley}, {Lodders}, {Saumon}, \&
  {Freedman}}]{fortney05}
{Fortney}, J.~J., {Marley}, M.~S., {Lodders}, K., {Saumon}, D., \& {Freedman},
  R. 2005, ApJL, 627, L69

\bibitem[{{Geballe} {et~al.}(2009){Geballe}, {Saumon}, {Golimowski}, {Leggett},
  {Marley}, \& {Noll}}]{geballe2009}
{Geballe}, T.~R., {Saumon}, D., {Golimowski}, D.~A., {Leggett}, S.~K.,
  {Marley}, M.~S., \& {Noll}, K.~S. 2009, \apj, 695, 844

\bibitem[{{Geballe} {et~al.}(2001{\natexlab{a}}){Geballe}, {Saumon}, {Leggett},
  {Knapp}, {Marley}, \& {Lodders}}]{geballe2001}
{Geballe}, T.~R., {Saumon}, D., {Leggett}, S.~K., {Knapp}, G.~R., {Marley},
  M.~S., \& {Lodders}, K. 2001{\natexlab{a}}, \apj, 556, 373

\bibitem[{{Geballe} {et~al.}(2001{\natexlab{b}}){Geballe}, {Saumon}, {Leggett},
  {Knapp}, {Marley}, \& {Lodders}}]{geballe01}
---. 2001{\natexlab{b}}, \apj, 556, 373

\bibitem[{{Gizis} {et~al.}(2012){Gizis}, {Faherty}, {Liu}, {Castro}, {Shaw},
  {Vrba}, {Harris}, {Aller}, \& {Deacon}}]{gizis12}
{Gizis}, J.~E., {Faherty}, J.~K., {Liu}, M.~C., {Castro}, P.~J., {Shaw}, J.~D.,
  {Vrba}, F.~J., {Harris}, H.~C., {Aller}, K.~M., \& {Deacon}, N.~R. 2012, \aj,
  144, 94

\bibitem[{{Goody} {et~al.}(1989){Goody}, {West}, {Chen}, \& {Crisp}}]{goody89}
{Goody}, R., {West}, R., {Chen}, L., \& {Crisp}, D. 1989, \jqsrt, 42, 539

\bibitem[{{Hubeny} \& {Burrows}(2007)}]{hubeny07}
{Hubeny}, I. \& {Burrows}, A. 2007, \apj, 669, 1248

\bibitem[{{Kirkpatrick} {et~al.}(2019){Kirkpatrick}, {Martin}, {Smart},
  {Cayago}, {Beichman}, {Marocco}, {Gelino}, {Faherty}, {Cushing}, {Schneider},
  {Mace}, {Tinney}, {Wright}, {Lowrance}, {Ingalls}, {Vrba}, {Munn}, {Dahm}, \&
  {McLean}}]{kirkpatrick2019}
{Kirkpatrick}, J.~D., {Martin}, E.~C., {Smart}, R.~L., {Cayago}, A.~J.,
  {Beichman}, C.~A., {Marocco}, F., {Gelino}, C.~R., {Faherty}, J.~K.,
  {Cushing}, M.~C., {Schneider}, A.~C., {Mace}, G.~N., {Tinney}, C.~G.,
  {Wright}, E.~L., {Lowrance}, P.~J., {Ingalls}, J.~G., {Vrba}, F.~J., {Munn},
  J.~A., {Dahm}, S.~E., \& {McLean}, I.~S. 2019, ApJS, 240, 19

\bibitem[{{Kitzmann} {et~al.}(2019){Kitzmann}, {Heng}, {Oreshenko}, {Grimm},
  {Apai}, {Bowler}, {Burgasser}, \& {Marley}}]{kitzmann19}
{Kitzmann}, D., {Heng}, K., {Oreshenko}, M., {Grimm}, S.~L., {Apai}, D.,
  {Bowler}, B.~P., {Burgasser}, A.~J., \& {Marley}, M.~S. 2019, arXiv e-prints,
  arXiv:1910.01070

\bibitem[{{Kostov} \& {Apai}(2013)}]{kostov13}
{Kostov}, V. \& {Apai}, D. 2013, ApJ, 762, 47

\bibitem[{{Kreidberg} {et~al.}(2014){Kreidberg}, {Bean}, {D{\'e}sert},
  {Benneke}, {Deming}, {Stevenson}, {Seager}, {Berta-Thompson}, {Seifahrt}, \&
  {Homeier}}]{kreidberg14}
{Kreidberg}, L., {Bean}, J.~L., {D{\'e}sert}, J.-M., {Benneke}, B., {Deming},
  D., {Stevenson}, K.~B., {Seager}, S., {Berta-Thompson}, Z., {Seifahrt}, A.,
  \& {Homeier}, D. 2014, Nature, 505, 69

\bibitem[{{Leggett} {et~al.}(2007{\natexlab{a}}){Leggett}, {Marley},
  {Freedman}, {Saumon}, {Liu}, {Geballe}, {Golimowski}, \&
  {Stephens}}]{legget07}
{Leggett}, S.~K., {Marley}, M.~S., {Freedman}, R., {Saumon}, D., {Liu}, M.~C.,
  {Geballe}, T.~R., {Golimowski}, D.~A., \& {Stephens}, D.~C.
  2007{\natexlab{a}}, \apj, 667, 537

\bibitem[{{Leggett} {et~al.}(2007{\natexlab{b}}){Leggett}, {Saumon}, {Marley},
  {Geballe}, {Golimowski}, {Stephens}, \& {Fan}}]{leggett2007b}
{Leggett}, S.~K., {Saumon}, D., {Marley}, M.~S., {Geballe}, T.~R.,
  {Golimowski}, D.~A., {Stephens}, D., \& {Fan}, X. 2007{\natexlab{b}}, \apj,
  655, 1079

\bibitem[{{Lew} {et~al.}(2016){Lew}, {Apai}, {Zhou}, {Schneider}, {Burgasser},
  {Karalidi}, {Yang}, {Marley}, {Cowan}, {Bedin}, {Metchev}, {Radigan}, \&
  {Lowrance}}]{lew16}
{Lew}, B.~W.~P., {Apai}, D., {Zhou}, Y., {Schneider}, G., {Burgasser}, A.~J.,
  {Karalidi}, T., {Yang}, H., {Marley}, M.~S., {Cowan}, N.~B., {Bedin}, L.~R.,
  {Metchev}, S.~A., {Radigan}, J., \& {Lowrance}, P.~J. 2016, ApJL, 829, L32

\bibitem[{Liebert \& Burgasser(2007)}]{liebert07}
Liebert, J. \& Burgasser, A.~J. 2007, \apj, 655, 522

\bibitem[{Line {et~al.}(2015)Line, Teske, Burningham, Fortney, \&
  Marley}]{line15}
Line, M.~R., Teske, J., Burningham, B., Fortney, J.~J., \& Marley, M.~S. 2015,
  \apj, 807, 183

\bibitem[{{Madhusudhan} {et~al.}(2016){Madhusudhan}, {Ag{\'u}ndez}, {Moses}, \&
  {Hu}}]{madhu16}
{Madhusudhan}, N., {Ag{\'u}ndez}, M., {Moses}, J.~I., \& {Hu}, Y. 2016, \ssr,
  205, 285

\bibitem[{{Marley} {et~al.}(2013){Marley}, {Ackerman}, {Cuzzi}, \&
  {Kitzmann}}]{marley13}
{Marley}, M.~S., {Ackerman}, A.~S., {Cuzzi}, J.~N., \& {Kitzmann}, D. {Clouds
  and Hazes in Exoplanet Atmospheres}, ed. S.~J. {Mackwell}, A.~A.
  {Simon-Miller}, J.~W. {Harder}, \& M.~A. {Bullock} (The University of Arizona
  Press), 367--391

\bibitem[{{Marley} \& {Robinson}(2015)}]{marley15}
{Marley}, M.~S. \& {Robinson}, T.~D. 2015, \araa, 53, 279

\bibitem[{{Marley} {et~al.}(2012){Marley}, {Saumon}, {Cushing}, {Ackerman},
  {Fortney}, \& {Freedman}}]{marley12}
{Marley}, M.~S., {Saumon}, D., {Cushing}, M., {Ackerman}, A.~S., {Fortney},
  J.~J., \& {Freedman}, R. 2012, \apj, 754, 135

\bibitem[{{Marley} {et~al.}(1996){Marley}, {Saumon}, {Guillot}, {Freedman},
  {Hubbard}, {Burrows}, \& {Lunine}}]{marley96}
{Marley}, M.~S., {Saumon}, D., {Guillot}, T., {Freedman}, R.~S., {Hubbard},
  W.~B., {Burrows}, A., \& {Lunine}, J.~I. 1996, Science, 272, 1919

\bibitem[{{Marley} {et~al.}(2018){Marley}, {Saumon}, {Morley}, \&
  {Fortney}}]{marley18}
{Marley}, M.~S., {Saumon}, D., {Morley}, C., \& {Fortney}, J.~J. 2018, {Sonora
  2018: Cloud-free, solar composition, solar C/O substellar atmosphere models
  and spectra [Data set]. Zenodo.},
  \url{http://doi.org/10.5281/zenodo.1309035}

\bibitem[{{Marley} {et~al.}(2021){Marley}, {Saumon}, {Visscher}, {Lupu},
  {Freedman}, {Morley}, {Fortney}, {Seay}, {Smith}, {Teal}, \&
  {Wang}}]{marley19}
{Marley}, M.~S., {Saumon}, D., {Visscher}, C., {Lupu}, R., {Freedman}, R.,
  {Morley}, M., {Fortney}, J.~J., {Seay}, C., {Smith}, A.~J.~R.~W., {Teal},
  D.~J., \& {Wang}, R. 2021, ApJ, accepted July 2021

\bibitem[{{Marley} {et~al.}(2002){Marley}, {Seager}, {Saumon}, {Lodders},
  {Ackerman}, {Freedman}, \& {Fan}}]{marley02}
{Marley}, M.~S., {Seager}, S., {Saumon}, D., {Lodders}, K., {Ackerman}, A.~S.,
  {Freedman}, R.~S., \& {Fan}, X. 2002, ApJ, 568, 335

\bibitem[{{Miles} {et~al.}(2020){Miles}, {Skemer}, {Morley}, {Marley},
  {Fortney}, {Allers}, {Faherty}, {Geballe}, {Visscher}, {Schneider}, {Lupu},
  {Freedman}, \& {Bjoraker}}]{miles20}
{Miles}, B.~E., {Skemer}, A. J.~I., {Morley}, C.~V., {Marley}, M.~S.,
  {Fortney}, J.~J., {Allers}, K.~N., {Faherty}, J.~K., {Geballe}, T.~R.,
  {Visscher}, C., {Schneider}, A.~C., {Lupu}, R., {Freedman}, R.~S., \&
  {Bjoraker}, G.~L. 2020, \aj, 160, 63

\bibitem[{{Morley} {et~al.}(2012){Morley}, {Fortney}, {Marley}, {Visscher},
  {Saumon}, \& {Leggett}}]{morley12}
{Morley}, C.~V., {Fortney}, J.~J., {Marley}, M.~S., {Visscher}, C., {Saumon},
  D., \& {Leggett}, S.~K. 2012, \apj, 756, 172

\bibitem[{Morley {et~al.}(2015)Morley, Fortney, Marley, Zahnle, Line, Kempton,
  Lewis, \& Cahoy}]{morley15}
Morley, C.~V., Fortney, J.~J., Marley, M.~S., Zahnle, K., Line, M., Kempton,
  E., Lewis, N., \& Cahoy, K. 2015, \apj, 815, 110

\bibitem[{{Morley et al}(\noop{3001}in preparation)}]{morley19}
{Morley et al}. \noop{3001}in preparation, to be submitted

\bibitem[{{Moses} {et~al.}(2016){Moses}, {Marley}, {Zahnle}, {Line}, {Fortney},
  {Barman}, {Visscher}, {Lewis}, \& {Wolff}}]{moses2016}
{Moses}, J.~I., {Marley}, M.~S., {Zahnle}, K., {Line}, M.~R., {Fortney}, J.~J.,
  {Barman}, T.~S., {Visscher}, C., {Lewis}, N.~K., \& {Wolff}, M.~J. 2016,
  \apj, 829, 66

\bibitem[{{Moses} {et~al.}(2011){Moses}, {Visscher}, {Fortney}, {Showman},
  {Lewis}, {Griffith}, {Klippenstein}, {Shabram}, {Friedson}, {Marley}, \&
  {Freedman}}]{moses11}
{Moses}, J.~I., {Visscher}, C., {Fortney}, J.~J., {Showman}, A.~P., {Lewis},
  N.~K., {Griffith}, C.~A., {Klippenstein}, S.~J., {Shabram}, M., {Friedson},
  A.~J., {Marley}, M.~S., \& {Freedman}, R.~S. 2011, \apj, 737, 15

\bibitem[{{Noll} {et~al.}(1997){Noll}, {Geballe}, \& {Marley}}]{noll97}
{Noll}, K.~S., {Geballe}, T.~R., \& {Marley}, M.~S. 1997, \apjl, 489, L87

\bibitem[{{Oppenheimer} {et~al.}(2013){Oppenheimer}, {Baranec}, {Beichman},
  {Brenner}, {Burruss}, {Cady}, {Crepp}, {Dekany}, {Fergus}, {Hale},
  {Hillenbrand}, {Hinkley}, {Hogg}, {King}, {Ligon}, {Lockhart}, {Nilsson},
  {Parry}, {Pueyo}, {Rice}, {Roberts}, {Roberts}, {Shao}, {Sivaramakrishnan},
  {Soummer}, {Truong}, {Vasisht}, {Veicht}, {Vescelus}, {Wallace}, {Zhai}, \&
  {Zimmerman}}]{oppenheimer13}
{Oppenheimer}, B.~R., {Baranec}, C., {Beichman}, C., {Brenner}, D., {Burruss},
  R., {Cady}, E., {Crepp}, J.~R., {Dekany}, R., {Fergus}, R., {Hale}, D.,
  {Hillenbrand}, L., {Hinkley}, S., {Hogg}, D.~W., {King}, D., {Ligon}, E.~R.,
  {Lockhart}, T., {Nilsson}, R., {Parry}, I.~R., {Pueyo}, L., {Rice}, E.,
  {Roberts}, J.~E., {Roberts}, L.~C., J., {Shao}, M., {Sivaramakrishnan}, A.,
  {Soummer}, R., {Truong}, T., {Vasisht}, G., {Veicht}, A., {Vescelus}, F.,
  {Wallace}, J.~K., {Zhai}, C., \& {Zimmerman}, N. 2013, \apj, 768, 24

\bibitem[{{Phillips} {et~al.}(2020){Phillips}, {Tremblin}, {Baraffe},
  {Chabrier}, {Allard}, {Spiegelman}, {Goyal}, {Drummond}, \&
  {H{\'e}brard}}]{phillips2020}
{Phillips}, M.~W., {Tremblin}, P., {Baraffe}, I., {Chabrier}, G., {Allard},
  N.~F., {Spiegelman}, F., {Goyal}, J.~M., {Drummond}, B., \& {H{\'e}brard}, E.
  2020, A\&A, 637, A38

\bibitem[{{Prinn} \& {Barshay}(1977)}]{prinn77}
{Prinn}, R.~G. \& {Barshay}, S.~S. 1977, Science, 198, 1031

\bibitem[{{Saumon} {et~al.}(2006){Saumon}, {Marley}, {Cushing}, {Leggett},
  {Roellig}, {Lodders}, \& {Freedman}}]{saumon06}
{Saumon}, D., {Marley}, M.~S., {Cushing}, M.~C., {Leggett}, S.~K., {Roellig},
  T.~L., {Lodders}, K., \& {Freedman}, R.~S. 2006, \apj, 647, 552

\bibitem[{{Saumon} {et~al.}(2007){Saumon}, {Marley}, {Leggett}, {Geballe},
  {Stephens}, {Golimowski}, {Cushing}, {Fan}, {Rayner}, {Lodders}, \&
  {Freedman}}]{saumon07}
{Saumon}, D., {Marley}, M.~S., {Leggett}, S.~K., {Geballe}, T.~R., {Stephens},
  D., {Golimowski}, D.~A., {Cushing}, M.~C., {Fan}, X., {Rayner}, J.~T.,
  {Lodders}, K., \& {Freedman}, R.~S. 2007, \apj, 656, 1136

\bibitem[{{Skemer} {et~al.}(2014){Skemer}, {Marley}, {Hinz}, {Morzinski},
  {Skrutskie}, {Leisenring}, {Close}, {Saumon}, {Bailey}, {Briguglio},
  {Defrere}, {Esposito}, {Follette}, {Hill}, {Males}, {Puglisi}, {Rodigas}, \&
  {Xompero}}]{skemer14}
{Skemer}, A.~J., {Marley}, M.~S., {Hinz}, P.~M., {Morzinski}, K.~M.,
  {Skrutskie}, M.~F., {Leisenring}, J.~M., {Close}, L.~M., {Saumon}, D.,
  {Bailey}, V.~P., {Briguglio}, R., {Defrere}, D., {Esposito}, S., {Follette},
  K.~B., {Hill}, J.~M., {Males}, J.~R., {Puglisi}, A., {Rodigas}, T.~J., \&
  {Xompero}, M. 2014, \apj, 792, 17

\bibitem[{{Tsai} {et~al.}(2018){Tsai}, {Kitzmann}, {Lyons}, {Mendon{\c{c}}a},
  {Grimm}, \& {Heng}}]{tsai2018}
{Tsai}, S.-M., {Kitzmann}, D., {Lyons}, J.~R., {Mendon{\c{c}}a}, J., {Grimm},
  S.~L., \& {Heng}, K. 2018, \apj, 862, 31

\bibitem[{Visscher(2012)}]{visscher2012}
Visscher, C. 2012, The Astrophysical Journal, 757, 5

\bibitem[{{Visscher} {et~al.}(2006){Visscher}, {Lodders}, \&
  {Fegley}}]{visscher2006}
{Visscher}, C., {Lodders}, K., \& {Fegley}, Bruce, J. 2006, \apj, 648, 1181

\bibitem[{{Visscher} \& {Moses}(2011)}]{visscher11}
{Visscher}, C. \& {Moses}, J.~I. 2011, \apj, 738, 72

\bibitem[{{Wang} {et~al.}(2014){Wang}, {Gierasch}, {Lunine}, \&
  {Mousis}}]{dong2014}
{Wang}, D., {Gierasch}, P., {Lunine}, J., \& {Mousis}, O. 2014, in AAS/Division
  for Planetary Sciences Meeting Abstracts, Vol.~46, AAS/Division for Planetary
  Sciences Meeting Abstracts \#46, 512.03

\bibitem[{{Yang} {et~al.}(2016){Yang}, {Apai}, {Marley}, {Karalidi}, {Flateau},
  {Showman}, {Metchev}, {Buenzli}, {Radigan}, {Artigau}, {Lowrance}, \&
  {Burgasser}}]{yang16}
{Yang}, H., {Apai}, D., {Marley}, M.~S., {Karalidi}, T., {Flateau}, D.,
  {Showman}, A.~P., {Metchev}, S., {Buenzli}, E., {Radigan}, J., {Artigau},
  {\'E}., {Lowrance}, P.~J., \& {Burgasser}, A.~J. 2016, ApJ, 826, 8

\bibitem[{{Zahnle} \& {Marley}(2014)}]{zahnle14}
{Zahnle}, K.~J. \& {Marley}, M.~S. 2014, \apj, 797, 41

\bibitem[{{Zhou} {et~al.}(2016){Zhou}, {Apai}, {Schneider}, {Marley}, \&
  {Showman}}]{zhou16}
{Zhou}, Y., {Apai}, D., {Schneider}, G.~H., {Marley}, M.~S., \& {Showman},
  A.~P. 2016, ApJ, 818, 176

\end{thebibliography}
\end{document}